\documentclass[manuscript,nonacm]{acmart}

\usepackage[linesnumbered]{algorithm2e}
\usepackage{afterpage}
\usepackage{enumitem}  

\usepackage{subcaption}

\AtBeginDocument{%
  }

\begin{document}

\title{Evaluating the Impact of Packet Scheduling and Congestion Control Algorithms on MPTCP Performance over Heterogeneous Networks}

\author{Dimitrios Dimopoulos}
\orcid{0000-0002-3722-2825}
\affiliation{%
  \institution{Lenovo Research}
  \city{Athens}
  \country{Greece}
}
\affiliation{%
  \institution{Department of Informatics and Telecommunications, National and Kapodistrian University of Athens}
  \city{Ilisia}
  \state{Athens}
  \country{Greece}
}
\email{ddimopoulos@lenovo.com}
\email{ddimopoulos@di.uoa.gr}

\author{Apostolis K. Salkintzis}
\orcid{0000-0002-8147-3285}
\affiliation{%
  \institution{Lenovo Research}
  \city{Athens}
  \country{Greece}
}
\email{salki@motorola.com}

\author{Dimitris Tsolkas}
\orcid{0000-0003-0301-3950}  
\affiliation{%
  \institution{Department of Informatics and Telecommunications, National and Kapodistrian University of Athens}
  \city{Ilisia}
  \state{Athens}
  \country{Greece}
}
\email{dtsolkas@di.uoa.gr}

\author{Nikos Passas}
\orcid{0000-0002-2112-899X}
\affiliation{%
  \institution{Department of Informatics and Telecommunications, National and Kapodistrian University of Athens}
  \city{Ilisia}
  \state{Athens}
  \country{Greece}
}
\email{passas@di.uoa.gr}

\author{Lazaros Merakos}
\orcid{0000-0003-4822-2393}
\affiliation{%
  \institution{Department of Informatics and Telecommunications, National and Kapodistrian University of Athens}
  \city{Ilisia}
  \state{Athens}
  \country{Greece}
}
\email{merakos@di.uoa.gr}

\renewcommand{\shortauthors}{D. Dimopoulos, et al.}

\begin{abstract}
  Modern mobile and stationary devices are equipped with multiple network interfaces aiming to provide wireless and wireline connectivity either in a local LAN or the Internet. Multipath TCP (MPTCP) protocol has been developed on top of legacy TCP to allow the simultaneous use of multiple network paths in the communication route between two end-systems. Although the combination of multiple paths is beneficial in case of links with similar network characteristics, MPTCP performance is challenged as heterogeneity among the used paths increases. This work provides an overview of the MPTCP protocol operation, analyzes the state-of-art packet scheduling and congestion control algorithms available in literature, and examines the impact of the various algorithm combinations on MPTCP performance, by conducting an extensive experimental evaluation under diverse path-heterogeneity conditions.
\end{abstract}





\maketitle



\section{Introduction}
Modern digital devices, from laptops and cell phones to large data centers, are usually equipped with multiple network interfaces (e.g., Ethernet, WiFi, LTE/5G, optical, Bluetooth, etc.). Multipath TCP (MPTCP) protocol \cite{rfc8684}, which builds upon and extends standard TCP, emerged from the necessity to provide data communication services over multiple paths, by jointly using the available network interfaces of a multi-homed device. Although new protocols arise (e.g, \textit{QUIC} \cite{rfc9000, rfc9369} along with its multipath extension \textit{MPQUIC} \cite{ietf-quic-multipath-16}), TCP is still the predominant protocol for critical applications, as well as the default standard in all operating systems. MPTCP enables the exchange of data traffic over multiple disjoint or common-bottleneck links allowing the communicating hosts to reach specific application goals; be it higher data rate, lower flow completion time, or resilience in case of link failure. To this end, MPTCP provides a clear benefit to the application between two communicating systems. However, the simultaneous use of paths with inherently different network characteristics poses a challenge on MPTCP's performance. In fact, network links are affected either by the constantly fluctuating demand in shared network resources (e.g., in the Internet) which often leads to network congestion, increased intermediate buffer queues and packet drops, or by the nature of the physical medium (e.g., the 5G wireless interface) which is susceptible to random packet loss. The presence of path heterogeneity, in conjunction with MPTCP's adherence to TCP principles for reliable and in-order bytestream delivery, may lead to head-of-line (HoL) blocking at the receiver, stalling the connection and degrading the overall performance. In an attempt to tackle the functional and performance issues arising by the presence of heterogeneous paths, research efforts focus on providing enhanced congestion control and optimal packet scheduling algorithms that increase the overall performance and are tailored to application demands. In this direction, we hereby provide a comprehensible guide on MPTCP protocol structure and functionality, analyze the theoretical aspects behind the operation of the state-of-art packet scheduling and congestion control algorithms, construct and present the detailed algorithmic steps in pseudo-code, and finally, perform an extensive performance assessment of the algorithms' combined operation under homogeneous and asymmetric path conditions. The remainder of the paper is structured as follows. Section \ref{sec:background} provides the necessary background on MPTCP protocol aspects and analyzes the best to date packet scheduling and congestion control algorithms. Going beyond theory, an extensive experimental evaluation has been conducted, providing insight on how MPTCP performs under different packet scheduling and congestion control mechanism combinations. To this end, Section \ref{sec:exper_method} describes the experimentation methodology, while Section \ref{sec:evaluation} presents the produced results. Section \ref{sec:discussion} discusses the performance results and provides insight on the best-performing congestion control and packet scheduling schemes, while Section \ref{sec:related_work} is dedicated to the corresponding works available in literature. Finally, Section \ref{sec:conclusion} concludes the manuscript.

\section{Background}
\label{sec:background}
This section is dedicated in describing the fundamental aspects of MPTCP, such as the protocol structure, the architectural details, as well as foundational knowledge behind the most salient modules comprising and dictating the operation of MPTCP; the Path Manager, the Packet Scheduler, and the Congestion Controller. Afterwards, the main logic, as well as the algorithmic steps of the state-of-art packet scheduling and congestion control algorithms are presented.

\subsection{MPTCP Theory}
\label{subsec:mptcp_theory}
\subsubsection{General principles}
\label{subsub:general_principles}
MPTCP has been designed on top of TCP, and as such, shares and extends much of its principles and structural details. The intention behind its development was the design of a protocol which provides the traditional single-path TCP transport layer services, such as connection-oriented, reliable, and in-order data delivery, while being able to simultaneously use multiple paths for the transmission of data segments. MPTCP is thus extending traditional TCP, and its operation is determined by specific attributes present within TCP header options; for instance, the TCP option-kind number \textit{30} has been reserved by IANA to indicate whether the TCP header options carry additional MPTCP related fields. While advancing the legacy TCP functionality, it was also fundamental for MPTCP to be able to fall back to TCP, if required. Thus, the new protocol should not incur any modifications to the overlay application which, as with traditional TCP, can establish a MPTCP connection through a single socket. Despite the use of multiple paths in lower layers, each socket supports a single connection between the application and the transport layer. Applications may continue using the existing TCP socket API for MPTCP connections; in that case MPTCP specific parameters are configured in the operating system. However, a new MPTCP socket is also available allowing applications to control MPTCP behavior via socket parameters.

\subsubsection{Subflows}
\label{subsub:subflows}
The MPTCP's foundational principle is the ability to route application traffic over multiple paths, and yet enable seamless communication between two end-systems, as if it were for a single-path TCP connection. The underlying paths carrying data packets, can be either distinct end-to-end physical links (i.e., "disjoint" paths), or paths which at some point share a common bottleneck link (e.g., "partially disjoint"). As with standard TCP, different paths are uniquely identified by a 4-tuple (source \& destination address/port pair). Each data flow carried on an individual path which belongs to the same MPTCP connection, is called \textit{subflow}. Each subflow resembles a standard single-path TCP connection; subflow connection establishment and termination follows the typical single-path TCP procedures. Consequently, a MPTCP connection can comprise one or more subflows, ultimately enabling application communication between two end-hosts over multiple paths. From application's perspective, there is only a single socket bind and accept call to the MPTCP connection. Each of the communicating multipath-capable hosts generates a token, based on the keys exchanged during the initial MPTCP connection initiation procedure, which uniquely identifies each MPTCP connection. Then, each time a new subflow needs to be created and be associated to an existing MPTCP connection, each host binds the 5-tuple of the TCP subflow to the local token of the connection. This allows any port pairs to be used for a connection, eliminating in essence the identification of a subflow based on the 4-tuple, and relying merely on the token, instead \cite{rfc8684}.

\subsubsection{Multipath-capable hosts}
\label{subsub:mpcapable_hosts}
The establishment of a multipath TCP connection requires that both communicating hosts have multipath capability. MPTCP is already supported by operating systems which base their function on recent versions of the linux kernel \cite{LinuxKernel:LK}. End-hosts negotiate their multipath capabilities (\textit{MP\_CAPABLE}) over a typical 3-way handshake, before being able to establish a MPTCP connection. If one of the hosts is not multipath-capable, or in case middleboxes intervene altering either the TCP header options and removing the MP\_CAPABLE field or the payload itself, or in case of MPTCP version mismatch, the connection falls back to regular TCP. This allows the establishment of a standard TCP connection, absent any manual intervention. 

\subsubsection{Meta- vs. Subflow-level}
\label{subsub:meta_sf_level}
In case of MPTCP, transport layer is logically segregated into two parts; the MPTCP "Meta-level" (or "Connection-level"), and the MPTCP "subflow-level". The MPTCP meta-level is responsible for maintaining a single connection point to the application layer, thus MPTCP is still perceived as a single-path TCP connection from application's point of view. The meta-level has all those mechanisms required to allow the presence of a single connection to the upper layer, yet able to distribute (or reassemble) application bytestream across (or from) multiple subflows. At subflow-level, each subflow is essentially handled as an individual single-path TCP connection between end-hosts.

\subsubsection{Protocol Structure}
\label{subsub:protocol_structure}
\begin{figure}[h]
\centering
\begin{minipage}[t]{.3\textwidth}
  \centering
  \includegraphics[width=\linewidth]{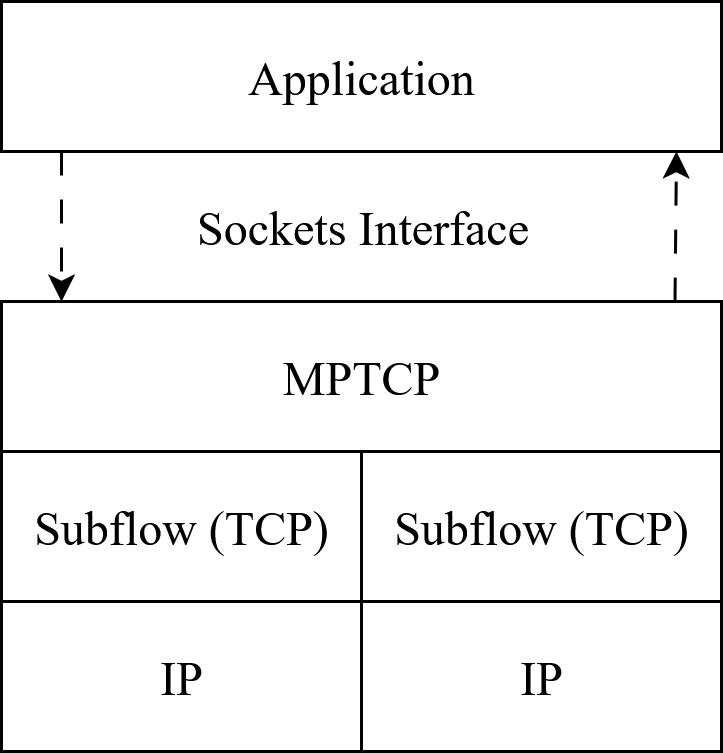}
  \captionof{figure}{MPTCP protocol stack (Source: IETF RFCs 8684 \cite{rfc8684}, 6897 \cite{rfc6897})}
  \label{fig:MPTCP_prot_stack}
  \Description[MPTCP protocol stack]{This figure depicts the MPTCP protocol stack, as defined within IETF RFCs 8684 and 6897.}
\end{minipage}
\hfill
\begin{minipage}[t]{.5\textwidth}
  \centering
  \includegraphics[width=\linewidth]{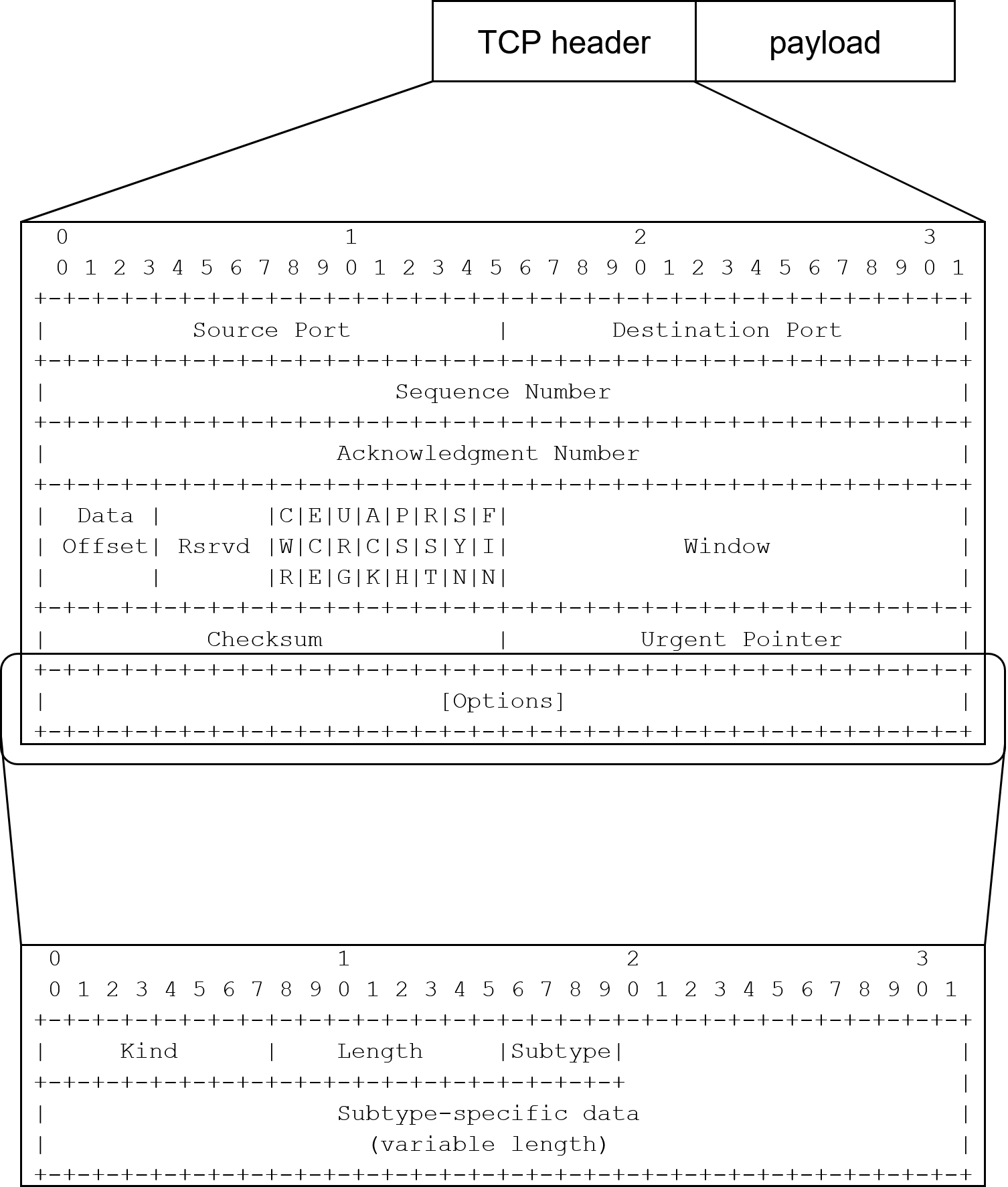}
  \captionof{figure}{MPTCP protocol header structure - MPTCP header is included into TCP header "Options" field}
  \label{fig:MPTCP_header}
  \Description[MPTCP protocol header structure]{This figure depicts the generic MPTCP protocol header structure. The figure illustrates how the MPTCP header is included as part of the standard TCP header's "Options" field.}
\end{minipage}
\end{figure}

Building upon and extending traditional TCP, MPTCP protocol operations are defined within TCP header "Options" field. There are different MPTCP fields set within TCP header "Options" field depending on the various MPTCP operations; i.e., \textit{MP\_CAPABLE} option is used during MPTCP connection initiation, the \textit{MP\_JOIN} option is used to associate additional subflows with an existing MPTCP connection, the \textit{ADD\_ADDR} option denotes the address advertisement operation which announces the availability of new interfaces at end-hosts, the Data Sequence Signal (\textit{DSS}) option carries the Data Sequence Mapping, etc. A general overview of the MPTCP protocol stack as well as the respective protocol structure, is depicted in Figures \ref{fig:MPTCP_prot_stack} and \ref{fig:MPTCP_header}, respectively.

\subsubsection{Path Manager}
\label{subsub:pm}
As mentioned earlier, each MPTCP subflow is uniquely identified by a 4-tuple (source \& destination address/port pair). The MPTCP \textit{Path Manager} module is responsible for advertising the availability or elimination of an IP addresses/ port pair of a multipath-capable host, for establishing and removing subflows between two end-systems, as well as for maintaining the MPTCP connection state. The MPTCP connection state includes details of the active and standby subflows constituting each MPTCP connection, and maps each subflow's 4-tuple to an \textit{Address ID}; the latter is generated and announced by a source host during MP\_JOIN or ADD\_ADDR operation to allow the receiver to refer to a source IP address/port pair implicitly, in case NAT protocol or other middleboxes conceal the actual source details.

\subsubsection{Packet Scheduler}
\label{subsub:ps}
The \textit{Packet Scheduling} functionality \cite{rfc6182} at the transmitting end is responsible for receiving the application bytestream, dividing application data into transport layer segments with appropriate connection-level sequence numbers, and transmitting the segments onto the available subflows according to the scheduler's own policy. The availability of the subflows is exposed to the scheduler by the path manager. At the receiving side, a corresponding functionality is in charge of appropriately ordering the received segments based on the connection-level sequence numbers, and delivering them in a bytestream fashion to the receiving application.

\subsubsection{Flow control}
\label{subsub:fc}
MPTCP inherits and extends standard TCP mechanisms. \textit{Flow control} guarantees that the transmitted application data will reach the receiving end in-order and flawlessly, while it also allows the receiver to regulate the sender's transmitting rate based on the receiving application processing capabilities; the transmitting rate is essentially controlled via the \textit{receive window (rwnd)} which is announced by the receiver to the sending host at regular intervals (i.e., with each cumulative ACK in TCP). The typical TCP flow control is maintained at subflow-level, where each individual subflow maintains its own \textit{sequence} and \textit{acknowledgment} numbers. However, the existing TCP flow control mechanism is not sufficient since although received packets may end up reassembled in correct order at subflow-level, they may not necessarily be appropriately ordered at MPTCP connection-level due to heterogeneous path conditions or retransmissions. For this reason, besides the subflow-specific sequence and acknowledgment numbers, flow control mechanism is extended at connection-level. \textit{Data Sequence Number (DSN)} and \textit{Data ACK} enable the appropriate \textit{Data Sequence Mapping} from the subflow sequence space to the data sequence space \cite{rfc8684}. This extended functionality allows for a cumulative data tracking at connection-level and ensures in-order delivery to the application.

\subsubsection{Congestion Control - the Reno approach}
\label{subsub:cca_reno}
The \textit{Congestion control} mechanism has been developed in the context of TCP \cite{rfc5681, rfc2914} to regulate the  transmission rate of each sender in an attempt to protect networks, and subsequently the Internet, from \textit{congestion collapse}. Since then, various congestion control algorithms have emerged, with CUBIC being the one most broadly utilized by Internet hosts, at the time of this publication. The remainder of this subsection provides the foundational aspects of congestion control as defined by the legacy TCP-Reno \cite{rfc5681}; CUBIC \cite{rfc9438} is analyzed in a dedicated subsection below \ref{subsub:cubic}. TCP's congestion control mechanism comprises the following algorithms (or "phases"): \textit{slow start}, \textit{congestion avoidance}, \textit{fast retransmit}, and \textit{fast recovery}; the first two are responsible for regulating the amount of data each TCP sender can inject into the network at any given moment, while the last two determine the way congestion control reacts to potential loss. Three important state variables influence the function of these algorithms: i) the congestion window (cwnd), which is essentially a sender-side limit on the amount of data that can be transmitted and, consequently be in-flight, at any moment, ii) the receive window (rwnd), which is part of flow control mechanism and constitutes a receiver-side limit on the amount of data the receiver can accept at any given time, and iii) the slow start threshold (ssthresh) which dictates whether congestion control operates according to slow start or congestion avoidance algorithm.
At any given moment, TCP must send no more than $min(cwnd, rwnd)$ data beyond the highest acknowledged sequence number; this formula guarantees that the amount of data transmitted at any instant neither incurs network congestion nor overloads the receive buffer. At any given time,the amount of bytes the sender is allowed to send next, is defined according to below formula:
\begin{equation} \label{eq:1}
 \#\_of\_bytes\_next = min(cwnd, rwnd) - FlightSize
\end{equation}
where \textit{FlightSize} is the amount of data in-flight, or in other words, the amount of outstanding data in the network.

It is also recommended that the initial value of ssthresh be set arbitrarily high to allow network conditions, rather than host limitations, dictate the sending rate. Nevertheless, sshthresh is adjusted to lower values in case congestion is detected. The selection of the algorithm to be employed by the congestion control mechanism, is based on the relation between the slow start threshold and the congestion window:
\begin{itemize}
    \item if $cwnd < ssthresh$, then \textit{slow start} algorithm is employed
    \item if $cwnd > ssthresh$, then \textit{congestion avoidance} algorithm is employed
    \item if $cwnd == ssthresh$, then either \textit{slow start} or \textit{congestion avoidance} algorithm is selected
\end{itemize}

\paragraph{Slow start} Slow start algorithm \cite{rfc5681} is executed at the beginning of a data transfer after a TCP connection has been successfully established, or when \textit{retransmission timeout (RTO)} timer elapses, indicating a loss. Slow start algorithm essentially allows sending-host's TCP to slowly probe the network to estimate network capacity, while avoiding network congestion induced by large data bursts. In slow start, the initial value of the congestion window immediately after the completion of the 3-way handshake is: 
\begin{itemize}
    \item if $SMSS > 2190\, bytes$:
    \begin{itemize}[label={}]
        \item $IW = 2 * SMSS\,bytes$ and NOT more than 2 segments
    \end{itemize}
    \item if $(SMSS > 1095 bytes)\, \&\&\, (SMSS <= 2190 bytes)$:
    \begin{itemize}[label={}]
        \item $IW = 3 * SMSS\, bytes$ and NOT more than 3 segments
    \end{itemize}
    \item if $SMSS <= 1095\, bytes$:
    \begin{itemize}[label={}]
        \item $IW = 4 * SMSS\, bytes$ and NOT more than 4 segments
    \end{itemize}
\end{itemize}

where $SMSS$ is the \textit{Sender Maximum Segment Size}, and $IW$ denotes the \textit{Initial Congestion Window}. During slow start, TCP increments the congestion window by at most $SMSS$ bytes for each ACK received that cumulatively acknowledges new data: 
\begin{equation} \label{eq:2}
    cwnd += SMSS
\end{equation}
A more conservative, yet secure against "ACK Division" approach \cite{savage1999tcp}, recommends that TCP increment congestion window according to the actual amount of bytes corresponding to new data, that are successfully acknowledged by the receiver; this transforms \ref{eq:2} to a more precise formula: 
\begin{equation} \label{eq:3}
    cwnd += min (N, SMSS)
\end{equation}
where \textit{N} is the number of previously unacknowledged bytes acknowledged in the incoming ACK. Incrementing the congestion window by at most one SMSS for every received ACK acknowledging new data, leads to an exponential growth of the congestion window, until the point it reaches slow-start threshold (\textit{ssthresh}); from that point onward, \textit{congestion avoidance} algorithm controls data transmission, provided that no loss is encountered.

\paragraph{Congestion avoidance} \cite{rfc5681} When $cwnd \ge sshthresh$, congestion control enters the congestion avoidance phase. During congestion avoidance, it is recommended that congestion window be incremented proportionally to the number of bytes acknowledged for new data, but no more than \textit{SMSS} bytes per RTT, as defined in equation \ref{eq:4}: 
\begin{equation} \label{eq:4}
    cwnd\, +=\, SMSS*SMSS/cwnd
\end{equation}
Equation \ref{eq:4} provides a byte-unit calculation of the congestion window during CA phase. An alternative to this, is the full-size segment equivalent provided through \ref{eq:3}. However, of vital importance here is the fact that the increment takes place per RTT, and not per ACK as is the case in slow start mode.

\paragraph{Loss detection due to RTO expiration} \cite{rfc5681}
When a TCP sender presumes that a segment is lost, by means of RTO expiration, and the segment has not yet been resent for the same reason (i.e., RTO expiration), \textit{ssthresh} is adjusted based on the below equation: 
\begin{equation} \label{eq:5}
    ssthresh\, =\, max\left(\frac{FlightSize}{2},\, 2*SMSS\right)
\end{equation}
If the lost segment has previously been sent as part of retransmission timer expiration, \textit{ssthresh} value remains unmodified. In addition, upon \textit{RTO} expiration, congestion window should be set as follows: 
\begin{equation} \label{eq:6}
    cwnd\, =\, 1*SMSS
\end{equation}

At this point, the TCP sender is expected to retransmit the dropped segment and then, using the slow start algorithm, to increase the congestion window from one segment to the new value of \textit{ssthresh} \ref{eq:5}. Once \textit{sshthresh} is reached, the algorithm switches to \textit{congestion avoidance} phase \cite{rfc2988}.

\paragraph{Fast Retransmit \& Fast Recovery} \cite{rfc5681}
Fast retransmit algorithm exploits ACKs to detect and repair loss. The receiver should respond with a duplicate-ACK when an out-of-order segment arrives, and should continue responding with duplicate ACKs for every new segment arriving, other than the one still missing. Duplicate-ACKs can in principle be implying a couple of conditions: 

\begin{enumerate}[label=\roman*.]
    \item the segment has been dropped in transit to the destination
    \item the network might have re-ordered the segments
    \item the network replicates ACKs or data segments (i.e., triggered by intermediate devices in the network path between two end nodes)
\end{enumerate}

Upon reception of the missing segment, the receiver is expected to immediately respond with an ACK, indicating that the received segment fills in the gap in the sequence space. Upon reception of the first and second duplicate ACKs, the sender (unless SACK mechanism is enabled\footnote{A sender using SACK \cite{rfc2018} must not send new data unless the incoming duplicate acknowledgment contains new SACK information.}) is expected to send a segment of new data, provided that: i) \textit{rwnd} allows, ii)  $FlightSize + min(N, SMSS) \le cwnd + 2*SMSS$, and iii) new data is available for transmission. However, \textit{cwnd} does not yet change, avoiding to reflect the new segments \cite{rfc3042}.

The reception of three duplicate ACKs denotes that a particular segment has potentially been lost. In that case, the sender sets \textit{ssthresh} according to equation \ref{eq:5}, retransmits immediately the missing segment without waiting for the RTO to expire, and sets \textit{cwnd} according to the below formula:
\begin{equation} \label{eq:7}
    cwnd = ssthresh + 3*SMSS
\end{equation}
 This immediate response to the detected loss is termed \textit{fast retransmit}. Equation \ref{eq:7} indicates that the congestion window is artificially inflated by the number of segments (i.e., three) that have left the network and which the receiver has buffered. Notable here is the fact that congestion control does not switch to the slow start phase; this is reasonable, since the reception of duplicate ACKs, if not excessive for different segments, indicates that the network can still deliver segments despite the perceived congestion. As a result, the reception of duplicate ACKs is considered a \textit{transient} congestion, contrary to the expiration of RTO which indicates severe congestion and requests for more drastic measures, enforced by entering the slow start phase. 

Following the \textit{fast retransmit}, congestion control enters the \textit{fast recovery} phase, until a normal (i.e., non-duplicate) ACK arrives, indicating the recovery of the segment which was supposed to be missing. In \textit{fast recovery} phase, for each additional duplicate ACK received after the third, congestion window is incremented by one SMSS, that is  $cwnd += SMSS$, per dup-ACK receipt. This increase artificially inflates the congestion window in order to reflect the segment that has left the network. Based on this increase, the sender can send $1*SMSS$ bytes of previously unsent data, given that the normal requisites hold true (e.g., new value of \textit{rwnd} and \textit{cwnd} allow, and new data is available for transmission). Upon reception of an ACK that acknowledges new data, beyond the sequence range of the missing segment, loss is deemed repaired and TCP sets $cwnd = ssthresh$; beyond that point, \textit{congestion avoidance} algorithm takes over to regulate \textit{cwnd} increase.

\paragraph{Final remarks}
The aforementioned congestion control operation is also known as Additive Increase - Multiplicative Decrease (AIMD). The additive increase describes the behavior of the congestion control algorithm within the congestion avoidance phase, where the congestion window increases linearly by at most $1*SMSS$ bytes per RTT. The multiplicative decrease is referred to loss detection by means of three duplicate ACKs' receipt, where the congestion window is halved, and then continues again to increase until the next congestion event is detected. The initial implementation of TCP (i.e., \textit{TCP Tahoe} \cite{enwiki:1308091666, rfc6349}), included the first three phases of congestion control, that is, the \textit{slow-start}, \textit{congestion avoidance}, and \textit{fast recovery}. Whenever a loss was detected by way of receipt of three duplicate ACKs, congestion control was retransmitting the packet which was supposed to be lost, and then entered slow start phase. A subsequent update of the congestion control mechanism, known as \textit{TCP Reno} \cite{rfc5681}, introduced the \textit{fast recovery} phase, where the detection of loss via three duplicate ACKs does not require entering the slow start phase after \textit{fast retransmit}, thus optimizing the performance by remaining in the \textit{congestion avoidance} phase. The entire congestion control mechanism described earlier, is based on the \textit{TCP Reno} IETF standard; however, several optimizations took place since then (e.g., \textit{TCP NewReno} \cite{rfc6582} which optimized the \textit{fast recovery} phase, and the successor of them, \textit{CUBIC} \cite{rfc9438}). The following subsections describe in more detail how the legacy congestion control mechanism is operating in the context of MPTCP, as well as new algorithms and variations of the existing ones.

\subsubsection{MPTCP Implementation}
\label{subsub:mptcp_implemenation}
As mentioned earlier, in case of MPTCP, the application maintains a single socket for each individual MPTCP connection, regardless of the number of subflows associated with this connection. The life cycle of each individual subflow adheres to the legacy single-path TCP principles; consequently each subflow maintains the same mechanisms and operates similarly to a standard TCP connection. On the sending node, each subflow maintains its own subflow-level send window, as well as its own congestion and flow control mechanisms, which enable TCP-like communication with the receiving host at subflow-level. Thus, each subflow is individually responsible for the reliable and in-order delivery of the segments it carries. When a loss is detected, the subflow-level congestion control mechanism of the sending node undertakes the task of retransmitting the missing segment, while the receiving node ensures that the segments will be rearranged in the appropriate order at subflow-level. For this purpose, each individual subflow makes use of a dedicated out-of-order queue and a corresponding receive buffer.

However, as explained earlier as part of MPTCP flow control, even if it is ensured that each individual subflow delivers the segments carried over it reliably and in-order, it is not guaranteed that, when combining the ordered segments from multiple subflows, the result will be the original application bytestream. The MPTCP packet scheduler at the sending node could, for instance, split a data flow into segments traversing multiple subflows with heterogeneous path characteristics; then, at the receiving node, the in-order re-arrangement of the segments at subflow-level does not provide any guarantee for in-order arrangement \textit{across} subflows. This task is assigned to the MPTCP connection-level out-of-order queue which ensures the reassembly of segments arriving from different subflows, reconstructing essentially the original bytestream in the MPTCP receive buffer, and delivering it to the receiving application. This is why MPTCP makes use of a new sequence space; the \textit{Data Sequence Number (DSN)} and \textit{Data Acknowledgment (Data-ACK)} at connection-level complement the subflow-level standard TCP's \textit{sequence (SN)} and \textit{acknowledgment numbers (ACK)}. It is a prerequisite that each segment be acknowledged both at subflow- and connection-level before being passed to the receiving application's socket; only then is the sending node's connection-level send window able to slide further to the right, allowing the transmission of more data.

If a segment is never acknowledged on a specific subflow, a subflow-level timeout occurs indicating severe congestion or complete failure of that specific subflow. In that case, the segment is copied into the retransmission queue which is read on priority by the scheduler, who injects the missing segment to another subflow as soon as an opening in the congestion window permits. All segments residing within the shared retransmission queue have higher priority over the segments within the MPTCP send buffer; only when all segments from the retransmission queue are scheduled, thus the retransmission queue becomes empty, does the scheduler pull segments from the shared send buffer. 

\begin{figure}[h]
  \centering
  \includegraphics[width=\linewidth]{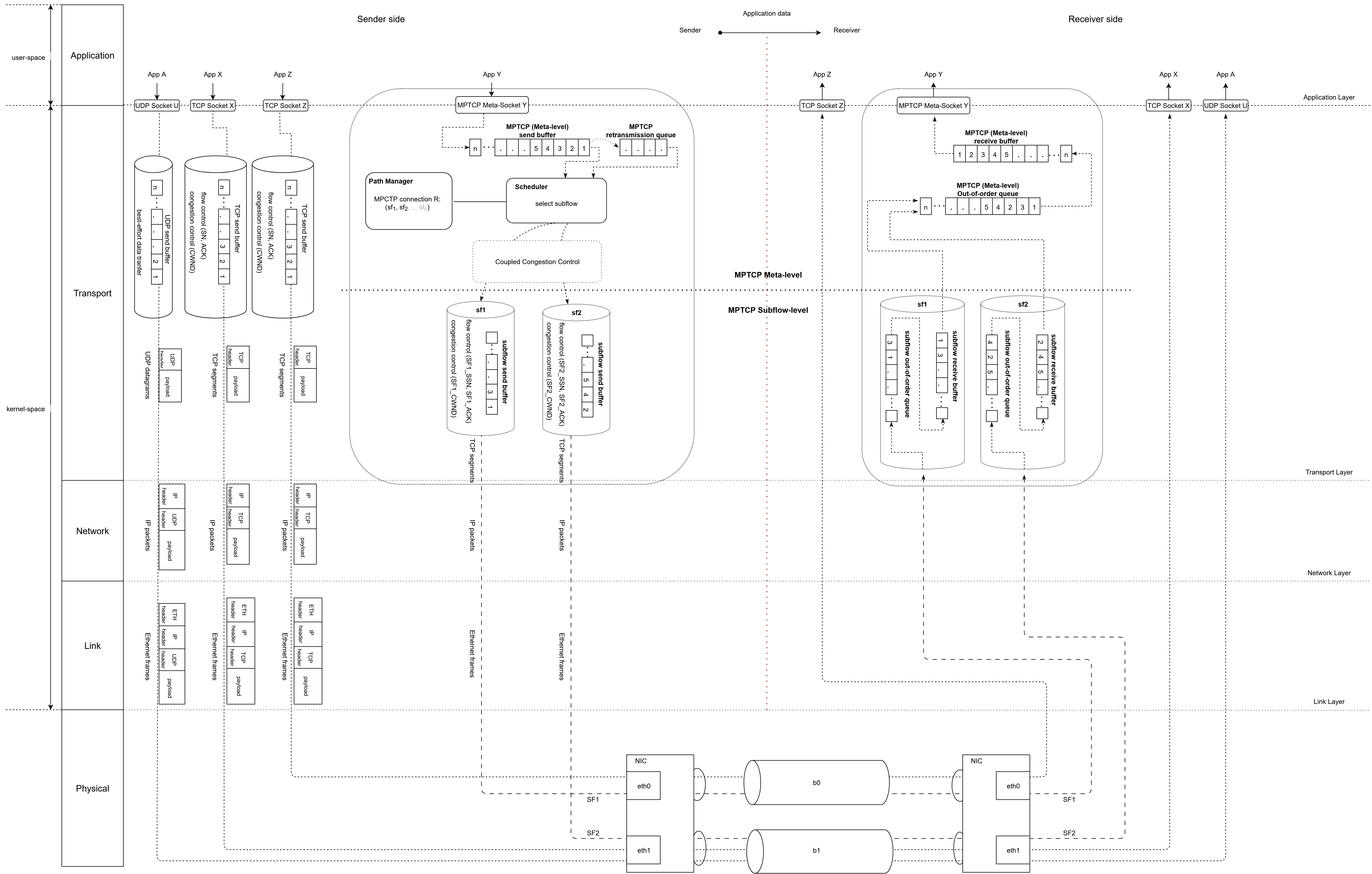}
  \caption{Indicative MPTCP service delivery model}
  \label{fig:MPTCP_service_delivery}
  \Description[MPTCP service delivery model]{This figure depicts the MPTCP service delivery model. It illustrates the internal MPTCP components, such as the \textit{path manager}, the \textit{packet scheduler}, the \textit{coupled congestion control} logical module, as well as the different buffers used on both sides of the MPTCP connection. It additionally presents the entire route followed by a segment belonging to a data flow of a MPTCP connection, between the sending and receiving application hosts.}
\end{figure}

A concise view of MPTCP protocol's end-to-end service delivery model, is depicted in Figure \ref{fig:MPTCP_service_delivery}. 


\subsection{Packet Scheduling Algorithms}
\label{subsec:psa}

Research efforts have yielded a multitude of packet scheduling algorithms, with each one being tailored to specific application demands. Besides the target application, packet schedulers can also be classified according to the inner techniques they employ to achieve their goals; be it heuristics or machine learning (ML) algorithms. This work seeks to present and assess the state-of-art \textit{heuristics-based} MPTCP packet scheduling algorithms available in literature, up to the time of this publication. A more exhaustive list of multipath-compatible packet schedulers, along with some classification options, can be studied through other works available in literature, references to which are provided throughout the current manuscript. A brief overview of the MPTCP packet schedulers investigated through this work, is provided in Table \ref{tab:psa}.

\begin{table*}[htb!]   
  \caption{MPTCP packet schedulers}
  \label{tab:psa}
  \begin{tabular}{lp{0.55in}p{0.5in}p{0.5in}p{3.3in}}  
    \toprule
    Scheduler & Full Name & Reference & Approach & Main concept\\
    \midrule
    \texttt{minRTT} & default & v0.96 \cite{raiciu2012hard, paasch2014experimental, githubminrtt, mptcporg, githubmptcp:v096} & reactive & This is the default MPTCP packet scheduler. Assigns segments to the subflow experiencing the lowest RTT, as long as CWND permits. \\
    \noalign{\hrule height 0.2pt}
    \texttt{BLEST} & Blocking Estimation & v0.96 \cite{ferlin2016blest, githubblest, mptcporg, githubmptcp:v096} & proactive & BLEST tries to estimate and eliminate HoL blocking, by assessing whether it is beneficial to schedule segments on a currently available subflow, or skip it, waiting for a more advantageous one. \\
    \noalign{\hrule height 0.2pt}
    \texttt{ECF} & Earliest \nobreak{Completion} First & v0.96 \cite{lim2017ecf, githubecf, mptcporg, githubmptcp:v096} & proactive & ECF prioritizes scheduling over the faster subflows, anticipating to achieve lowest flow completion time. ECF may elect not to use a slower path, in an attempt to increase utilization of the faster ones. \\
    \noalign{\hrule height 0.2pt}
    \texttt{RR} & roundrobin & v0.96 \cite{paasch2014experimental, githubrr, mptcporg, githubmptcp:v096} & blind (ack-clocked) & Experimental scheduler. RR transmits traffic in a round-robin fashion. Tunable configuration following either real round-robin scheduling or ACK-clocked behavior. \\
    \noalign{\hrule height 0.2pt}
    \texttt{LLHD} & Low \nobreak{Latency} and High Data Rate & v0.96 \cite{lubna2022low, githubllhd} & reactive & LLHD bases its decisions on the combined measurement of real-time throughput, delay, and packet loss rate, rather than relying on a single parameter. LLHD adjusts scheduling reactively according to the perceived network conditions. \\
    \noalign{\hrule height 0.2pt}
    \texttt{ReMP TCP} & redundant & v0.96 \cite{frommgen2016remp, mptcporg, githubmptcp:v096} & blind & Sends each data segment on all available subflows. It guarantees lowest per-packet latency and increased reliability, at the expense of limited goodput and higher flow completion time. \\

    \bottomrule
  \end{tabular}
\end{table*}

\subsubsection{MPTCP Default (minRTT)}
\label{subsub:minrtt}
minRTT is the default MPTCP packet scheduler. It selects the subflow with the lowest RTT which has space in its congestion window (\textit{cwnd}), to transmit the next segment(s). If the subflow experiencing the lowest RTT among the available subflows (i.e., among all subflows established and not marked as 'backup') does not have space in its cwnd, minRTT selects the subflow with the second lowest RTT that has sufficient cwnd. minRTT scheduler may optionally use some additional mechanisms to optimize its operation when the MPTCP connection is receive-window limited; that is, the receive window is blocked by one or more segments not yet acknowledged (e.g., missing segments or segments traversing a slow subflow), which are in turn causing HoL blocking at the receiving end. In such cases, \textit{opportunistic retransmission and penalization} \cite{raiciu2012hard} technique showcases some additional benefit. \textit{Opportunistic retransmission} mechanism is responsible for retransmitting a not-yet-acknowledged segment over a faster subflow, in anticipation for expedited delivery and unblocking of the receive window. In addition, the slow subflow is \textit{penalized} by halving its congestion window and setting \textit{ssthresh} to the new cwnd value. While those mechanisms provide some short-term advantage, their benefit is questionable in the long run; the reactive approach those mechanisms follow, limits artificially the congestion window of the slow path, causing it to increase slowly until the next HoL blocking occurrence. This continuous penalization approach has a detrimental effect on the slow subflows, the bandwidth of which is significantly underutilized.

The algorithmic steps of \textit{minRTT} packet scheduling algorithm are presented in detail in \ref{alg:minRTT}.

\subsubsection{BLocking ESTimation-based (BLEST)}
\label{subsub:blest}
BLEST adapts proactively its scheduling decisions by estimating whether injecting segments on a slow subflow would incur HoL blocking when the fast subflow becomes available again. It tries to limit, in essence, the amount of time a fast subflow remains underutilized due to receive window (and the mirror-effect of MPTCP send window) blocking. To achieve this, BLEST estimates the amount of data that would be sent on a fast subflow during the duration of a slow subflow (\textit{RTTs}). If this calculated amount of data is larger than the usable MPTCP send window \cite{rfc9293}, then potential scheduling on the slow subflow would block the entire MPTCP send window. Instead, BLEST selects to skip the currently available slow subflow, waiting for the faster one to become available again (i.e., by delivering some segments and opening some space in its congestion window). Using this core idea, BLEST tackles the HoL blocking issue which is present during scheduling across heterogeneous networks or paths, at the expense of underutilizing the slower paths in case their use is deemed detrimental in the long run. BLEST uses minRTT algorithm \ref{subsub:minrtt} to identify the subflow experiencing the lowest RTT at any given moment before applying its sophisticated scheduling logic. A correction factor lambda ($\lambda$) adjusts the scale of X (see algorithmic steps in \ref{alg:BLEST}), controlling essentially how strict or relaxed will the decision on selecting the slower path be; an increase in the value of $\lambda$ induced by a HoL blocking observation, incurs more stringent requirements on the selection of the slow subflow $x_{s}$.

The algorithmic steps of \textit{BLEST} packet scheduling algorithm are presented in detail in \ref{alg:BLEST}.

\subsubsection{Earliest Completion First (ECF)}
\label{subsub:ecf}

ECF aims to reduce the completion time of a data flow traversing heterogeneous paths. It achieves this by ensuring that the completion of a flow is not delayed by transmission over a slower path. Instead, ECF prioritizes scheduling on the fastest (in terms of RTT) path as long as it is available for sending; that is, having available space in its congestion window. Unlike minRTT, ECF may elect not to send on a slower path and rather wait for a faster one to become available, if this is estimated to be beneficial for the total flow completion time. Thus, a slower path can only be used as long as it does not delay the completion of the entire flow. The estimation of completion times is calculated using not only path RTTs, but also an estimation of paths' bandwidth incorporated within the congestion window, and the size of the connection-level send buffer. ECF uses a margin $\delta$ which incorporates the standard deviation of each path's RTT (denoted by $\sigma$) to compensate for RTT and CWND variabilities, a hysteresis constant $\beta$ which limits frequent switchover between subflows, as well as parameter \textit{k} which holds the number of bytes remaining to be sent in order to complete the transfer of the data flow.

The algorithmic steps of \textit{ECF} packet scheduling algorithm are presented in detail in \ref{alg:ECF}.

\subsubsection{Round-Robin (RR)}
\label{subsub:rr}

The round-robin scheduler selects the subflows to schedule segments on, in a round-robin fashion \cite{paasch2014experimental}. While this technique might offer load-balancing and equal use of the available subflow resources in case of homogeneous paths, it is detrimental in case of highly heterogeneous paths. Segments scheduled in a round-robin fashion over paths with different network characteristics will end up out-of-order at the receiving node, causing HoL blocking, limiting the receive window, and delaying the whole data flow completion. Such conditions have a domino effect, increasing retransmissions, and stalling the entire connection. RR scheduler's default operation is \textit{congestion-window limited}, that is, trying to fill in the entire congestion window of each subflow; as a result, RR becomes \textit{ack-clocked} \cite{jacobson1988congestion} in case of bulk data transfers, waiting, in essence, for ACKs to open up some space in the subflow's congestion window before the scheduler is able to inject new segments on that subflow again. However, the MPTCP development community has provided the option to tune RR's behavior through configurable parameters which control both the number of consecutive segments to be transmitted in each subflow before switching to another, as well as whether RR will leave some open space in the congestion window to retain its real round-robin scheduling scheme, avoiding the ack-clocked behavior \cite{mptcporg}. 

The algorithmic steps of \textit{RR} packet scheduling algorithm are presented in detail in \ref{alg:RR}.

\subsubsection{Low Latency and High Data Rate (LLHD)}
\label{subsub:llhd}

Instead of relying on a single parameter to devise the next scheduling action, LLHD seeks to combine all the available network parameters (i.e., delay, path loss, and available bandwidth) and adapt quickly to the constantly changing network conditions. For this purpose, LLHD tries to optimize a utility function $\gamma$, which embodies the various network parameters as follows: 

\begin{equation} \label{eq:8}
    \gamma=GP_{N} + \beta \times \frac{1}{RTT_{N}}
\end{equation}

where $\beta$ is a balancing factor, and $GP_{N}$\footnote{$Goodput = Throughput - Losses$} and $RTT_{N}$ are the normalized goodput and RTT, respectively, which are defined as follows:

$GP_{N} = \frac{GP}{GP_{max}}$, \, $RTT_{N} = \frac{RTT}{RTT_{max}}$

where $GP$ and $RTT$ are the goodput and RTT of $sf_{i}$, respectively; similarly, $GP_{max}$ and $RTT_{max}$ are the maximum goodput and RTT values among all subflows, respectively. The algorithm considers available all those subflows with sufficient congestion window (cwnd). The subflow that maximizes the utility function $\gamma$ is the one to be selected for scheduling at any given time. Subflows that have been marked as "backup" are used only when no normal subflows are available. In addition, the algorithm examines whether a subflow is \textit{unavailable} (e.g., no longer usable), or \textit{temporarily unavailable} (e.g., no available space in its cwnd) for data transfer. The LLHD algorithm presented in \ref{alg:LLHD}, is based on the respective paper \cite{lubna2022low}. However, the set of steps is simplified; multiple conditions are examined within the actual code published in GitHub \cite{githubllhd}, controlling how the temporarily unavailable subflows, as well as any backup ones, are actually used upon each call of the algorithm.  

The algorithmic steps of \textit{LLHD} packet scheduling algorithm are presented in detail in \ref{alg:LLHD}.

\subsubsection{ReMP TCP (Redundant)}
\label{subsub:redundant}

In multipath scheduling, redundancy is realized by sending duplicate segments on all available subflows. This technique is profoundly inefficient in terms of bandwidth utilization; it might however be applied to reduce per-packet latency and any induced retransmissions in case of interactive bandwidth-tolerant applications, which are though limited by stringent latency and jitter requirements. Using redundant packet scheduling, the fastest path is the one determining the effective end-to-end latency. In addition, redundancy contributes to reduced RTT, minimizing delays induced by excessive retransmissions; this can be achieved since segments arriving at the receiving node via the faster path will be acknowledged faster compared to how their duplicates (carried over a slower and even lossy path) would. According to its authors, \textit{ReMP TCP} can efficiently handle queuing delays, latency variations (jitter), and heterogeneous path performance, exchanging, in essence,  bandwidth for reduced latency in existing best-effort networks. If a subflow is not available at the time of redundant scheduling (i.e., in case cwnd does not permit, or in case of link unavailability or failure), a segment is injected only on the available subflow.

The algorithmic steps of \textit{ReMP TCP} packet scheduling algorithm are presented in detail in \ref{alg:ReMPTCP}.


\subsection{Congestion Control Algorithms}
\label{subsec:cca}

\paragraph{Uncoupled Congestion Control.} As with standard TCP, MPTCP also makes use of the congestion control mechanism to regulate the amount of outstanding data, in an attempt to detect and recover from network congestion. The standard TCP congestion control mechanism is applied individually to each MPTCP subflow, managing the rate at which each subflow can inject segments into the network. Thus, each MPTCP subflow maintains its own congestion window, operating essentially as an individual single-path TCP connection. While such an approach is functional, it leads to an unfair share of a bottleneck link's bandwidth. For instance, when two subflows belonging to the same MPTCP connection coexist alongside a standard single-path TCP connection in a bottleneck link, the MPTCP connection would occupy 2/3 of the bottleneck link's bandwidth. Applying such a congestion control mechanism for broader utilization, would provide an unfair advantage to MPTCP connections over standard TCP ones, allowing the latter to be allocated only a small fraction of the available bandwidth, and eventually leading them to resource starvation. A multipath congestion control mechanism with the aforementioned characteristics is called \textit{uncoupled}, since the congestion window of each subflow scales individually, absent any MPTCP connection-level limit.
\paragraph{Coupled Congestion control.} On the contrary, one of the main principles to be considered during the design of TCP congestion control algorithms, is to ensure \textit{fairness} among the TCP connections which are competing for bandwidth \cite{rfc2914}. The same principle is carried over to the MPTCP congestion control design \cite{rfc6356, rfc5033, rfc8684, rfc6182}, ensuring that each MPTCP connection will be allocated the same amount of bandwidth as if it were a standard TCP connection; in the above example, the MPTCP connection would be allocated 1/2 of the bottleneck bandwidth, however many subflows comprising it. Such a congestion control algorithm, which imposes an upper bound on the aggregate amount of outstanding data at MPTCP connection-level, is called \textit{coupled}. 

Table \ref{tab:cca} provides a concise description of the state-of-art congestion control algorithms used in MPTCP, inherited either directly from TCP (uncoupled), or adapted and new ones focusing on fairness (coupled).


\begin{table*}[htb!]    
  \caption{MPTCP Congestion Control Algorithms (CCA)}
  \label{tab:cca}
  \resizebox{\textwidth}{!}{
  \begin{tabular}{p{0.5in}p{0.6in}p{0.5in}p{0.6in}p{3.3in}}  
    \toprule
    CCA & Full Name & Reference & Approach & Main concept\\
    \midrule
    \texttt{CUBIC} & - & v0.96 \cite{10.1145/1400097.1400105, rfc9438, githubcubic, githubmptcp:v096} & uncoupled (loss-based) & CUBIC replaces the linear additive increase phase of Reno-based AIMD CCAs, with a cubic function. This allows fast convergence toward the actual link bandwidth in high-BDP networks.\\
    \noalign{\hrule height 0.2pt}
    \texttt{Coupled (LIA)} & Linked Increases Algorithm  & v0.96 \cite{wischik2011design, rfc6356, githublia, githubmptcp:v096} & coupled (loss-based) & LIA is the default MPTCP congestion control algorithm. It has been the first coupled congestion control scheme to ensure MPTCP's incentive and fairness to single-path TCP flows. It modifies the additive increase part of Reno-based AIMD algorithm. \\
    \noalign{\hrule height 0.2pt}
    \texttt{OLIA} & Opportunistic Linked-Increases Algorithm & v0.96 \cite{10.1109/TNET.2013.2274462, khalili-mptcp-congestion-control-05, githubolia, githubmptcp:v096} & coupled (loss-based) & OLIA has been proposed as an optimization alternative to LIA, managing to eliminate the trade-off between responsiveness and optimal congestion balancing; thus, providing a pareto-optimal solution that combines both simultaneously. As with LIA, the proposed optimization is applied to the additive increase part of the AIMD algorithm. \\
    \noalign{\hrule height 0.2pt}
    \texttt{BALIA} & Balanced \nobreak{Linked Adaptation} & v0.96 \cite{peng2014multipath, walid-mptcp-congestion-control-04, githubbalia, githubmptcp:v096} & coupled (loss-based) & BALIA reveals an inevitable tradeoff between friendliness, responsiveness, and window oscillation of congestion control schemes. Basing its design on an analytical framework, BALIA proves to reach a unique equilibrium point, managing to mitigate the tradeoff and provide balanced performance. BALIA adjusts the entire AIMD algorithm controlling the Congestion Avoidance (CA) phase. \\
    \noalign{\hrule height 0.2pt}
    \texttt{wVegas} & weighted Vegas & v0.96 \cite{cao2012delay, xu-mptcp-congestion-control-05, githubwvegas, githubmptcp:v096} & coupled (delay-based) & wVegas is a coupled delay-based alternative of standard TCP-Vegas for MPTCP. Contrary to loss-based CCAs that associate packet loss with congestion, wVegas interprets queuing delay as congestion signal. Using weights to equally distribute the congestion cost among subflows, wVegas ensures that each sublow allocates a fair share of the bottleneck bandwidth. Maintaining a moderate buffer occupancy in link queues, wVegas accounts for congestion proactively, thus preventing excessive packet loss. \\
    \noalign{\hrule height 0.2pt}
    \texttt{BBR} & Bottleneck Bandwidth and Round-trip propagation time & v0.96 \cite{cardwell2016bbr, cardwell-iccrg-bbr-congestion-control-00, cardwell-iccrg-bbr-congestion-control-02, githubbbr, githubmptcp:v096, googlegithubbbr} & uncoupled (congestion-based) & BBR-v1 \cite{cardwell-iccrg-bbr-congestion-control-00} is neither a loss- nor a purely delay-based scheme. Although inspired by the delay-based approach of TCP-Vegas, BBR follows a unique method to identify the optimal operating point, by accurately estimating the bottleneck bandwidth and base RTT. Based on these estimations and appropriate traffic pacing adjustments, BBR achieves maximum bandwidth utilization at the lowest possible round-trip delay, eliminating intermediate queue formation. To derive these estimations, BBR implements a state transition model, cycling over four states (Startup, Drain, Probe\_BW, and Probe\_RTT). BBR's congestion-proactive approach leads to sustained high bandwidth utilization, low RTT, and infrequent loss. \\
    \noalign{\hrule height 0.2pt}
    \texttt{C-MPBBR} & Coupled Multipath BBR & v0.96 \cite{mahmud2020coupled, githubcmpbbr} & coupled (congestion-based) & C-MPBBR is a coupled alternative of BBR for MPTCP. It exploits BBR's existing BtlBw and DelRt parameters to ensure multipath benefit, as well as fairness among single- and multi-path flows traversing a common bottleneck. C-MPBBR achieves the first goal by closing any subflows not contributing sufficiently to the multipath connection. Fairness is established by leveraging BtlBw estimations to identify subflows on the same bottleneck, and dividing the measured bandwidth equally among them. \\

    \bottomrule
  \end{tabular}
}
\end{table*}

\subsubsection{CUBIC}
\label{subsub:cubic}

The initial AIMD approach, as specified and applied in legacy \textit{TCP Reno} and \textit{NewReno} algorithms, was following a linear increase of the congestion window (cwnd) during the congestion avoidance phase, that is, an increase of the cwnd by one segment per RTT. While this approach is suitable for low BDP (bandwidth-delay product) links, there is a noticeable drawback when used in long-fat networks (i.e., with links experiencing high bandwidth and round-trip time delays). In such cases, the available bandwidth is underutilized since the linear increase of the congestion window takes too long to reach the link's saturation point, that is, the link's full capacity. As a result, the amount of time needed to reach the link's full capacity may be much higher than the one a flow may take to complete. For instance, using a 10Gbps/100ms round-trip delay link, the congestion window would reach the link's full capacity in $\sim{1.4}$h.

The above computation is based on S. Floyd's et al. calculation of the round-trip times required until full congestion window is reached following a loss \cite{floyd2000comparison}, that is:

\begin{equation} \label{eq:9}
    \frac{\beta}{a}W+1 \quad \textnormal{round-trip times}
\end{equation}

where \textit{W} is the congestion window value just before the loss event occurs, and \textit{a} and \textit{$\beta$} are AIMD's additive-increase and multiplicative-decrease factors, respectively. In the above example, using traditional TCP Reno and NewReno where $AIMD(a, \beta) = AIMD(1, 1/2)$, it would take approximately (given that: $1 \, Byte \approx{10 \, bits}$, $MSS \approx{1000 \, Bytes}$):

$RTT*\frac{\beta}{a}W = 0.1*\frac{1}{2}10^5 = 5000$\, sec. for the congestion window to reach its full capacity. To cope with such long-distance and high-latency links, CUBIC replaces AIMD's additive increase alorithm with a cubic function, the window growth of which, is provided in equation \ref{eq:10}:

\begin{equation} \label{eq:10}
    W(t) = C(t-K)^3 + W_{max}
\end{equation}

where \textit{C} is a CUBIC parameter determining the aggressiveness of the protocol against competing flows which may be using other congestion control schemes, \textit{t} is the elapsed time from the last window reduction, $W_{max}$ is the congestion window value when the loss occurred, and \textit{K} is the time period that the above function takes to increase $W$ to $W_{max}$ when there is no further loss event, and is calculated by the following equation:

\begin{equation} \label{eq:11}
    K=\sqrt[3]{\frac{W_{max}\beta}{C}}
\end{equation}

CUBIC heuristically sets constant $C=0.4$, and the AIMD decrease factor $\beta=0.2$, to adapt for fairness with other TCP flows, and attain a balance between protocol stability and convergence speed. CUBIC algorithm essentially computes the target value of the congestion window growth during the next RTT: $W(t+RTT)$, and then compares this target value with the growth of the congestion window that standard TCP would achieve within the same time interval. If the current congestion window is less than the estimated standard TCP's congestion window (calculated in terms of time \textit{t}, where t is the elapsed time after the loss event), then CUBIC operates in \textit{TCP mode} where the current congestion window is updated according to standard TCP's AIMD; this allows CUBIC to perform equally to TCP Reno in low BDP networks. If the current congestion window is greater than the one TCP would reach within the same time interval, then the protocol operates either in the \textit{concave} or the \textit{convex} region, depending on whether the current congestion window value is less or greater than $W_{max}$, respectively. Within both the concave and convex regions, CUBIC increments the congestion window (cwnd) by $\frac{W(t+RTT)-cwnd}{cwnd}$. The \textit{concave} profile allows the algorithm to approach quickly the area close to link's saturation point, but then slow down to remain long-enough in this area which is expected to provide the highest bandwidth utilization. In case network conditions have improved since the last congestion event (i.e., the reception of a duplicate ACK), the protocol is switching to the \textit{convex} region (also known as \textit{maximum probing phase}) where it seeks to identify the new bandwidth saturation point; it does this following the convex function's exponential nature, in which the algorithm prompts for the new $W_{max}$ initially slowly, and then more aggressively.

The advantage of CUBIC compared to standard TCP is that, while maintaining the standard TCP principles, it manages to converge faster in case of high BDP links. It is also unique in the way it approaches the saturation point, by combining the concave and convex profiles encompassed within a single cubic function. This way, it approaches quickly the last epoch's $W_{max}$, then prompts slowly for the new saturation point close to the plateau created around $W_{max}$, and in case the available bandwidth has increased since the previous epoch, it increases the congestion window's growth rate more aggressively to find the new maximum value. Another key property of CUBIC is the fact that the congestion window increase rate is independent of RTTs outside the TCP-mode; this characteristic allows flows with different RTTs to maintain similar congestion window sizes, thus further contributing to the protocol's fairness.

Since CUBIC is based on the AIMD congestion control scheme, where the congestion window decreases\footnote{To avoid any confusion between the multiplicative decrease factor as defined within \cite{10.1145/1400097.1400105} ($\beta = 0.2$) and the one specified within \cite{rfc9438} ($\beta=0.7$), we have to clarify that both are using the same decrease factor. The RFC updates \textit{ssthresh} by exploiting the decrease factor directly, i.e. by:  $ssthresh = cwnd * \beta$, while the paper calculates it using: $ssthresh = (1-\beta)*cwnd$.} rapidly upon an identified loss and then grows linearly, it is considered a \textit{loss-based} congestion control algorithm. From MPTCP perspective, since CUBIC does not incorporate any multipath-specific logic to account for fairness between MPTCP and standard TCP connections, it is considered \textit{uncoupled}. A condensed version of CUBIC pseudo-code is presented in the Appendix, based on the original source \cite{10.1145/1400097.1400105}. Some more recent updates can be found within RFC 9438 \cite{rfc9438}.

The algorithmic steps of \textit{CUBIC} congestion control algorithm are presented in detail in \ref{alg:CUBIC}.


\subsubsection{Coupled}
\label{subsub:lia}

Coupled (also called \textit{Linked Increases Algorithm (LIA)}) \cite{wischik2011design, rfc6356}, has been the first coupled congestion control algorithm for Multipath TCP that efficiently tackles the fairness issues arising when single- and multi-path flows concurrently share the same bottleneck link resources. Coupled congestion control has been therefore introduced to satisfy the fairness principles that the respective standard TCP congestion control schemes lack. 

The bottleneck fairness properties that any multipath congestion control scheme should employ, are succinctly summarized in the following three goals \cite{rfc6356}:

\begin{itemize}
 \item Goal \#1 (Improve Throughput): A multipath flow should at least achieve the performance of a single path flow on the best of the paths
  available to it.
  \item Goal \#2 (Do no harm): A multipath flow should not allocate more capacity from the resources shared when traversing multiple paths, than if it were a single flow using only one of these paths. This guarantees fairness against other competing flows.
  \item Goal \#3 (Balance congestion): A multipath flow is expected to remove traffic from its most congested paths, subject to
  meeting the first two goals.
\end{itemize}

On the basis of these goals, three main approaches have been studied in literature in an attempt to identify the most appropriate multipath congestion control coupling scheme \cite{honda2009multipath}. The first approach suggests the use of a common congestion window, shared between the subflows of a multipath connection. While this approach sounds reasonable, it may lead to suboptimal performance of the multipath connection, since individual subflows traversing distinct paths usually experience different RTT characteristics, which in turn lead to varying ACK receipts and, consequently, to differing congestion window increase and decrease intervals. As a result, there is no perfect synchronization among subflows regarding the congestion window increase and decrease events. Furthermore, a packet loss event on a particular subflow would decrease the shared congestion window, degrading the performance of the entire multipath connection. The second approach includes the detection of shared bottleneck links, and ensuring that no more than one subflow will traverse such a link. While that approach was not that convincing a couple of years ago \cite{zhang2004transport, honda2009multipath}, recent works provide the means for accurate detection of shared bottlenecks \cite{cardwell2016bbr}, enabling also efficient coupling approaches \cite{mahmud2020coupled}.       
The third approach considers individual congestion windows for each subflow of a multipath connection, where each subflow adjusts its own congestion window according to a weight, such that the entire multipath flow (comprising multiple subflows) has the same aggressiveness as a standard TCP flow \cite{honda2009multipath}. Building upon this approach, the work pursued in \cite{honda2009multipath, wischik2011design} suggests to modify the legacy additive increase algorithm of the congestion avoidance phase, as follows: 

\begin{itemize}
 \item upon receipt of an ACK, increase congestion window by:  $\frac{\alpha}{w_r}$
  \item upon loss detection, decrease the congestion window by:  $\frac{w_r}{2}$
\end{itemize}

where $w_r$ is the congestion window of subflow $r$. According to \cite{honda2009multipath}, the relation between the increase parameter $\alpha$ and each subflow's individual weight factor $D$ is denoted by $\alpha = D^2$ (a). A weight factor of $D = \frac{1}{n}$ (b), where $n$ is the number of subflows belonging to a multipath connection, would couple the throughput of a multipath connection to the equivalent of a standard TCP connection within an RTT interval. Consequently, according to (a) and (b), the increase parameter should be: $\alpha = \frac{1}{\sqrt{n}}$. While this approach ensures fairness among single- and multi-path flows traversing a shared bottleneck link where flows experience the same RTTs, it may lead to suboptimal resource utilization when the flows traverse multi-hop bottleneck paths encompassing varying RTTs and packet loss rates \cite{wischik2011design}.

This is where \textit{Coupled} congestion control algorithm kicks in, managing to equalize the multipath flow's aggregate bandwidth with the one a regular TCP flow would get on the best path available to the multipath flow. Coupled algorithm enables the multipath flow to first estimate the target rate of a regular TCP flow based on currently measured path conditions (i.e., RTT, packet loss rate). Then, it calculates the increase parameter $\alpha$ which controls the overall aggressiveness of the protocol and assists in reaching the desirable target rate. Similarly to EWTCP \cite{wischik2011design}, Coupled modifies the additive increase algorithm of the congestion avoidance phase, while maintaining the legacy TCP's multiplicative decrease, fast retransmit, and fast recovery schemes \cite{rfc6356, rfc5681}. 

More specifically\footnote{Details on the specific equations which determine the congestion window increase and decrease steps, are provided within the algorithm in \ref{alg:Coupled}.}, 
\begin{itemize}
 \item upon ACK receipt on $subflow_r$, increase the congestion window $w_r$ by:  $\min(\frac{\alpha}{w_{total}}, \frac{1}{w_r})$ 
 \item upon loss detection on $subflow_r$, decrease the congestion window $w_r$ by:  $\frac{w_r}{2}$
\end{itemize}

Noteworthy here, is the fact that Coupled algorithm's design encompasses an adaptive behavior based on the level of statistical multiplexing. In cases of high statistical multiplexing, where the multipath flow coexists alongside single-path flows, the multipath flow does not influence the link loss rates; in such scenarios, MPTCP will achieve the same throughput as standard TCP on the best path. However, when low statistical multiplexing is present, in which case the multipath flow influences the loss rate of the path, MPTCP's throughput is guaranteed to be higher than that of a single-path TCP flow on the best path; in that case, the achieved multipath throughput is expected to be equal to the sum of the idle paths' individual throughput, thus, fully aggregating the available bandwidth.

Contrary to EWTCP \cite{wischik2011design} where the increase parameter $\alpha$ was inversely proportional to the square root of the number of subflows, Coupled bases its decisions on how TCP would perform on the best path available at each RTT interval; since path conditions in real networks are subject to constant and dynamic changes, Coupled needs to compute $\alpha$\footnote{$alpha$ calculation formula: The calculation of $\alpha$ within algorithm's step \#2 in \ref{alg:Coupled}, is derived by equalizing the rate of the multipath flow with the rate of a TCP running on the best path, and solving for $\alpha$.} once per RTT or upon packet loss detection\footnotemark \cite{rfc6356}.  

\footnotetext{The pseudo-code steps showcase the congestion window increment and $\alpha$ computation per ACK; in actual implementations, however, such computations are performed once per RTT or upon packet loss detection.}

Since Coupled CCA couples only the congestion window increases, it satisfies Goals \#1 \& \#2. To achieve Goal \#3, which essentially implies perfect resource pooling \cite{wischik2008resource}, would require to couple also the multiplicative decrease phase, while ensuring that no traffic would be scheduled on links with higher loss rates. According to LIA's authors, given the Internet's dynamic path conditions, such a restriction would lead the algorithm to \textit{flappiness} between paths and insufficient probing of lossy paths, thus failing to utilize them quickly once they recover. The notion behind LIA's operation is to allocate congestion windows to the subflows, such that $p_i * cwnd_i$ is constant, for all $i$ ($p_i$ denotes the packet loss rate experienced by $sf_i$). This ensures that equal congestion windows will be allocated among subflows in case they experience equal loss rates, while higher congestion window values will be progressively allocated to less-lossy subflows when their packet loss rates differ.

The algorithmic steps of \textit{LIA} congestion control algorithm are presented in detail in \ref{alg:Coupled}.


\subsubsection{Opportunistic Linked-Increases Algorithm (OLIA)}
\label{subsub:olia}

LIA's shortcoming in fulfilling the third principle of multipath congestion control, has gathered attention. Some more complex scenarios examining MPTCP connections' fairness to regular TCP flows under LIA congestion control, identified a couple of issues: i) the fact that upgrading TCP users to MPTCP has a negative impact to the remaining regular TCP flows without any added benefit to the MPTCP ones, and ii) that MPTCP is over-aggressive to regular TCP flows \cite{10.1109/TNET.2013.2274462}. Scientific research on this field, revealed that it is feasible to attain all three multipath congestion control goals simultaneously. As a result, \textit{OLIA} congestion control algorithm has been proposed as an alternative to LIA. Contrary to LIA, OLIA's design eliminates the trade-off between responsiveness and optimal congestion balancing, managing to provide a pareto-optimal solution which achieves both simultaneously.  

Similarly to LIA, OLIA couples only the additive increase algorithm of the congestion avoidance phase; in case of loss, the multiplicative decrease, as well as the fast retransmit and fast recovery, retain the legacy TCP NewReno approach.

OLIA maintains a couple of variables, such as the set of paths over which subflows of the MPTCP connection have been established (\textit{all\_paths}), a subset of "all\_paths" containing all presumably best paths (\textit{best\_paths}), a subset of "all\_paths" including paths with the largest congestion windows (\textit{max\_w\_paths}), and a subset of "all\_paths" which also belong to "best\_paths" but are not among those exhibiting largest congestion windows (\textit{collected\_paths}). Exploiting these variables, the increase part of OLIA is defined as follows:

\begin{itemize}
 \item for each ACK received on path r, increase the congestion window ($w_r$) by:
    \begin{equation} \label{eq:olia_AI}
        \frac{w_r/rtt_r^2}{\left (\displaystyle \sum_{p \in R_u} w_p/rtt_p \right)^2} + \frac{a_r}{w_r}
    \end{equation}
    where $p$ denotes a path on the set of "all\_paths" $R_u$, and $w_p$ and $rtt_p$ denote the congestion window and the round-trip time of path $r$, respectively. Equation \ref{eq:olia_AI} comprises two terms: the first term ensures Pareto optimality in resource pooling (satisfying MPTCP Goal \#3: "Balance congestion"), while the second one, encompassing $\alpha_r$, guarantees responsiveness and non-flappiness (satisfying MPTCP Goal \#1: "Improve throughput", and Goal \#2: "Do no harm") \cite{10.1109/TNET.2013.2274462}
    \BlankLine
    $a_r$ is calculated as follows:
    \begin{itemize}
        \item if $r$ is within the \textit{collected\_paths}, then: $a_r = \frac{1/num\_of\_paths}{|collected\_paths|}$
        \item if $r$ is within the \textit{max\_w\_paths} and if \textit{collected\_paths} is not empty, then: $a_r = -\frac{1/num\_of\_paths}{|max\_w\_paths|}$
        \item otherwise, $a_r = 0$.
    \end{itemize}
    \BlankLine
 \item for each loss on path r, decrease the congestion window ($w_r$) by: $\frac{w_r}{2}$.
\end{itemize}

OLIA adapts the congestion window increases as a function of the underlying paths' RTTs, thus efficiently compensates for different RTTs. Analytical results and testbed experiments prove that OLIA provides fairness among coexisting TCP and MPTCP flows, as well as  optimal congestion balancing, while remaining as responsive and non-flappy as LIA.

The algorithmic steps of \textit{OLIA} congestion control algorithm are presented in detail in \ref{alg:OLIA}.


\subsubsection{Balanced Linked Adaptation (BALIA)}
\label{subsub:balia}

A subsequent work has identified that OLIA also suffers a shortcoming, that is, its limited responsiveness to dynamic network conditions when the used paths experience similar round-trip times (RTTs). To confront OLIA's weakness, meticulous research conducted in the context of \cite{peng2014multipath} devised BALIA which, similarly to its predecessors LIA and OLIA, is a window-based coupled congestion control algorithm designed for MPTCP. This work identified an inevitable tradeoff between friendliness, responsiveness, and window oscillation of congestion control schemes, and proposed an analytical framework that enables the aforementioned properties' assessment for any congestion control algorithm, already from the design phase. 

Leveraging the designed analytical framework to prove BALIA's unique equilibrium point and asymptotical stability, and confirming it through experimental evaluation, BALIA's authors managed to successfully mitigate the tradeoff and provide balanced performance. As with LIA and OLIA, BALIA modifies the additive increase (AI) algorithm of the congestion avoidance phase, while maintaining the TCP NewReno fast retransmit \& recovery algorithms. In case of loss, BALIA adjusts the TCP NewReno multiplicative decrease (MD) algorithm, multiplying it by a factor in the range of [1, 1.5]. In case of single path use, BALIA reduces to the standard TCP NewReno increment and decrement algorithms.

The algorithmic steps of \textit{BALIA} congestion control algorithm are presented in detail in \ref{alg:BALIA}.


\subsubsection{Weighted Vegas (wVegas)}
\label{subsub:wvegas}


An alternative approach to the aforementioned loss-based congestion control schemes (i.e., LIA, OLIA, BALIA), is \textit{wVegas} which bases its decisions on queuing delay measurements. wVegas builds upon the respective standard TCP congestion control algorithm TCP-Vegas, and is adapted to serve the fairness requirements of multipath environments. Contrary to other coupled multipath congestion control schemes (e.g., LIA, OLIA, and BALIA) that estimate network congestion through packet losses, wVegas interprets queuing delay as congestion signal, thus being more sensitive to network condition changes in an attempt to achieve fine-grained load balancing. wVegas focuses on satisfying the "Congestion Equality Principle" \cite{cao2012delay}, where all flows attempt to allocate a fair share of the available bottleneck bandwidth via an equal share of the congestion cost. For this purpose, wVegas assigns a weight to each subflow belonging to a multipath flow, and adaptively adjusts this weight according to the Congestion Equality Principle. The weight quantifies, in essence, the aggressiveness of each subflow on acquiring part of the available bandwidth. Subflows established on less congested paths are expected to be assigned a larger weight value, allowing them to compete for bandwidth more aggressively; this competition is expected to increase the underlying path's congestion, which will in turn lead to decrease of the subflow's weight value on that path. This cycle is executed for all subflows traversing the various paths until congestion equilibrium is reached; at that point network resources will be fairly shared by all flows. 

Diving deeper into the design principles of the algorithm, wVegas is essentially assigning a fixed parameter $\alpha=10$ to each one of the multipath flows. This parameter determines the total number of bytes backlogged in the network for all subflows belonging to a MPTCP flow \cite{cao2012delay, xu-mptcp-congestion-control-05}. Each subflow (e.g., assuming a subflow on path $r$) belonging to a broader multipath flow, is then assigned a portion of this value (e.g., $\alpha_r$) according to the weight corresponding to this subflow $w_r$. The weight is calculated as the portion of a subflow's rate to the overall rate of the entire multipath flow, as depicted in \ref{eq:wvegas_weight}. 

The \textit{rate} and \textit{weight} of a subflow $r$ is respectively provided by the below equations: 

\begin{equation} \label{eq:wvegas_rate}
    rate_r = cwnd_r / rtt_r
\end{equation}
\begin{equation} \label{eq:wvegas_weight}
    weight_r = rate_r /  \textstyle \sum_{i}(rate_i)
\end{equation}
where $\sum_{i}(rate\_i)$ denotes the aggregate rate achieved by all subflows of the multipath flow.

The portion of $\alpha$\footnotemark, which then corresponds to a subflow on path $r$, $\alpha_r$, is calculated as follows:

\begin{equation} \label{eq:wvegas_a_r}
    \alpha_r = weight_r * \alpha
\end{equation}

\footnotetext{Parameter $\alpha$ is associated with a MPTCP flow, while $a_r$ is associated with a specific subflow $r$ of that MPTCP connection. Parameter $\alpha$ is referred to as $total\_alpha$ within the algorithmic steps in \ref{alg:wVegas}. Similarly, the aggregate rate of all MPTCP subflows, denoted as $\textstyle \sum_{i}(rate_i)$ in \ref{eq:wvegas_weight}, is referred to as $total\_rate$ into the algorithmic steps.}

Within wVegas, $\alpha_r$ serves as a threshold which controls whether the equilibrium rate should be updated, and whether the congestion window of the respective subflow established on path $r$ will increase or decrease at the end of each RTT interval. If the currently measured rate deviates from the expected equilibrium rate beyond $\alpha_r$, then the congestion window is decreased at the end of the RTT; otherwise it is increased. The aforementioned rate difference is calculated within the wVegas' weight adjustment algorithm as follows:

\begin{equation} \label{eq:wvegas_diff}
    diff = \left(\frac{cwnd}{baseRTT} - \frac{cwnd}{rtt} \right) \cdot baseRTT
\end{equation}

Consequently, wVegas adjusts parameter $\alpha$ dynamically, thereby influencing the transmission rate of the corresponding subflow and managing to equalize congestion on the path. Normalized $\alpha$, that is $\frac{\alpha_r}{\alpha}$, is essentially the weight factor $w_r$ applied on the subflow established on path $r$. The packet queuing delay measurement employed by wVegas is meant to provide a more accurate estimation of the underlying path's congestion compared to loss-based techniques. Since wVegas maintains a moderate buffer occupancy in link queues, it accounts for congestion proactively and rarely leads to packet losses. On the contrary, loss-based congestion control schemes try to fill in intermediate queues before a loss is detected and backoff procedure begins. However, wVegas experiences some shortcomings, such as its reliance on accurate RTT measurements, its less aggressive behavior when competing with loss-based algorithms, as well as its limited efficiency on high BDP paths.

The algorithmic steps of \textit{wVegas} congestion control algorithm are presented in detail in \ref{alg:wVegas}.


\subsubsection{Bottleneck Bandwidth and Round-trip propagation time (BBR)}
\label{subsub:bbr}

Traditional loss-based congestion control algorithms cycle through the process of filling in the bottleneck link and any intermediate queues, and then responding to the detected bufferbloat-induced loss. Deviating from such loss-based schemes, \textit{BBR} follows a different approach, inspired by the delay-based control logic proposed within TCP-Vegas \cite{10.1145/190809.190317}. Contrary to the operation of loss-based schemes, BBR attempts to identify a bottleneck link's optimal operating point. Such a point entails maximum utilization of the underlying link's capacity at the lowest round-trip delay. At each moment, a sender operating at this optimal point, transmits at the bottleneck link rate, while simultaneously ensuring that no intermediate queues are formed. This prevents round-trip delay increases and minimizes the packet loss probability. 

For this purpose, BBR makes use of two parameters: \textit{BtlBw} (bottleneck bandwidth) and \textit{RTprop} (round-trip propagation time); the first tracking bottleneck link's capacity, and the latter the round-trip delay. Rtprop reflects the base RTT and is used to measure any subsequent round-trip delay updates. Given a fixed data path, RTprop remains constant up to the point where intermediate queues begin to be formed; from that point onward, an increase in RTT indicates that the pipe is already full and the sending rate exceeds bottleneck bandwidth. An increase in RTprop while BtlBw remains constant, allows BBR to infer that intermediate queues have started to be filled in. Bottleneck bandwidth (BtlBw) is used to probe for any updates in the underlying bottleneck capacity by comparing it with the actual delivery rate (DelRt). BBR attempts to estimate the bottleneck link's optimal operating point by trying to estimate these two parameters. The constraints reflected through these parameters are intrinsically contradicting, meaning that the operating point where one can be measured, hinders the measurement of the other. Measuring RTprop, for instance, requires that inflight packets remain below BDP, so that the link is underutilized. On the other hand, bottleneck bandwidth estimations request for probing above the known BDP, which could lie in the region where queues start to be formed, in case bottleneck capacity has not changed. 

In order for a link to be fully utilized at each moment, BDP amount of data should be inflight, where $BDP = BtlBw * Rtprop$. BtlBw controls the sending rate and ensures full bandwidth utilization, while Rtprop guarantees an always full pipe by managing the volume of inflight data; an amount of data less than the optimal would cause link starvation, while an excess would overfill the pipe and lead to developing queues. To accurately measure BtlBw and RTprop, BBR implements a state transition model, cycling over different states, such as the \textit{Startup}, \textit{Drain}, \textit{Probe\_BW}, and \textit{Probe\_RTT}. During Startup phase, a flow's congestion window increases exponentially, similarly to that of loss-based congestion control schemes; however the congestion window increase with BBR is smoother and the algorithm does not back off in case of packet loss or delay increase, maintaining its robustness toward discovering the available bandwidth. The Startup phase is over once the bottleneck bandwidth is considered to be successfully detected. Then, BBR enters the Drain phase, the purpose of which is to drain any queues formed during Startup. The Drain phase lasts one RTT interval, managing to clean up any filled queues rapidly. Once BBR estimates that the queue is empty and the link is sufficiently filled with BDP data inflight, it leaves Drain and enters Probe\_BW phase. BBR consumes the majority of its time in this phase, keeping a BDP of data inflight, paced at a rate equal to the BtlBw estimate. The \textit{pacing\_gain} parameter used in this phase controls the way the algorithm probes for updates in BtlBw estimation. More specifically, Probe\_Bw phase consists of eight-cycle rounds, each lasting an RTprop interval. A full Probe\_BW round involves cycling through pacing\_gain values in the following order: $\{5/4, 3/4, 1, 1, 1, 1, 1, 1\}$ \cite{cardwell2016bbr}. According to this pacing\_gain sequence, BBR spends one round exploring whether bottleneck bandwidth has increased, thus probing at a rate equal to 5/4 of the current estimate. Then, pacing\_gain decreases equally to the initial increase portion (3/4), draining any queues formed during the previous round. BBR spends the next six rounds "cruising" at BtlBw rate, thus achieving the highest link utilization with minimum delay. 

If probing for higher bandwidth leads to higher delivery rate, BBR updates its bottleneck bandwidth estimation and adapts traffic transmission rate at the new delivery rate. If probing for higher bandwidth is not accompanied by delivery rate increase, it means that the current transmission rate already matches the bottleneck link's capacity, and the excess data transmitted during BtlBw probing leads to queue formation, which is also perceptible through a subsequent increase in RTprop.  

In case RTprop estimate has not been updated by a lower RTT measurement within 10 seconds of operation into the Probe\_BW phase, BBR enters Probe\_RTT phase. The intention of this phase is to obtain an accurate measurement of the base RTT by reducing the congestion window (cwnd) to a very small value (i.e., four packets). After remaining in Probe\_RTT phase for at least 200ms and one RTT \cite{cardwell2016bbr}, BBR enters either Startup or Probe\_BW phase, depending of the BtlBw estimate; if BtlBw indicates a full pipe, BBR transitions to Probe\_BW phase; otherwise to Startup. 

Another important parameter used in BBR\footnotemark is the \textit{cwnd\_gain} which bounds the amount of data inflight in each state; cwnd\_gain is different from the pacing\_gain, in that the first allows for an upper bound of inflight data, while the latter controls the sending rate and consequently influences the amount of outstanding data occupying the link.   

What distinguishes BBR from AIMD algorithms, is the fact that it is proactive to loss, operating at the point where intermediate queues start to be formed, whereas AIMD schemes operate at the saturation point where queues are already full, merely reacting to detected loss. BBR's subtle traffic pacing at a rate equal to bottleneck link's bandwidth, differs from traditional AIMD algorithms whose sending scheme follows a bursty pattern within each RTT interval \cite{GHustonBBR}.

In the context of multipath communications, BBR is deemed an uncoupled congestion control scheme. In the following subsection, a coupled alternative, based on BBR, is presented (C-MPBBR).

\footnotetext{It is important to mention here that throughout our work, any reference to BBR is referred to \textit{BBR v1} \cite{cardwell2016bbr, cardwell-iccrg-bbr-congestion-control-00}, which is also available in the older MPTCPv0 linux kernels. However, BBR is still under development at the time of this publication. IETF details in regard to BBR are referred to BBR v2 \cite{cardwell-iccrg-bbr-congestion-control-02}.}

The algorithmic steps\footnotemark of \textit{BBR} congestion control algorithm are presented in detail in \ref{alg:BBRv1}.
\footnotetext{An abstract version of BBR algorithm's pseudo-code is provided in \ref{alg:BBRv1}, as originally published for BBR\_v1 within \cite{cardwell-iccrg-bbr-congestion-control-00}. BBR actually spans more than a thousand lines of C code, details of which can be found within \cite{cheng-iccrg-delivery-rate-estimation-02, cardwell-iccrg-bbr-congestion-control-00, cardwell-iccrg-bbr-congestion-control-02, githubbbr}.}


\subsubsection{Coupled Multipath BBR (C-MPBBR)}
\label{subsub:cmpbbr}

C-MPBBR arose from an attempt to create a coupled congestion control algorithm for MPTCP, on the basis of BBR. The main goals C-MPBBR aims at fulfilling, are the following:

\begin{enumerate}[label=\roman*.]
    \item Goal \#1: to incentivize the use of MPTCP over standrard single-path TCP connections, which implies that MPTCP performs at least as well as a single-path TCP (SPTCP) flow on the best path, and
    \item Goal \#2: to reinforce fairness among flows consuming the resources of a shared bottleneck.
\end{enumerate}

To achieve Goal \#1, C-MPBBR continuously measures the bottleneck bandwidth (BtlBw) and the actual delivery rate (DelRt) using BBR's existing parameters. If the algorithm identifies that, during five consecutive ProbeBw rounds, the aggregate throughput of the MPTCP subflows is below that of a single-path flow on the best path, it closes the subflow experiencing the lowest performance. The algorithm performs the same operation for as long as the same condition holds true, up until the point where only a single subflow remains available; at that point, an MPTCP connection comprising one subflow is equivallent to a SPTCP flow, acquiring and utilizing the same amount of resources. 

C-MPBBR adheres to Goal \#2 by ensuring that the aggregate amount of bandwidth allocated to the subflows of a MPTCP connection equals that of a SPTCP flow, when sharing the same bottleneck link. C-MPBBR satisfies fairness among single- and multi-path flows, by exploiting the BtlBw parameter value. The BtlBw estimations, which are measured individually by each subflow, allow the algorithm to infer whether the subflows of a MPTCP connection traverse the same bottleneck link. In that case, the BtlBw value measured by an individual subflow is divided by the number of subflows sharing the same bottleneck link, so that multipath flows are allocated their fair share when competing against single-path flows for bottleneck resources. 

C-MPBBR is a worth-researching coupled congestion control scheme, as it manages to combine the state-of-art technology inherited from BBR, along with MPTCP's \textit{aggregation benefit} \cite{KasparOslo} and \textit{fairness} concepts \cite{rfc2914, rfc9743}; initial tests \cite{mahmud2020coupled} showcased that it fulfills both goals when compared to coupled loss-based congestion control schemes (i.e., LIA, OLIA, BALIA), or other BBR-based variants. For this reason, we selected to include C-MPBBR within our own experimentation campaign, and verify its behavior both compared to other congestion control schemes, as well as in combination with various MPTCP packet scheduling algorithms.  

The pseudo-code determining \textit{C-MPBBR's} operation toward fulfilling its incentive and fairness goals, is presented in \ref{alg:C-MPBBR}.


\section{Experimentation Methodology}
\label{sec:exper_method}
This section provides details on the methodology followed which laid the groundwork for conducting the experiments. It initially presents the multipath test-environment's topology, the tools used, and the configuration applied; it then continues with the experimentation campaign's design and the set of experiments executed. Finally, it delineates the main performance metrics against which the packet scheduling and congestion control algorithms have been assessed. 

\subsection{Multipath Environment Setup \& Configuration}
\label{subsec:setup_config}



\begin{figure}[h]
  \centering
  \includegraphics[scale=0.2]{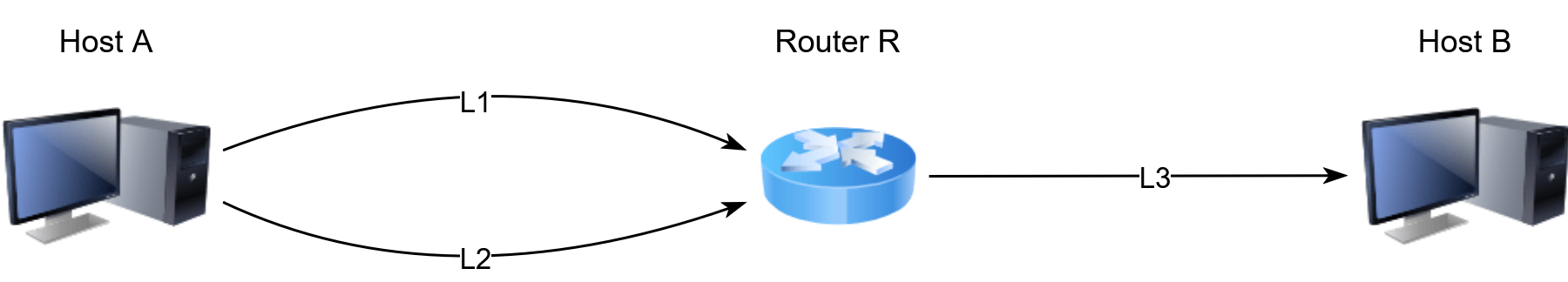}
  \caption{Multipath test environment topology}
  \label{fig:Mininet_arch}
  \Description[Mininet topology]{Two multipath-capable hosts (A, B) are connected via an intermediate router (R). More specifically, source Host A is connected to R over two disjoint paths (L1, L2) and the router is then connected to destination Host B over path L3. The MPTCP connection established between Hosts A and B, comprises two subflows (SF1, SF2), that is, one subflow is established over path L1 (i.e., SF1) and another over path L2 (i.e., SF2). The two subflows are eventually traversing the same path L3 towards destination host (Host B).}
\end{figure}

\paragraph{Mininet topology}
Figure \ref{fig:Mininet_arch} depicts our Mininet-based multipath topology comprising two multipath-capable hosts (A, B) which are connected via an intermediate router (R). More specifically, source host (i.e., Host A) is connected to R over two disjoint paths (L1, L2), and R is then connected to the destination host (i.e., Host B) over path L3. The MPTCP connection established between Hosts A and B, comprises two subflows (SF1, SF2), that is, one subflow is established over paths L1, L3 (i.e., SF1), and another subflow over paths L2, L3 (i.e., SF2). The two subflows are partially disjoint, since both are eventually traversing the same path L3 towards destination host (Host B).

The aforementioned topology has been selected based on the \textit{hybrid-access} network as depicted within \cite{rfc8041} (Fig.2: \textit{Hybrid Access Network}), in an attempt to have a simple and explainable configuration that guarantees reliable performance results for the algorithms under study. In our hybrid-access topology, subflows SF1 (over paths L1, L3) and SF2 (over paths L2, L3) are established between the two interfaces available in Host A (e.g., eth0, eth1) and the single interface in Host B (e.g., eth0).


\paragraph{HW/OS details} The above-mentioned multipath topology is emulated using Mininet v2.3.0 \cite{10.1145/1868447.1868466, mininetorg, githubmininet} and Python v3.10 \cite{python310}. The Mininet environment is instantiated on top of Linux 22.04.5 LTS (Jammy Jellyfish) OS \cite{CanonicalUbuntu}\footnote{Noteworthy here is the fact that the deployment and execution of Mininet in a VM produced weird results, non-deterministic under identical test iterations, which forced us to execute Mininet directly on the OS.}, which is in turn running on a Dell OptiPlex-7050 server, equipped with an Intel Core i5-6500 CPU @ 3.20 GHz, 16 GiB RAM, and 238 GiB SSD. To be able to execute Mininet with MPTCP capabilities, we loaded the out-of-tree MPTCP linux kernel v5.4.230 \cite{LinuxKernel:LK} (MPTCP v0.96 \cite{githubmptcp:v096}) which we have further customized in order to be able to collect kernel parameter values, such as the MPTCP out-of-order queue \cite{githubOFOrbtree394, githubOFOpackets460}. Another reason that forced us patch the legacy MPTCP kernel, was the fact that we needed to port all packet scheduling and congestion control algorithms to the same MPTCP version (v0.96) for compatibility and fair comparison purposes; to this end, some modifications were also required to the legacy kernel code basis.

\paragraph{MPTCP version} The MPTCPv0 linux kernel version, which we selected to use for our experiments, has been recently frozen. MPTCPv1 efforts are ongoing to develop MPTCP in compliance with the latest IETF RFC \cite{rfc8684}. The reason behind selecting the older MPCTP linux kernel version was the fact that the \text{out-of-tree} version (MPCTPv0 \cite{githubmptcpv0}) was mature enough and already included most of the MPTCP packet scheduling and congestion control schemes as loadable kernel modules. The new upstream version (MPTCPv1 \cite{githubmptcpv1}) facilitates algorithms' development in user- rather than kernel-space (e.g., via eBPF); thus, MPTCPv1 lacked the needed algorithms, offering only the default minRTT packet scheduler.

\paragraph{MPTCP configuration} Delving into TCP configuration details, MPTCP can be enabled via the \texttt{mptcp\_enabled} parameter, and the desired packet scheduling and congestion control algorithm can be adjusted using the \texttt{mptcp\_scheduler} and \texttt{tcp\_congestion\_control} linux parameters, respectively \cite{mptcporg}. In our setup, the MPTCP default path manager, \textit{fullmesh} has been used throughout the entire experimentation campaign (configured via \texttt{mptcp\_path\_manager} parameter).  In addition, we have disabled \texttt{mptcp\_checksum} since our Mininet experimentation platform operates in a controlled environment, absent any middleboxes which could potentially alter the payload \cite{rfc8684}. Furthermore, \texttt{tcp\_no\_metrics\_save} has been enabled, flushing metrics from the route cache once a connection is closed \cite{HammarMScKarlstad, LinuxgazetteClearCache}. Except these, all the remaining TCP parameters maintained their default values, allowing a performance evaluation as closest to a normal setup as possible; thus, TCP autotuning, and selective acknowledgement (SACK) features have been kept enabled. Noteworthy here is the fact that autotuning has been experimentally associated with MPTCP performance degradation and lower aggregation benefit, as well as with slower congestion window increase within slow-start phase, when compared to the equivalent manual setup to the same maximum buffer limit \cite{10.1145/2535372.2535403}; however, we have selected to keep it enabled in our environment to emulate the default setup's use case. Another important condition to be considered in any such multipath performance evaluation scenario, either emulated or real-world, is to guarantee that the end-hosts are not receive-window limited. This means that the send and receive buffers should be allocated enough memory so that flow control is not the limiting factor in the experiment, but rather the congestion control algorithm via the congestion-window. Thus, it has been suggested that, for MPTCP, the maximum send and receive buffer size be set to a value of $2*BDP$, that is, $ 2*\sum_{i=1}^{n}
bw_i*RTT_{max}$, where $bw_i$ is the bandwidth of subflow $i$, and $RTT_{max}$ the highest RTT value measured across all subflows \cite{barre2011implementation, raiciu2012hard, 10.1145/2535372.2535403}. The logic behind this formula is that one BDP of memory is required to keep sending while waiting for a packet sent on the slowest path to be delivered; to keep sending when a path is used for fast retransmission, another BDP of buffer size is required \cite{raiciu2012hard}. In our system, where autotuning has been enabled, the maximum send and receive buffer size (\texttt{wmem} and \texttt{rmem}, respectively) have been set by default to 16MiB which is considered to be sufficient for our experiments and in accordance with the proposed value of \textit{at least} $2*BDP$ \cite{barre2011implementation}. Details on the exact bandwidth, delay, and packet loss rate value ranges of each scenario is provided later in this section, but for a quick calculation, our experiments consider two subflows with a maximum capacity of 100 Mbps each, an $RTT_{max}$ of 50ms, and a maximum packet loss rate of 2\%; these values require a buffer size of at least 1.25 MiB, which is covered by the existing autotuning upper limit.  

\paragraph{Loadable modules} The packet scheduling and congestion control algorithms can be loaded in the linux kernel at runtime using \texttt{insmod} or \texttt{modprobe} commands, without the need to modify, rebuild, and install a new kernel. As a result, some work was initially required to port all algorithms to the common running-kernel basis (i.e., MPTCP v0.96 \cite{githubmptcp:v096}). Consequently, we had to modify the legacy code for LLHD packet scheduler \cite{githubllhd}, and C-MPBBR congestion control algorithm \cite{githubcmpbbr}.

\paragraph{Additional tools} Concerning the set of tools used to conduct the experiments, we have exploited \texttt{iperf} v2.1.5 to generate traffic between the two Mininet hosts \cite{iPerfpage, iPerf3github}, \texttt{matplotlib} v3.10.0 for visualization of the collected results \cite{Hunter:2007}, as well as the \texttt{tcp\_probe} linux feature which allowed us to record the MPTCP connection state and retrieve kernel metrics during test execution \cite{TCPprobe, EnableTCPprobe}.

\subsection{Traffic methodology}
\label{subsec:traffic_method}
Using iperf tool, Host A (i.e., the iperf client) establishes a MPTCP connection with Host B (i.e., the iperf server). Afterwards, Host A sends traffic over the two subflows (SF1, and SF2) toward Host B, for a predetermined duration of 30 sec., trying to fill in the available link bandwidth; this type of traffic emulates bulk data transfer between two hosts. While this method would be considered equivalent to \textit{timed transfers} (i.e., performance tests measuring the time needed to complete a 10MB tranfer between two end nodes), there have been some deviations reported in literature \cite{brakmo1996end}. However, in this work we focus on \textit{"length transfers"} \cite{brakmo1996end}, leaving "timed transfer" verification for future work.

\subsection{Design of Experimentation Scenarios}
\label{subsec:exper_scenarios}

The main idea behind the design of the experimentation scenarios has been the quest for a universally superior congestion control and packet scheduling algorithm combination, which could potentially be suitable for generic use, regardless of the underlying path conditions. And if the experimentation campaign would not come up with such a distinguishable combination, it would still be of interest to identify which packet scheduling and congestion control algorithm combination is the most appropriate under specific network conditions.

In an attempt to assess the various packet scheduling and congestion control algorithm combinations under various path heterogeneity conditions, the Mininet emulator has been used. Execution in Mininet is performed over the real protocol stack instead of a simulated protocol model. Furthermore, each test iteration has been executed using a static network configuration for the entire test duration (30 sec.), which in addition to the absence of background traffic, made our simple network setup ideal for assessing the combined use of algorithms, eliminating the effect of other factors on the performance outcome. This allowed us to repeat identical experiments and yet expect deterministic results, while providing us with the flexibility to modify the underlying path conditions and cover a wide range of path heterogeneity scenarios.

Our experimentation campaign included scenarios of variable path heterogeneity and severity in terms of capacity, round-trip delay, and packet loss rate, as well as all possible combinations of the packet scheduling and congestion control algorithms included in \ref{tab:psa}\footnotemark and \ref{tab:cca}. This has been in accordance with the qualitative and quantitative factors that should be considered in such experiments, as mentioned in \cite{10.1145/2535372.2535403}.  

\footnotetext{ReMP TCP has been excluded from our experiments since, by duplicating packets, it clearly incurs sub-optimal performance in terms of goodput and overall flow completion time.}

The entire set of experiments is depicted in Table \ref{tab:experiment_scenarios}.

\begin{table}[htb!]
  \caption{Experimentation Scenarios}
  \label{tab:experiment_scenarios}  
  \begin{tabular}{lccc}
    \toprule
        Subflow & Capacity (Mbps) & Round-trip delay (ms) & Packet loss rate (\%)\\
    \midrule
       SF1 & [100] & [0, 5, 10] & [0, 1, 2]\\
       SF2 & [100, 75, 50, 5] & [0, 2, 5, 20, 50] & [0, 0.05, 0.5, 2]\\
  \bottomrule
\end{tabular}
\end{table}

In Table \ref{tab:experiment_scenarios}, SF1 and SF2 traffic characteristics are referred to the underlying Mininet link conditions (L1 and L2, respectively) over which the subflows have been established. Link L3 is statically configured to a $BW/RTT/PLR$ value of $2Gbps/0ms/0\%$ across the entire set of experiments. Based on the range of values described in Table \ref{tab:experiment_scenarios} which determine the underlying path characteristics, we have initially divided our experimentation campaign in five major scenario families: \textit{homogeneous}, \textit{mild\_heterogeneity}, \textit{intense\_heterogeneity}, \textit{very\_intense\_heterogeneity}, and finally  \textit{mixed\_heterogeneity}. Each scenario family, is composed of up to seven scenarios, where heterogeneity is assessed over individual traffic characteristics (i.e., only bandwidth-induced heterogeneity, while round-trip delay and packet loss maintain equal values among the two subflows), as well as over combinations of them. In all scenario families except the last one, one subflow (i.e., SF1) maintains static path configuration across all scenarios of the scenario family, while heterogeneity is induced by degrading path conditions of the second subflow (i.e., SF2 is inferior to SF1); in the scenarios belonging to the mixed\_heterogeneity scenario family, both subflows experience mixed path characteristics, thus not permitting any a priori assumption as to which subflow is expected to perform better and how the applied algorithms would react (e.g., SF1 is established over a bottleneck link of $BW/RTT/PLR$ equal to $100Mbps/10ms/0\%$, while SF2 is over a $100Mbps/2ms/0.5\%$). These five scenario families produced twenty-nine (29) scenarios in total. Each scenario is executed against all combinations of five selected MPTCP packet schedulers (minRTT, BLEST, ECF, RR\footnotemark, LLHD) and seven congestion control algorithms (LIA, OLIA, BALIA, wVegas, Cubic, BBR, C-MPBBR); for each specific packet scheduling and congestion control algorithm combination, we execute the same scenario five times, to ensure that the outcome of each test is not affected by any transient issues. The total number of test iterations is given by the following formula: 
\begin{equation} \label{eq:num_tests}
    \#\_tests = (\#\_scenarios) * (\#\_schedulers) * (\#\_CCAs) * (\#\_iterations\_per\_scenario)
\end{equation}
which produces 5,075 test instances; as mentioned earlier, each test instance has a duration of 30 sec. 

\footnotetext{Round-robin packet scheduling algorithm maintained its default parameter values throughout the entire experimentation trial; that is, the number of consecutive segments sent on a subflow before switching to the other subflow is kept by default to one ($\texttt{num\_segments}=1$), and the scheduler tries to fill in the congestion window ($\texttt{cwnd\_limited=true}$), which means that upon filling up the congestion window, becomes \textit{ack-clocked} \cite{mptcporg}.}

\subsection{Performance Metrics}
\label{subsec:perf_metrics}

Inspired by the work performed in \cite{cech2020analyzing}, we selected to provide a combined view of the per-subflow goodput, the per-subflow smoothed RTT (SRTT), the MPTCP out-of-order (OFO) queue size, as well as the per-subflow number of retransmissions for each packet scheduling algorithm. 

\paragraph{Goodput and Retransmissions}
Our performance evaluation results illustrate, among others, the achieved goodput which is essentially derived by subtracting the protocol overhead (i.e., packet headers) from throughput computation. While \texttt{iperf} tool provides a metric of the achieved goodput, this suffers two limitations: i) iperf tool provides the aggregate goodput achieved over both subflows of the MPTCP connection, while our intention was to depict how each packet scheduler performs, hence how traffic is actually distributed among the two subflows, and ii) iperf's goodput scale is calculated based on divisions by a factor of 1000. For these reasons, we have selected to analyze the per-subflow goodput, as well as to retrieve the per-subflow number of retransmissions by capturing and analyzing every single packet. The python-based \texttt{pyshark} tool \cite{githubPyShark} has been used for this purpose, allowing us to calculate goodput based on each individual packet's payload; our extracted per-subflow goodput scale has been calculated through divisions by a factor of 1024 (e.g., $1\,KiB=1024\,bits$). For instance, iperf goodput expressed in Mbps using SI units (e.g., KB, MB) would be calculated by the formula: $GP (Mbps) = (Bytes / duration)*8 / (1000*1000)$, while our calculations based on binary units (e.g., KiB, MiB) are given by the formula: $GP (Mbps) = (Bytes / duration)*8 / (1024*1024)$, where $GP$ stands for \textit{goodput}, and \textit{duration} denotes the duration of each test instance (30 sec.). Thus, our goodput results might be even more conservative than those iperf produced. \textit{Pyshark} allows the distinction between subflows via common \texttt{tshark} filters; this makes the calculation of per-subflow goodput and the respective number of retransmissions easy.

\paragraph{SRTT and OFO queue}
The SRTT \cite{rfc6298} as well as the OFO queue size values have been collected by tracing the respective linux kernel parameters. Noteworthy here is mentioning that for as long as the kernel tracing feature has been enabled (that is, for the 30 sec. duration of each test instance's execution), a huge amount of kernel logs containing metrics had been produced; thus, as part of a post-processing step, we randomly sampled the collected metrics in order to limit, for instance, the out-of-order queue value samples to two per millisecond, that is, $1 \,sample \,/\, 500us$. Part of the collected metrics has also been the congestion window value of each subflow; however, we have selected to skip any congestion window plots and to focus on basic performance metrics.

Besides providing a direct performance comparison between the packet scheduling algorithms under a specific traffic scenario and under the presence of a single congestion control algorithm, the main focus of this work has been to assess the packet scheduling and congestion control algorithms across all traffic scenarios. For this reason, as it will be explained in detail within Section \ref{sec:evaluation}, a couple of additional assessment metrics have been used, such as the \textit{normalized goodput} achieved by each packet scheduler across a scenario family; a \textit{score} definition is also provided within Section \ref{sec:evaluation}, wherever deemed necessary. Finally, despite the fact that our traffic pattern resembles bulk transfer, we considered it useful to provide an overview of the per-packet delay which is an important metric in case of interactive applications. As such, we calculate the per-packet delay based on the following formulas:

\begin{enumerate}[label=\roman*.]
    \item we first calculate the number of packets-per-second (pps):
        \begin{equation} \label{eq:pps_metric}
            pps = total\_Bytes / (MSS * test\_duration)
        \end{equation} where $MSS=1514\,Bytes$ and test duration is $30\,sec.$
    \item then, the per-packet-delay in milliseconds is given by: 
        \begin{equation} \label{eq:ppd_metric}
            per\_packet\_delay\, (ms) = (1 / pps) * 1000
        \end{equation}
\end{enumerate}
Another performance metric to be also explained in detail within Section \ref{sec:evaluation}, is the \textit{coefficient of variation (CV)} which is given by $\frac{\sigma}{\mu}$, and is used when calculating the congestion control algorithm scores as a measurement of the extent of variability in relation to the mean.


\section{Evaluation}
\label{sec:evaluation}
This section is dedicated to the overall evaluation of the packet scheduling and congestion control algorithms. We first present a subset of figures, illustrating how the MPTCP packet scheduling algorithms performed across different congestion control algorithm (CCA) combinations, under some representative scenarios which resemble real-world network conditions. Then, we define two new metrics to assess the performance of the packet scheduling and the congestion control algorithms, that is, the \textit{PS\_score} and the \textit{CCA\_score}, respectively. The \textit{PS\_score} metric is based on the normalized goodput in order to assess the performance level of each packet scheduling algorithm within each scenario family. The \textit{CCA\_score} metric, which is based on the normalized \textit{PS\_score}, is used to assess the performance of the congestion control algorithms across all sccenario families. This metric provides the means to quantify the performance of each CCA through the normalized score that the underlying packet scheduling algorithms achieved using the respective CCA. The \textit{CCA\_score} indicates, in essence, how much "space" each CCA provides to the underlying packet scheduling algorithms to unveil their sophisticated scheduling logic. If a congestion control algorithm suppresses the congestion window of a subflow, then the packet scheduling algorithm cannot perform any better than applying its logic on the resources that the CCA makes available to the scheduler; this means that a bad congestion control algorithmic logic is anticipated to incur a poor performance outcome, regardless of the packet scheduling algorithm's intrinsic properties. 

Once we identify the "best" three congestion control algorithms, we proceed with a fine-grained assessment of the various packet scheduling algorithms when combined with each of the highest-score CCAs. The \textit{normalized goodput} and the \textit{per-packet delay} are the main metrics against which all packet scheduling algorithms are evaluated. The \texttt{Matplotlib} python library has been used for the visualization of the results \cite{Hunter:2007}.

Noteworthy here is that we have deliberately decided to maintain different axes' scale when plotting the values of some parameters across the various CCAs, such as the SRTT, the MPTCP OFO queue size, and the number of retransmissions. This allows to increase the granularity and, consequently, the visibility of the results, preventing subtle variations from being shadowed by any neighboring CCA's more coarse results within the same figure.

\subsection{Cross-CCA packet scheduling overview}
\label{subsec:cross_cca_ps_overview}
From the 1,015 unique tests executed (5,075 tests / 5 iterations per test), we have selected to present in this work a subset of the most representative scenarios. To this end, we hereby present: 
\begin{enumerate}[label=\roman*.]
    \item a scenario involving homogeneous paths,
    \item a scenario involving paths experiencing intense heterogeneity, and
    \item a scenario involving path conditions of mixed heterogeneity.
\end{enumerate}

\begin{figure}[htb!]
  \centering
  \includegraphics[width=\linewidth]{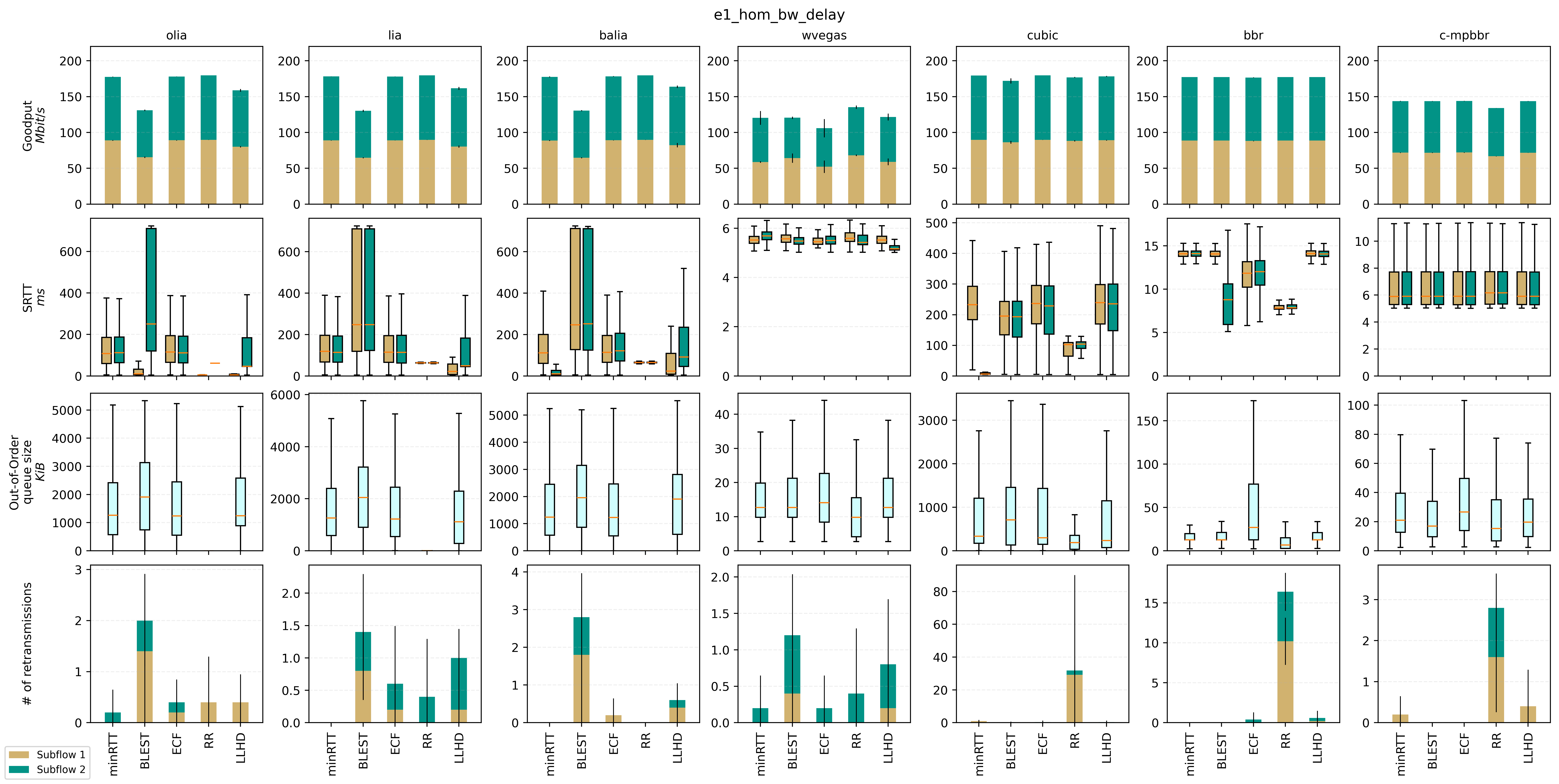}
  \caption{Homogeneous scenario - BW/RTT/PLR - $SF_1: 100\,Mbps/5\,ms/0\%,\, SF_2: 100\,Mbps/5\,ms/0\%$}
  \label{fig:e1_hom}
  \Description[Homogeneous scenario's performance figure]{Homogeneous scenario's performance figure, depicting how each scheduler performs under the various congestion control algorithm schemes in terms of goodput, SRTT, OFO queue size, and number of retransmissions.}
\end{figure}

\paragraph{Homogeneous scenario}
Figure \ref{fig:e1_hom} presents the results of a scenario derived from the \textit{homogeneous} scenario family, where both subflows traverse paths which experience common networking characteristics. When an equal delay of 5ms is present on both paths, minRTT, ECF, and RR outperform LLHD and BLEST under LIA CC, and its variations OLIA and BALIA. All schedulers follow the same performance pattern across these three CCs. It is evident that even when the CCA (i.e., LIA/OLIA/BALIA) leaves the congestion window of both subflows open, BLEST, which bases its decisions on send window blocking estimations, selects to equally limit the use of both subflows, underutilizing the available bandwidth of both paths, and eventually underperforming. wVegas, which is a delay-based CC, is highly sensitive to delay, interpreting it to congestion and capping aggressively the congestion window of both subflows; the aggregate goodput achieved by all schedulers is limited compared to the one achieved under the use of other CCAs. On the contrary, the use of Cubic, which is a loss-based uncoupled CCA, provides a clear benefit to all schedulers. BLEST’s performance improvements can be credited to the uncoupled nature of Cubic, that is, there is no upper bound for the aggregate congestion window of the two subflows, as is the case for coupled CCAs. Thus, each of the subflows maintains its individual congestion window which is treated by the employed CCA as if it were for a singe path flow belonging to a normal TCP connection. This makes MPTCP unfair to single path TCP flows, since each MPTCP subflow can allocate a portion of BW equal to that of a single path TCP flow, ultimately leading single path TCP flows to starvation when single path TCP and MPTCP subflows coexist in the same bottleneck link. In this sub-scenario, BBR allows all schedulers to reach high and stable performance; it is also noticeable that BBR’s accurate estimations of the bottleneck bandwidth and RTT lead to scheduling decisions which incur the lowest SRTT and very low OFO queue occupancy across all the employed CCAs. C-MPBBR, which is a coupled variant of BBR for MPTCP (ensuring that MPTCP has a clear benefit over and is fair to SP-TCP), incurs a lower goodput aggregation to all schedulers’ performance, when compared to uncoupled BBR. The fairness C-MPBBR achieves across the two MPTCP subflows traversing the same bottleneck link, can be clearly seen in the measured SRTT of the subflows for each scheduler; however, its coupled nature and fairness benefits come at the cost of a more restricted aggregate congestion window which limits the performance of all schedulers over the two subflows. The overall $e1\_hom\_bw\_delay$ scenario's verdict can be summarized as follows:
\begin{itemize}
    \item minRTT, ECF, and RR exhibit high performance under LIA/OLIA/BALIA CCAs
    \item BLEST is sensitive to delay and underperforms the rest when any one of the LIA/OLIA/BALIA is the employed CCA
    \item wVegas is a delay-based CCA and reacts more aggressively to delay, restricting the available congestion window, and leading all schedulers to low goodput aggregation over the two paths
    \item all schedulers perform comparably well under Cubic
    \item all schedulers perform comparably well and stably under BBR
    \item C-MPBBR bounds the coupled congestion window in favor of fairness and MPTCP’s benefit over SPTCP; this has a considerable impact on the performance of all schedulers when compared to that achieved with C-MPBBR’s rivals (i.e., uncoupled BBR and Cubic)
\end{itemize} 

\begin{figure}[htb!]
  \centering
  \includegraphics[width=\linewidth]{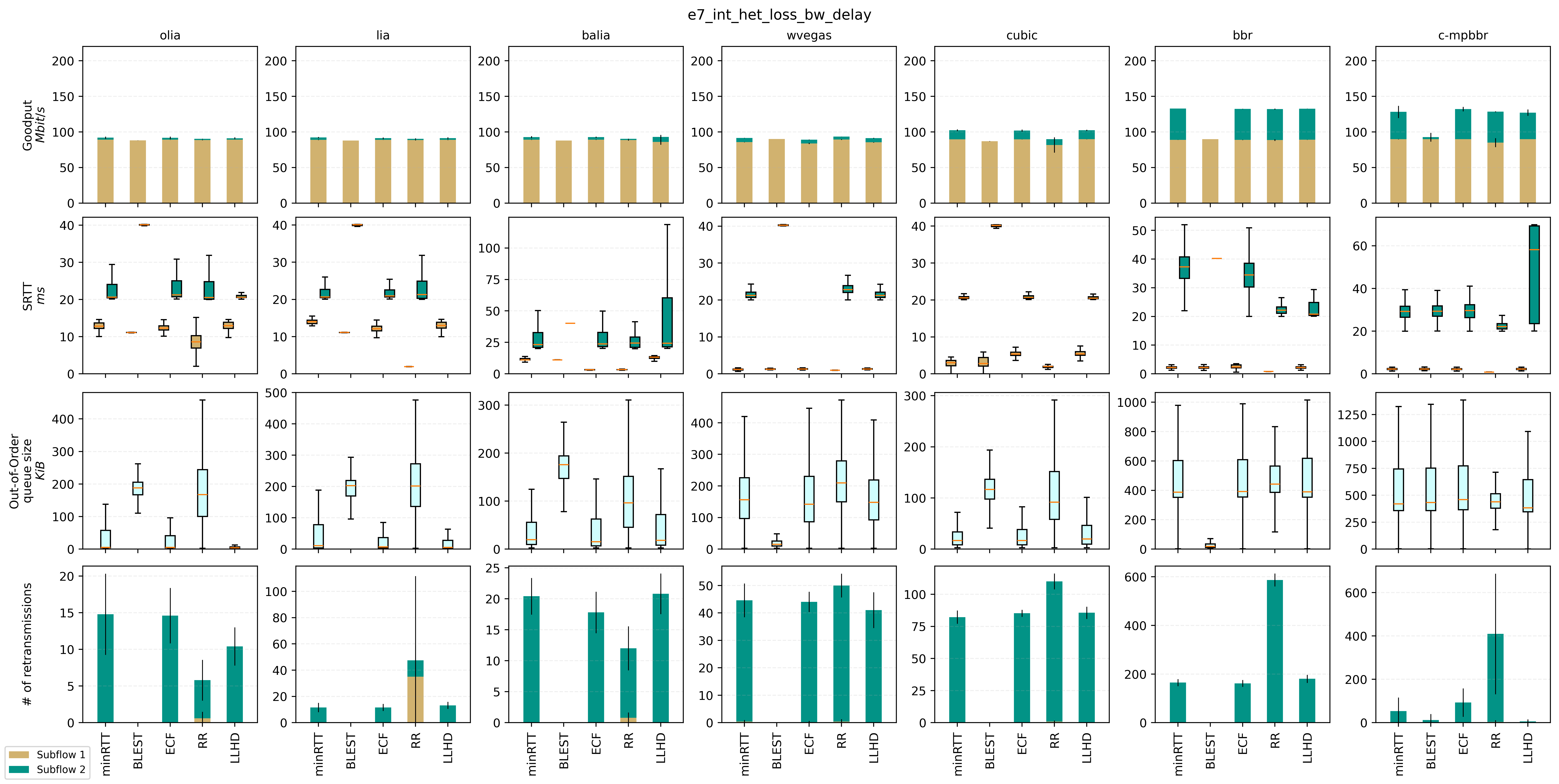}
  \caption{Intense heterogeneity scenario - BW/RTT/PLR - $SF_1: 100\,Mbps/0\,ms/0\%,\, SF_2: 50\,Mbps/20\,ms/0.5\%$}
  \label{fig:e7_int}
  \Description[Intense-heterogeneity scenario's performance figure]{Intense-heterogeneity scenario's performance figure, depicting how each scheduler performs under the various congestion control algorithm schemes in terms of goodput, SRTT, OFO queue size, and number of retransmissions.}
\end{figure}

\paragraph{Intense\_heterogeneity scenario}
Figure \ref{fig:e7_int} presents the results of a scenario derived from the \textit{intense\_heterogeneity} scenario family, where the two subflows traverse paths with intense-heterogeneity networking characteristics. The additional presence of delay on a path (i.e., SF2) on top of intense bandwidth restrictions and packet loss, has a detrimental effect on the goodput of all schedulers, when combined with any one of the traditional loss- and delay-based CCAs. All schedulers fail to efficiently utilize the available bandwidth of the affected path (SF2) when combined with any one of the LIA/OLIA/BALIA/wVegas/Cubic; this result is mainly credited to the effects of the congestion control approaches that the aforementioned CCAs follow; that is, increasing the congestion window during the CA (congestion avoidance) phase, which in turn allows intermediate buffers to be filled up, and then trying to tackle the congestion they created by drastically decreasing the congestion window, which deteriorates the already adverse conditions even further. Under such circumstances, schedulers cannot perform any better than trying to efficiently utilize the resources that the CCA makes available; CCA has a great impact (either positive or negative) on the final performance. The craftsmanship in the design of BBR and C-MPBBR, lies in the way they estimate the bottleneck bandwidth and the round-trip propagation time, thus pacing traffic (via the cwnd) at the highest possible rate which, however, does not lead to buffer formation; when buffers start being occupied, BBR can detect the RTT increase and adjust the rate efficiently until the queue is drained and a lower RTT is measured. This different logic allows BBR to sustain traffic rate close to Kleinrock’s optimal operating point, avoiding the post-congestion control operation of traditional CCAs (AIMD). BBR and C-MPBBR are rather adjusting traffic smoothly by avoiding high queue occupancies and the subsequently increased congestion; this allows all schedulers to unleash their full potential and apply their scheduling logic towards high goodput rates. In this sub-scenario, minRTT, ECF, RR, and LLHD perform comparably well, achieving close to optimal goodput; BLEST is not limited by the CCA decisions, but rather applies its algorithmic logic, which is based on limiting or eliminating the use of a slow path if this is anticipated to cause HoL and, consequently, MPTCP send window blocking (due to inflight data carried over the slower path). According to this logic, BLEST selects not to use the affected path (SF2), underutilizing its available bandwidth and thus exhibiting the lowest performance among schedulers. The overall $e7\_int\_het\_loss\_bw\_delay$ scenario's verdict can be summarized as follows:
\begin{itemize}
    \item the decisions of traditional delay- and loss-based CCAs (i.e., LIA/OLIA/BALIA/wVegas/Cubic) have a severe impact on any scheduler’s performance; when these CCAs have to deal with intense delay in conjunction with capped bandwidth and packet loss, their decisions have a disastrous effect on the final goodput, irrespective of the employed scheduler
    \item BBR (and its MPTCP coupled variant C-MPBBR) are designed in a way that does not aggressively suppress the congestion window of an already adversely impacted flow, thus allowing schedulers to achieve and sustain high goodput
    \item BLEST avoids the use of the slow path, either because the congestion window of the respective subflow is suppressed by a traditional CCA (i.e., LIA/OLIA/BALIA/wVegas/Cubic), or because of its own algorithmic logic to avoid slow paths in case this is expected to cause send window blocking
    \item RR experiences the highest number of retransmissions among schedulers
\end{itemize}

\begin{figure}[htb!]
  \centering
  \includegraphics[width=\linewidth]{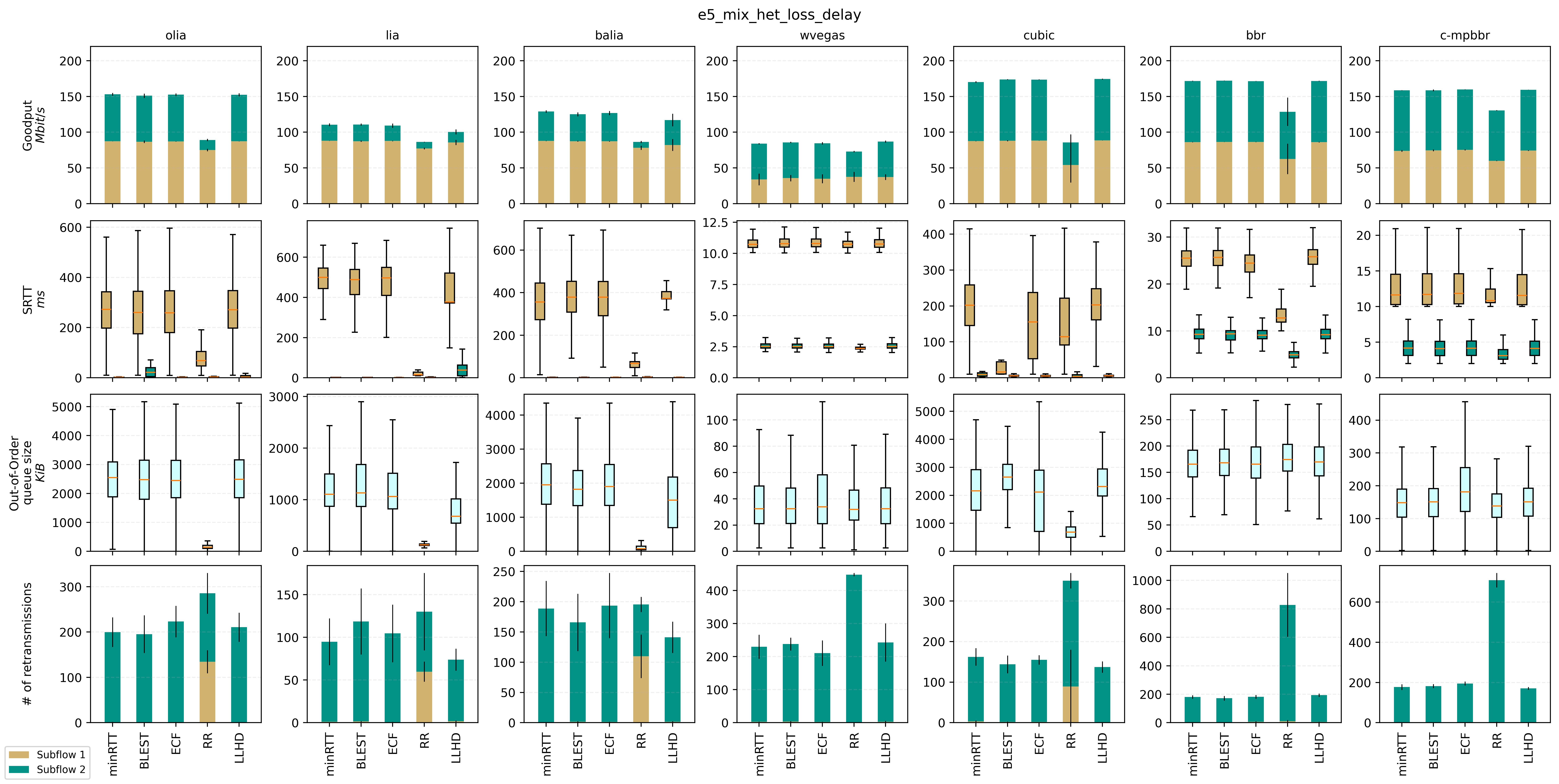}
  \caption{Mixed heterogeneity scenario - BW/RTT/PLR - $SF_1: 100\,Mbps/10\,ms/0\%,\, SF_2: 100\,Mbps/2\,ms/0.5\%$}
  \label{fig:e5_mix}
  \Description[Mixed-heterogeneity scenario's performance figure]{Mixed-heterogeneity scenario's performance figure, depicting how each scheduler performs under the various congestion control algorithm schemes in terms of goodput, SRTT, OFO queue size, and number of retransmissions.}
\end{figure}

\paragraph{Mixed\_heterogeneity scenario}
Figure \ref{fig:e5_mix} presents the results of a scenario derived from the \textit{mixed\_heterogeneity} scenario family, where the two subflows traverse paths experiencing mixed-heterogeneity networking conditions\footnotemark.
\footnotetext{Mixed heterogeneity scenarios emulate complex networking conditions where each of the paths is superior to the other when considering only a single factor; be it the allocated bandwidth, or the experienced RTT, or the packet loss rate. In such cases, however, it is sometimes not evident beforehand which path is eventually going to render an overall better performance.}
In this particular mixed delay- and loss-induced heterogeneity sub-scenario, there is a clear distinction between schedulers’ achieved goodput across CCAs, illustrating the effect that the various CCAs may have in scheduling. The highest scheduling performance is achieved when schedulers are combined with BBR; in that case, all schedulers (except for RR) achieve comparably high goodput over both subflows. Equally high goodput rates (to those observed with BBR) are reached by schedulers in case of uncoupled loss-based CCA Cubic; RR however is severely degraded. The next higher goodput rates are achieved when schedulers are combined with C-MPBBR; all (except RR) are performing comparably high, but they experience some goodput restriction over the higher-delay path (SF1), as compared to that achieved with BBR. This can be credited to the restrictions that C-MPBBR’s coupled nature imposes on the subflows that are considered to be traversing the same bottleneck path. With the couped loss-based OLIA, the goodput achieved over SF2 is slightly restricted (compared to the one achieved with BBR / C-MPBBR / Cubic), since the coupled nature of OLIA in conjunction with its more aggressive reaction to the packet loss detected over SF2, lead to further cwnd restrictions. The same goodput pattern among schedulers, with a higher goodput restriction over SF2, is also visible with BALIA, while with LIA, SF2’s goodput is limited even further. Finally, wVegas has a detrimental effect on schedulers' performance; all schedulers underperform over both subflows, being able to eventually reach an aggregate goodput which is less than a half of the one achieved with BBR. The overall $e5\_mix\_het\_loss\_delay$ scenario's verdict can be summarized as follows:
\begin{itemize}
    \item minRTT, BLEST, ECF, and LLHD perform comparably to one another, within each CCA combination
    \item BBR, C-MPBBR, and Cubic provide ideal conditions to the underlying schedulers, allowing them to achieve very high goodput rates
    \item OLIA is the third best (in terms of schedulers’ performance) CCA, allowing all schedulers to reach high goodput rates; goodput over the lossy path (SF2) is restricted more due to the loss-based nature of OLIA which caps the cwnd to tackle congestion, not being able to distinguish random loss from actual link congestion
    \item BALIA and LIA limit the SF2 cwnd further, leading all schedulers to moderate performance
    \item wVegas has a detrimental effect upon all schedulers' performance, due to its delay-based congestion inference
    \item RR experiences the highest number of retransmissions and underperforms, regardless of CCA selection
\end{itemize}

\subsection{PS and CCA score metrics}
\label{subsec:ps_cca_score_metrics}
As mentioned earlier, to evaluate the performance of the various packet scheduling and congestion control algorithms, we have introduced two new metrics; the \textit{PS\_score} and the \textit{CCA\_score}, respectively. 

\paragraph{PS\_score}
For each scheduler, we first calculate the \textit{average goodput} it achieved in each sub-scenario of each broader scenario family, over the five identical execution runs. Then, we calculate a \textit{PS\_score} for each packet scheduler per scenario family; the \textit{PS\_score} is actually the \textit{normalized aggregate goodput} for each scheduler in each scenario family, and is given by the formula:
\begin{equation} \label{eq:ps_score_metric}
    PS\_score = \sum_{\#subscen}^{}\left(\frac{subscen\_avg\_GP}{\#subscen\_in\_ScenFam\,*\,theoretical\_max\_gp}\right)
\end{equation}
where \textit{\#subscen} denotes the number of sub-scenarios included in each respective scenario family, \textit{subscen\_avg\_GP} is the average goodput achieved over the five identical iterations of each respective sub-scenario, \textit{\#subscen\_in\_ScenFam} denotes the number of sub-scenarios included within a broader scenario family, and the \textit{theoretical\_max\_gp} denotes the theoretical maximum aggregate goodput over the two subflows that is, $200\, Mbps$. This way, $PS\_score \in [0,1)$ .

\paragraph{CCA\_score}
A similar approach has been followed in order to define a metric which would allow us to assess the performance of the congestion control algorithms. We first present how each congestion control algorithm performed within each broader scenario family. This kind of assessment is essentially showcasing how the packet scheduling algorithms performed under each respective congestion control algorithm within each broader scenario family. Thus, each CCA's \textit{CCA\_score\_per\_ScenFam} is computed based on the average \textit{PS\_score} of the packet scheduling algorithms when combined with the respective CCA. The \textit{CCA\_score\_per\_ScenFam} is provided by the following formula:
\begin{equation} \label{eq:cca_score_per_ScenFam_metric}
    CCA\_score\_per\_ScenFam = \sum_{\#schedulers}^{}\left(\frac{PS\_score}{\#schedulers}\right)
\end{equation}
where \textit{\#schedulers} denotes the number of all packet scheduling algorithms, and \textit{PS\_score} is the respective score each packet scheduling algorithm received for this scenario family. For this metric, it also applies that $CCA\_score\_per\_ScenFam \in [0,1)$.

\subsection{Congestion control assessment}
\label{subsec:cca_assess}

\begin{figure}[htb!]
  \centering
  \includegraphics[scale=0.5]{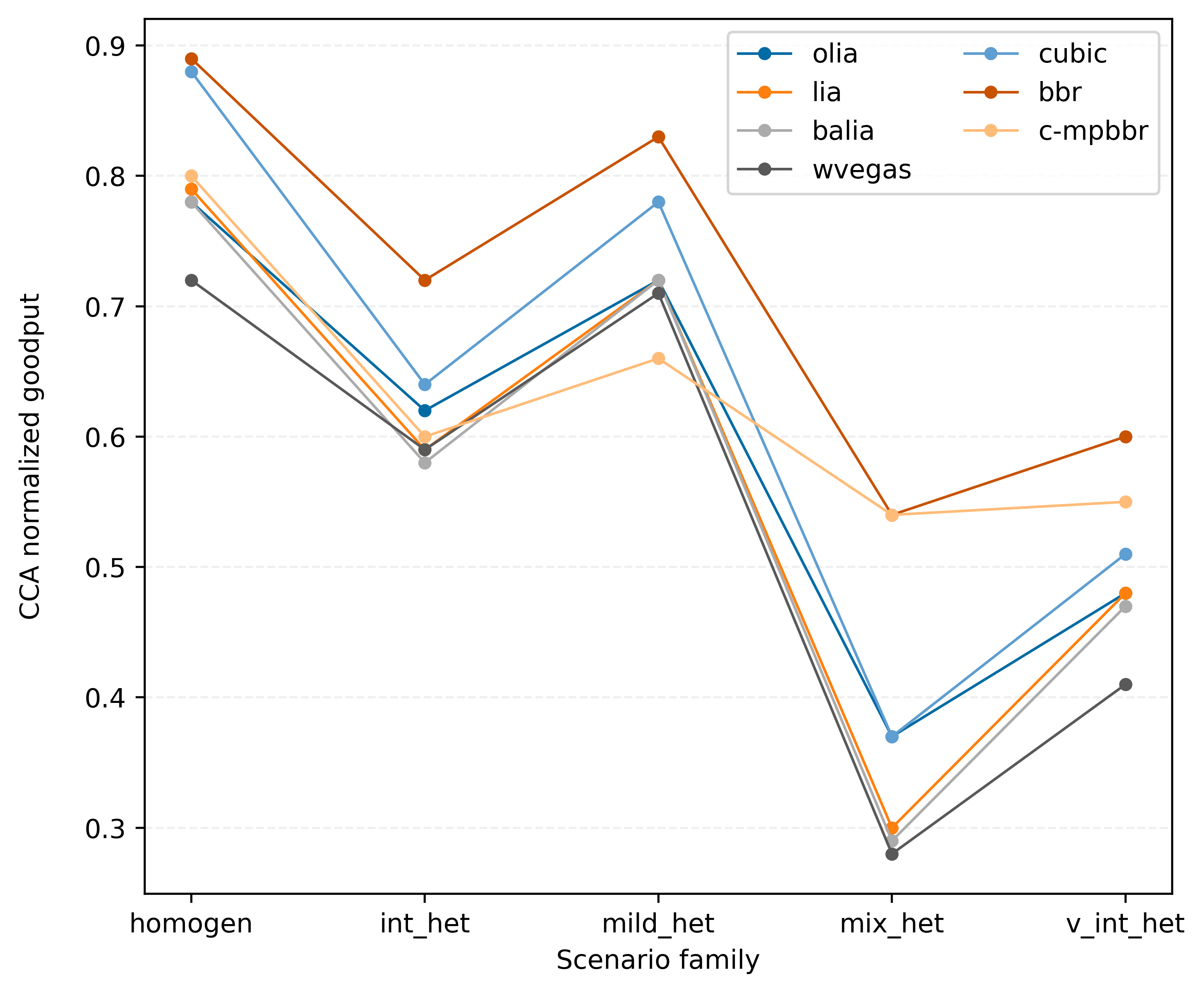}
  \caption{CCA score per scenario family}
  \label{fig:CCA_score_per_scen_fam}
  \Description[CCA score per scenario family]{This figure illustrates the score achieved by each congestion control algorithm in each scenario family.}
\end{figure}

The $CCA\_score\_per\_ScenFam$ reached by each congestion control algorithm in each scenario family, is depicted in Figure \ref{fig:CCA_score_per_scen_fam}.

Once an overview had been obtained on how each congestion control algorithm performed within each scenario family, it was interesting to identify the three best-performing CCAs across all scenario families. To this end, we defined the \textit{CCA\_score} which is essentially an average over the scores each congestion control algorithm has been assigned for its performance within each scenario family. \textit{CCA\_score} is consequently a \textit{cross-scenario-family} score indication, given by the formula:
\begin{equation} \label{eq:cca_score_metric}
    CCA\_score = \sum_{\#ScenFam}^{}\left(\frac{CCA\_score\_per\_ScenFam}{\#ScenFam}\right)
\end{equation}
where \textit{\#ScenFam} is the number of the different scenario families. 

\begin{figure}[htb!]
\centering
\begin{minipage}[t]{.45\textwidth}
  \centering
  \includegraphics[width=\linewidth]{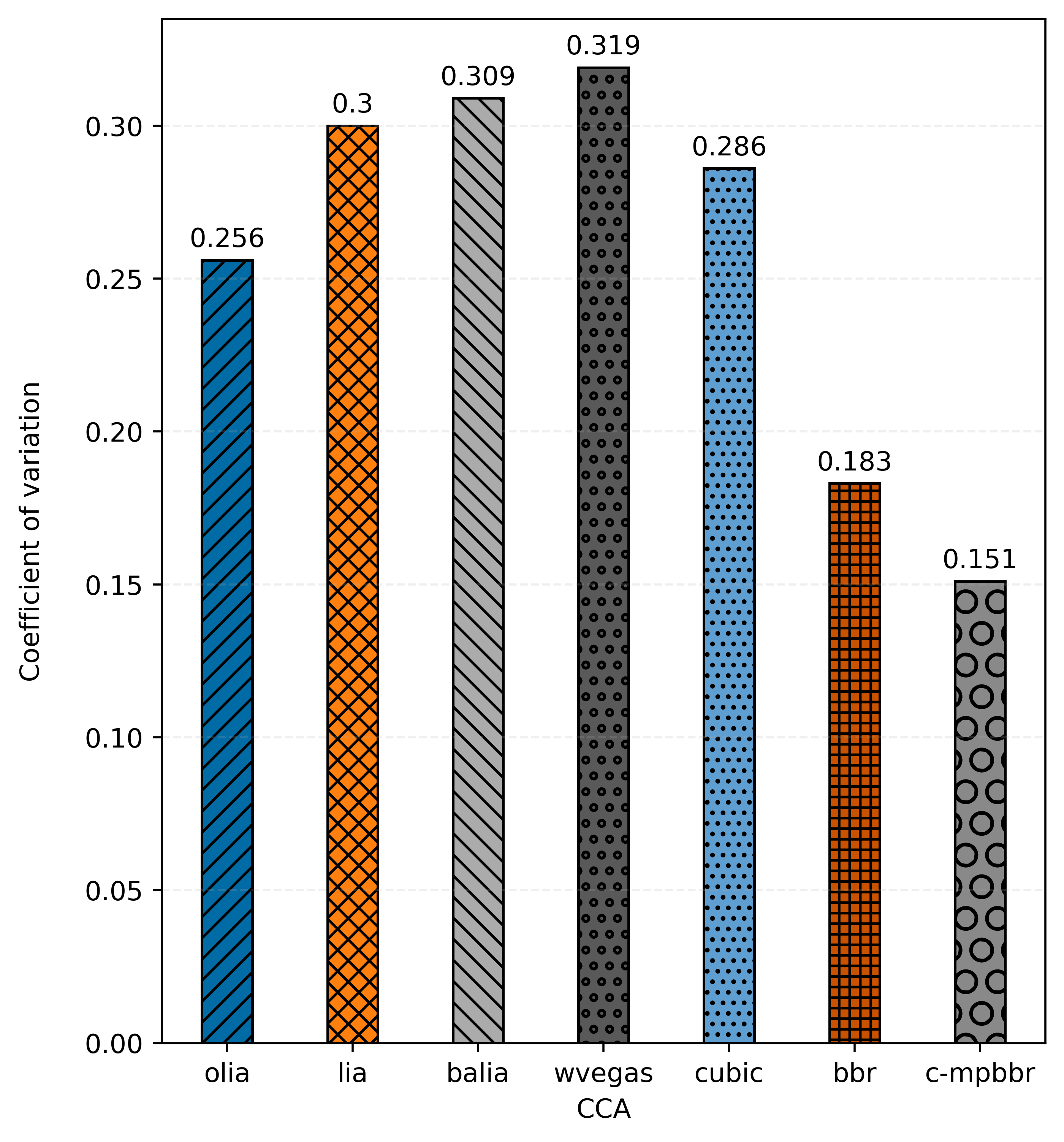}
  \captionof{figure}{CCA\_score CV}
  \label{fig:CCA_score_CV}
  \Description[CCA\_score coefficient of variation]{This figure depicts the coefficient of variation for each congestion control algorithm's CCA\_score.}
\end{minipage}
\hfill
\begin{minipage}[t]{.45\textwidth}
  \centering
  \includegraphics[width=\linewidth]{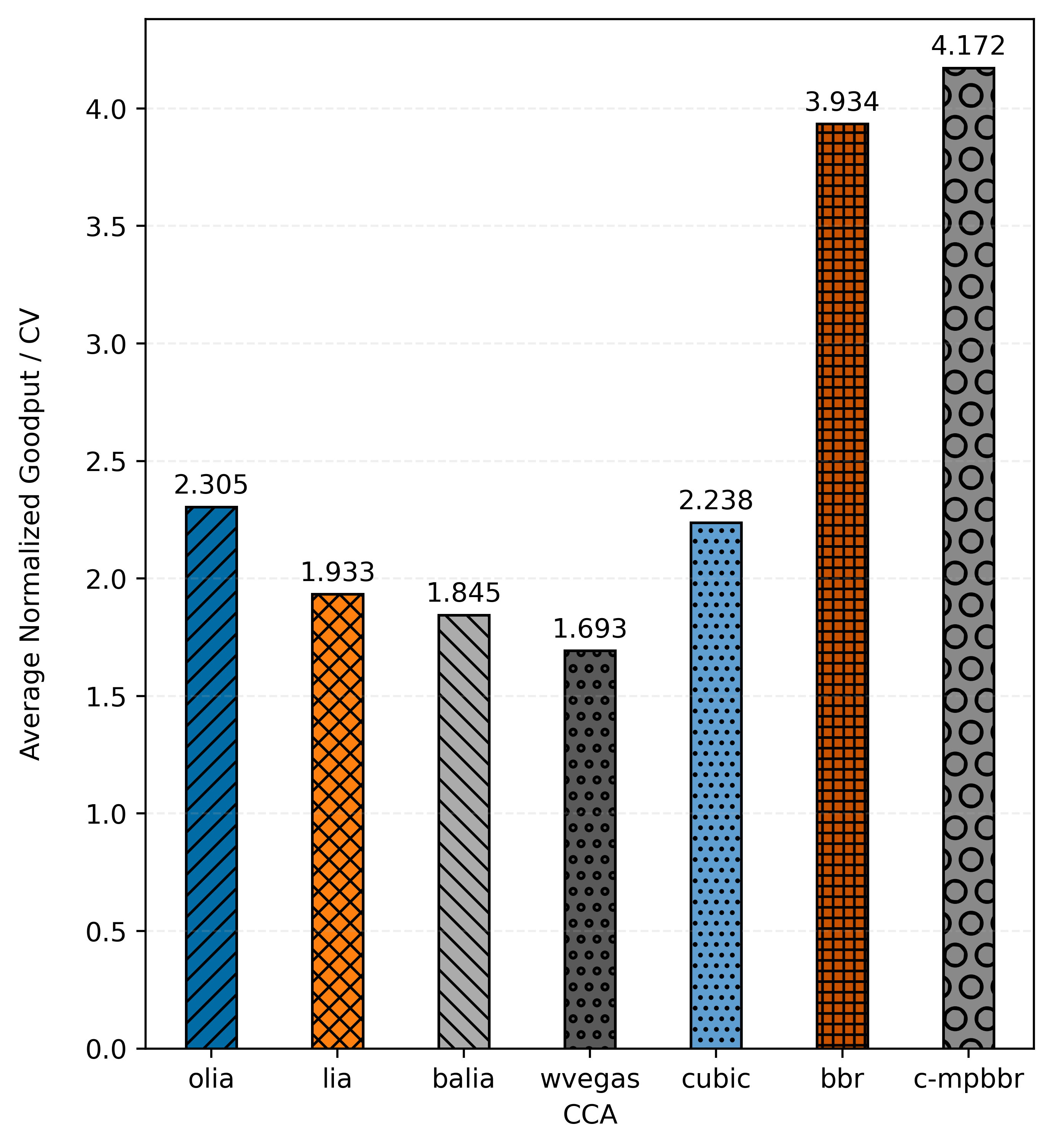}
  \captionof{figure}{CCA\_overall\_score}
  \label{fig:CCA_overall_score}
  \Description[CCA\_overall\_score]{This figure depicts the overall score achieved by each congestion control algorithm.}
\end{minipage}
\end{figure}

The drawback, however, of the above-mentioned metric is the fact that it does not consider how poorly each congestion control algorithm performed in its worst-case scenario family. Thus, to allow also for performance stability across the various scenario families, we combined the \textit{CCA\_score} with its respective \textit{coefficient of variation (CV)}, which provides insight on the extent of variability in relation to the mean. Figure \ref{fig:CCA_score_CV} illustrates the coefficient of variation for each individual \textit{CCA\_score}.

The lower the coefficient of variation, the higher the performance stability of each CCA across the various sub-scenarios. Dividing each CCA's \textit{CCA\_score} by its respective coefficient-of-variation value, we derive an \textit{overall score} for each congestion control algorithm, which allowed us to identify the best three in terms of goodput and performance stability across scenario families. The \textit{CCA\_overall\_score} is then given by the below formula:
\begin{equation} \label{eq:cca_overall_score_metric}
    CCA\_overall\_score = \frac{CCA\_score}{CCA\_CV}
\end{equation}

Figure \ref{fig:CCA_overall_score} illustrates each congestion control algorithm's \textit{CCA\_overall\_score}. This way, we have been then able to identify the three best-performing congestion control algorithms: \textbf{C-MPBBR}, \textbf{BBR}, and \textbf{OLIA}.

\subsection{Packet scheduling assessment - Goodput}
\label{subsec:psa_assess_gp}

\begin{figure}[htb!]
  \centering
  \includegraphics[width=\linewidth]{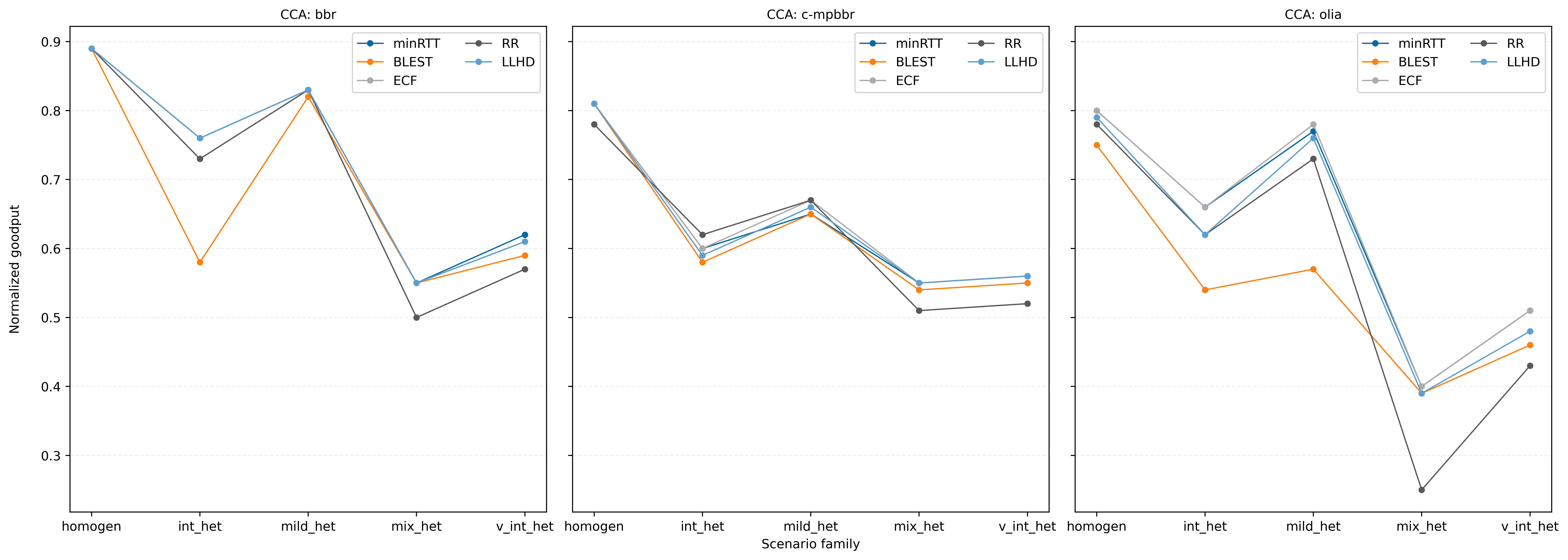}
  \caption{PS\_score}
  \label{fig:PS_score}
  \Description[PS\_score]{This figure depicts each packet scheduling algorithm's performance score per CCA, and per scenario family.}
\end{figure}

Having identified the three best-performing CCAs, we have then been able to delve into the packet scheduling algorithms' performance evaluation. The \textit{PS\_score} metric is indicative on how each scheduler performed within each scenario family, under each of the best three congestion control algorithms. Figure \ref{fig:PS_score} illustrates each packet scheduling algorithm's performance score per CCA, and per scenario family.

\begin{figure}[htb!]
  \centering
  \includegraphics[width=\linewidth]{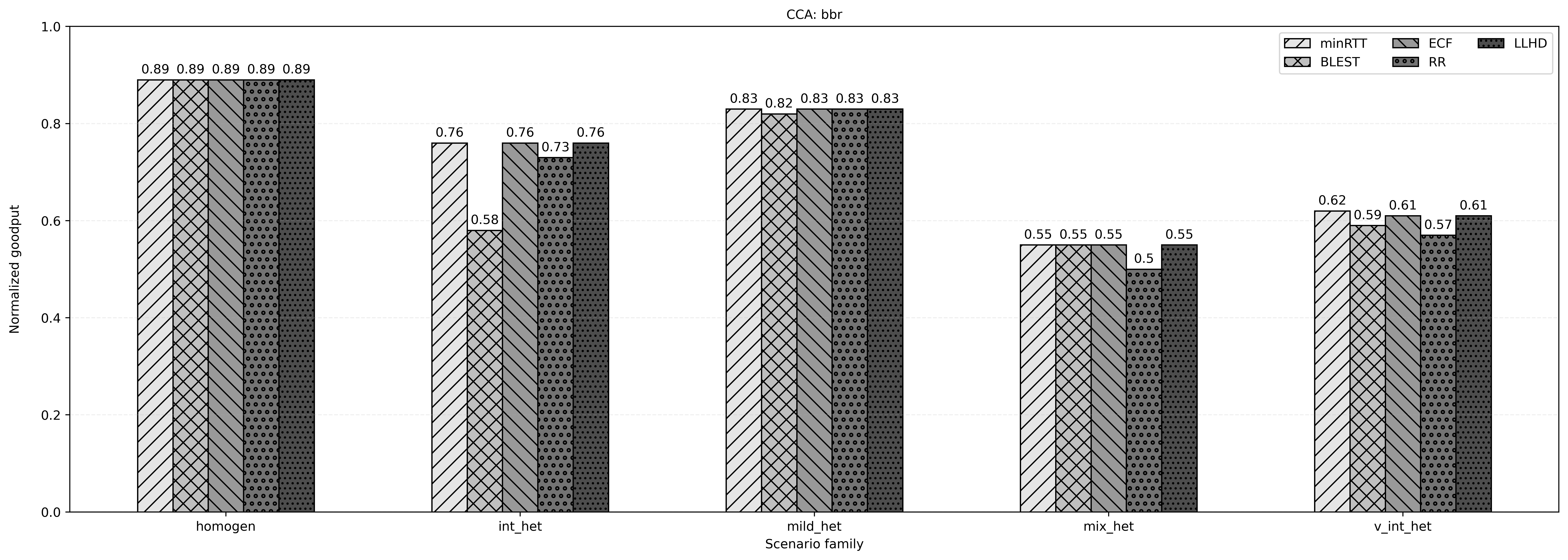}
  \caption{PS\_score per scenario family for BBR}
  \label{fig:PS_score_per_ScenFam_BBR}
  \Description[PS\_score\_per\_ScenFam\_BBR]{This figure depicts each packet scheduling algorithms performance score per scenario family, when combined with BBR congestion control algorithm.}
\end{figure}

\begin{figure}[htb!]
  \centering
  \includegraphics[width=\linewidth]{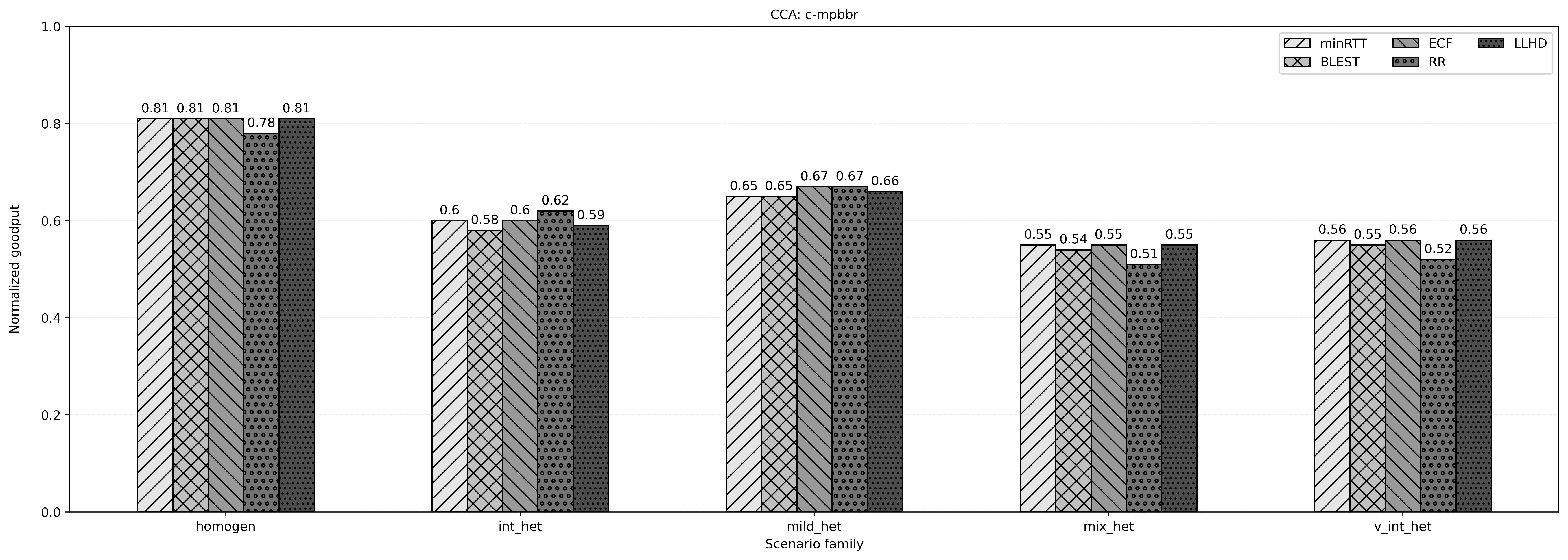}
  \caption{PS\_score per scenario family for C-MPBBR}
  \label{fig:PS_score_per_ScenFam_C-MPBBR}
  \Description[PS\_score\_per\_ScenFam\_C-MPBBR]{This figure depicts each packet scheduling algorithms performance score per scenario family, when combined with C-MPBBR congestion control algorithm.}
\end{figure}

\begin{figure}[htb!]
  \centering
  \includegraphics[width=\linewidth]{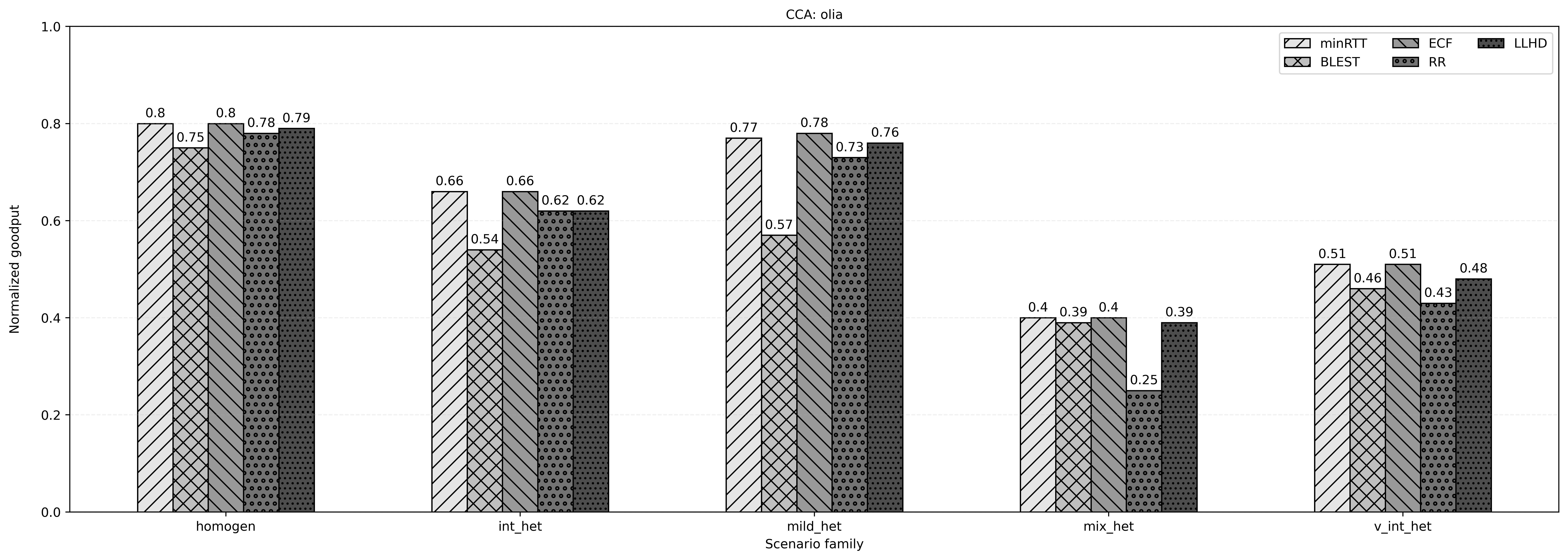}
  \caption{PS\_score per scenario family for OLIA}
  \label{fig:PS_score_per_ScenFam_OLIA}
  \Description[PS\_score\_per\_ScenFam\_OLIA]{This figure depicts each packet scheduling algorithms performance score per scenario family, when combined with OLIA congestion control algorithm.}
\end{figure}

A more granular \textit{PS\_score} analysis per scenario family is provided for each one of the BBR, C-MPBBR, and OLIA CCAs, within Figures \ref{fig:PS_score_per_ScenFam_BBR}, \ref{fig:PS_score_per_ScenFam_C-MPBBR}, and \ref{fig:PS_score_per_ScenFam_OLIA}, respectively.

\begin{figure}[htb!]
\centering
\begin{minipage}[t]{.3\textwidth}
  \centering
  \includegraphics[width=\linewidth]{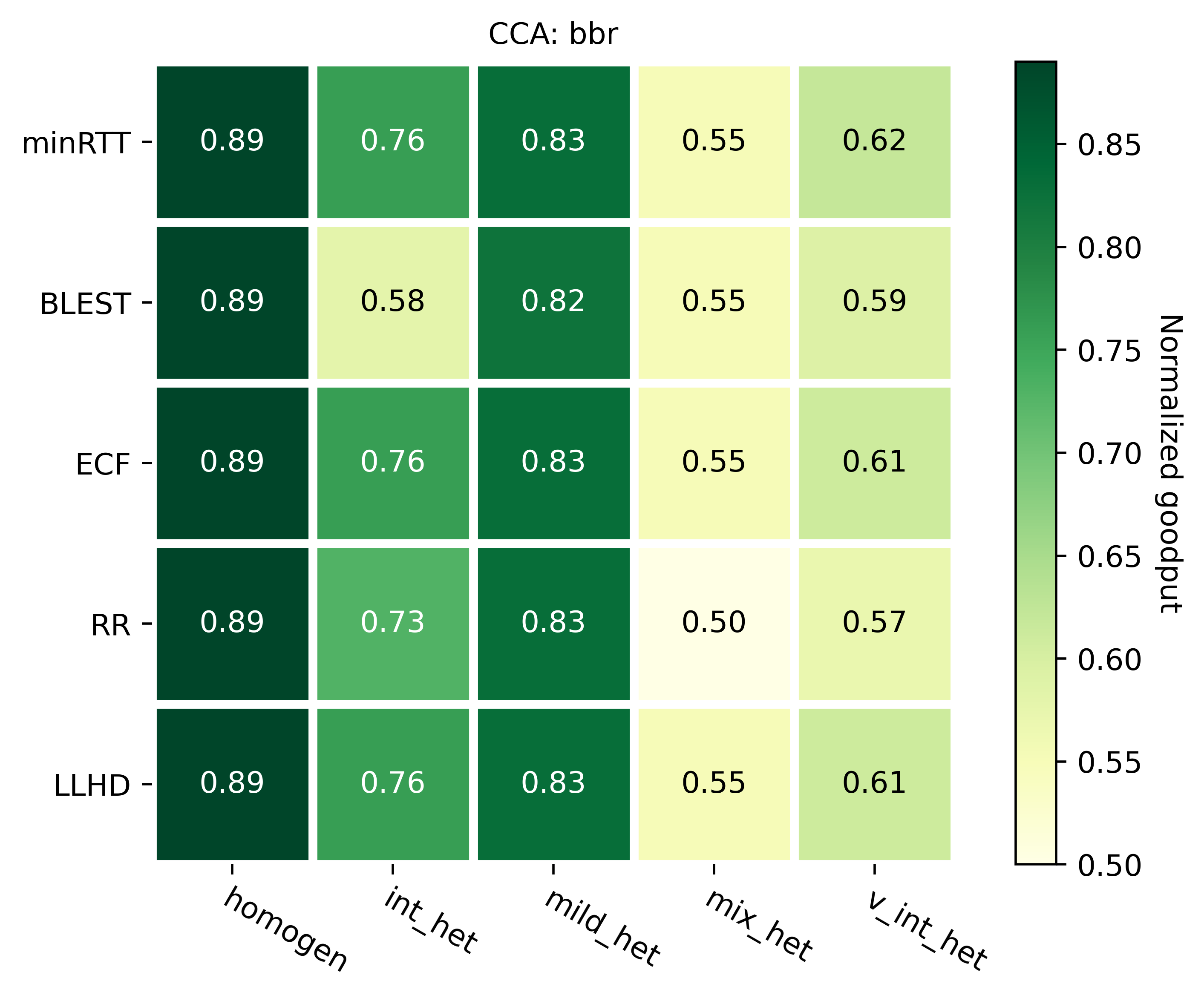}
  \captionof{figure}{PS\_score heatmap per scenario family for BBR}
  \label{fig:PS_score_heatmap_per_ScenFam_BBR}
  \Description[PS\_score\_heatmap\_per\_ScenFam\_BBR]{This figure provides an alternative view of each packet scheduling algorithm's score per scenario family for BBR, via a \textit{PS\_score} heatmap.}
\end{minipage}
\hfill
\begin{minipage}[t]{.3\textwidth}
  \centering
  \includegraphics[width=\linewidth]{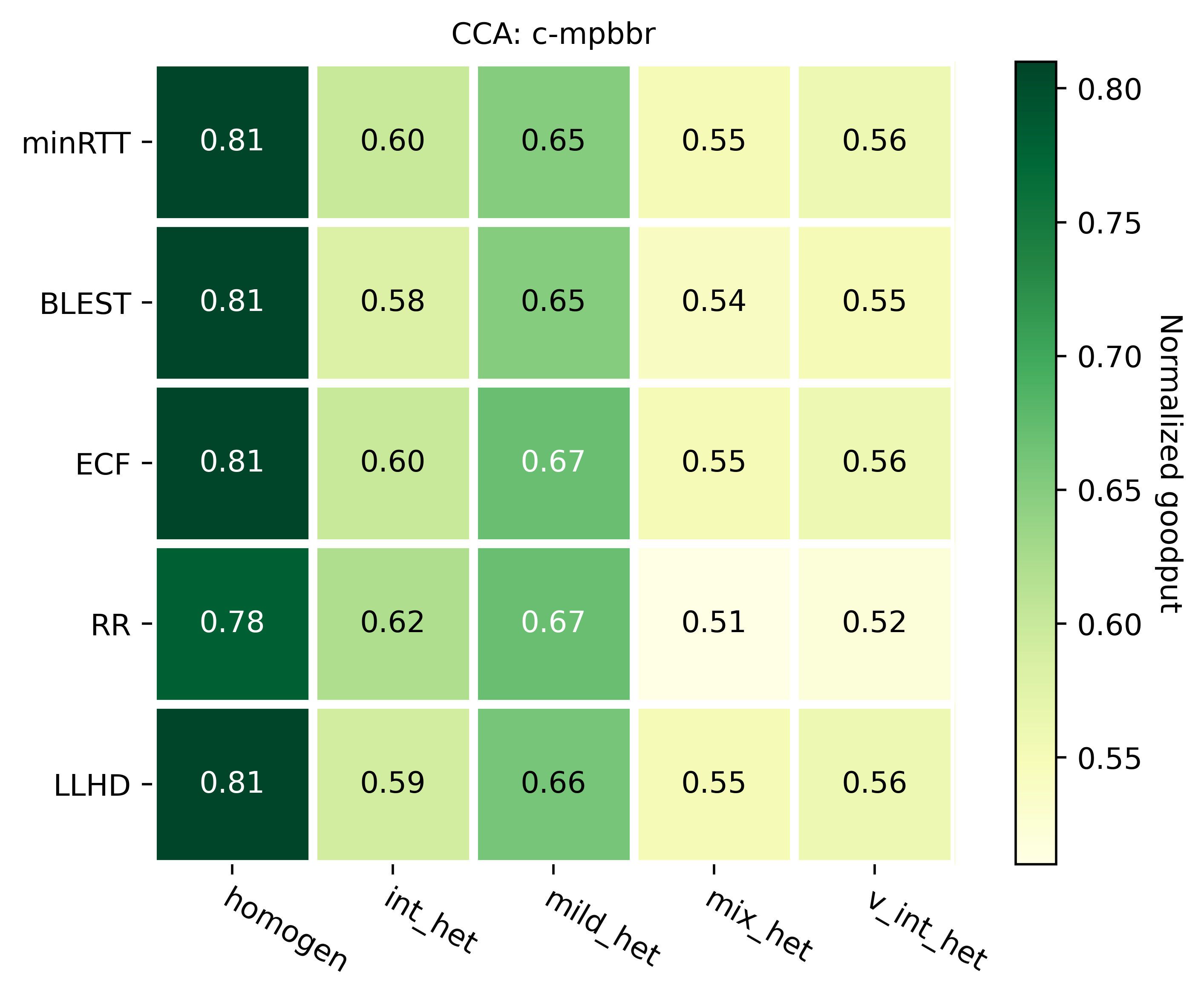}
  \captionof{figure}{PS\_score heatmap per scenario family for C-MPBBR}
  \label{fig:PS_score_heatmap_per_ScenFam_C-MPBBR}
  \Description[PS\_score\_heatmap\_per\_ScenFam\_C-MPBBR]{This figure provides an alternative view of each packet scheduling algorithm's score per scenario family for C-MPBBR, via a \textit{PS\_score} heatmap.}
\end{minipage}
\hfill
\begin{minipage}[t]{.3\textwidth}
  \centering
  \includegraphics[width=\linewidth]{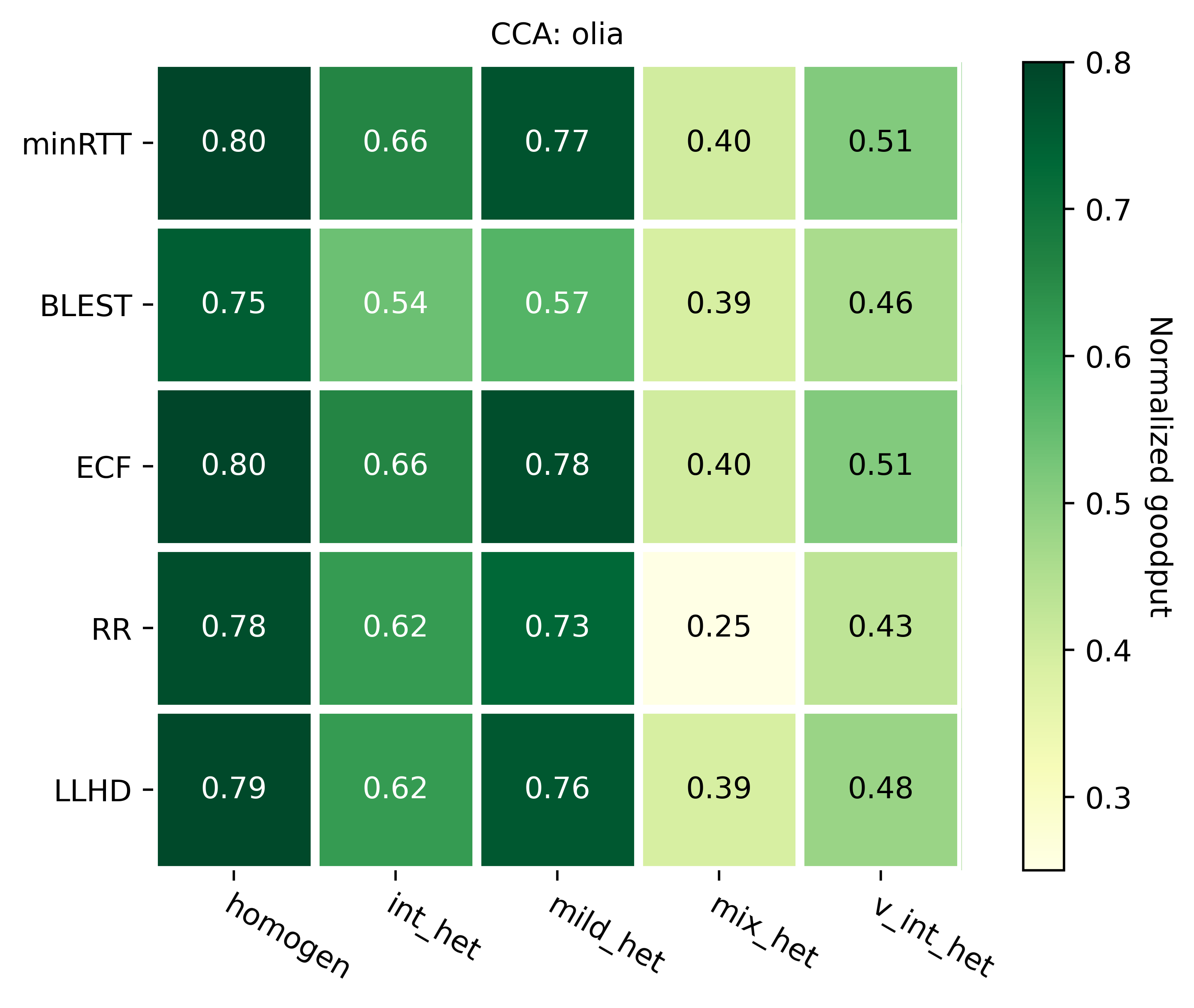}
  \captionof{figure}{PS\_score heatmap per scenario family for OLIA}
  \label{fig:PS_score_heatmap_per_ScenFam_OLIA}
  \Description[PS\_score\_heatmap\_per\_ScenFam\_OLIA]{This figure provides an alternative view of each packet scheduling algorithm's score per scenario family for OLIA, via a \textit{PS\_score} heatmap.}
\end{minipage}
\end{figure}

An alternative view of each packet scheduling algorithm's score per scenario family for each of the top three CCAs, is provided via the \textit{PS\_score} heatmaps showcased within Figures \ref{fig:PS_score_heatmap_per_ScenFam_BBR}, \ref{fig:PS_score_heatmap_per_ScenFam_C-MPBBR}, and \ref{fig:PS_score_heatmap_per_ScenFam_OLIA}, respectively.

\begin{figure}[htb!]
  \centering
  \includegraphics[width=\linewidth]{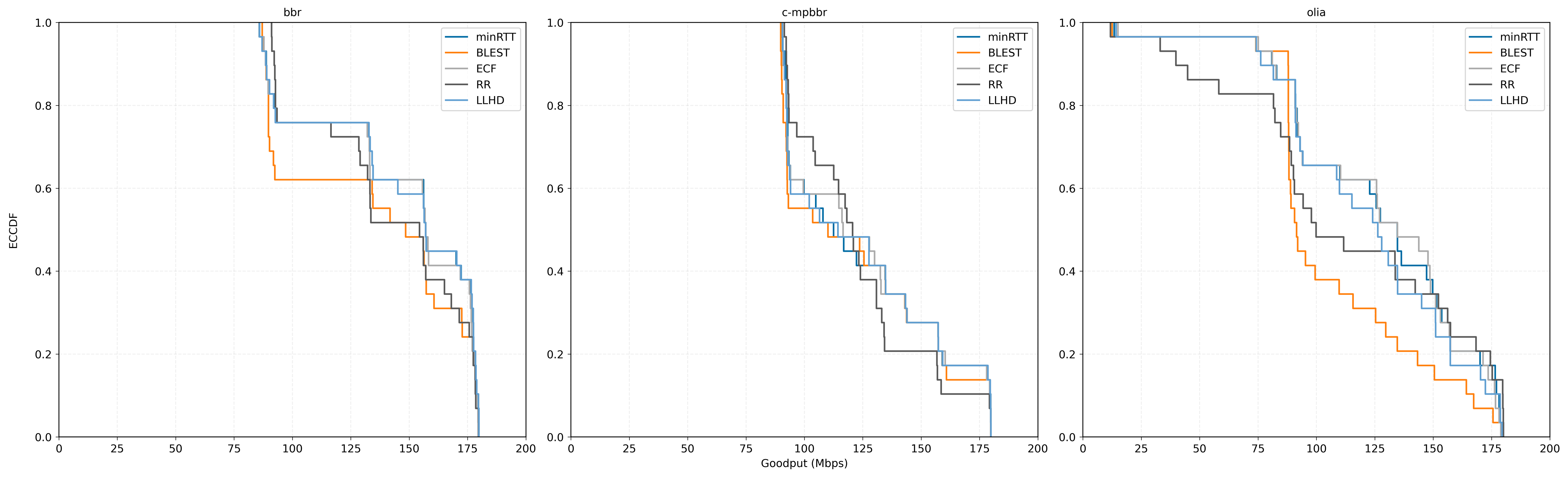}
  \caption{ECCDF depicting packet scheduler goodput probability across scenario families, for BBR, C-MPBBR, and OLIA}
  \label{fig:ECCDF_PS_goodput_probability_JointFig}
  \Description[ECCDF\_PS\_goodput\_probability]{This figure illustrates the ECCDF which depicts packet scheduler goodput probability across scenario families, for BBR, C-MPBBR, and OLIA.}
\end{figure}

One of the main purposes of this work, has been to provide a clear view on how each packet scheduler performs across all scenario families for each of the best-performing CCA. To this end, Figure \ref{fig:ECCDF_PS_goodput_probability_JointFig} illustrates the \textit{Empirical Complementary Cumulative Distribution Function (ECCDF)}, which depicts each scheduler's probability to perform above a specific level of goodput, for each of the top-three CCAs (i.e., BBR, C-MPBBR, and OLIA); ECCDF considers goodput probability across all scenario families and sub-scenarios.

\subsection{Packet scheduling assessment - Per-Packet Delay}
\label{subsec:psa_assess_ppd}
\begin{figure}[htb!]
  \centering
  \includegraphics[width=\linewidth]{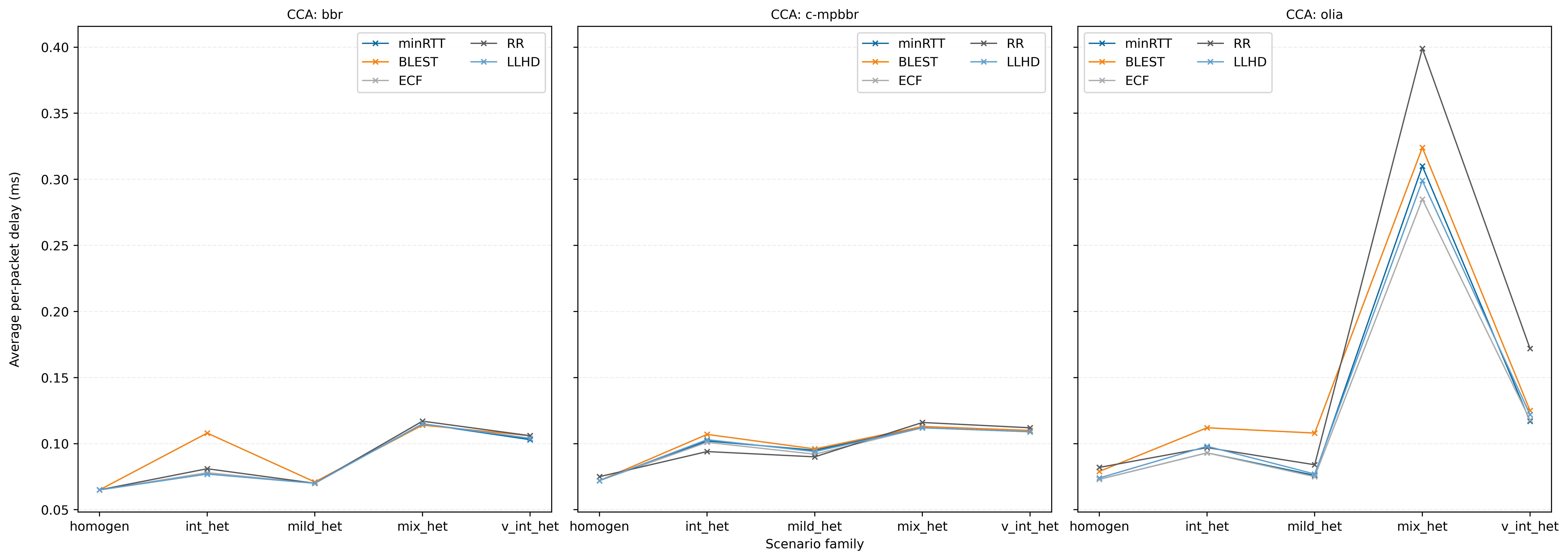}
  \caption{Qualitative view of each PSA's PPD per scenario family for BBR, C-MPBBR, and OLIA}
  \label{fig:PPD_Qualitative}
  \Description[PPD\_Qualitative]{This figure provides a qualitative view of each packet scheduling algorithm's (PSA) per-packet delay (PPD) per scenario family for BBR, C-MPBBR, and OLIA .}
\end{figure}

\begin{figure}[htb!]
  \centering
  \includegraphics[width=\linewidth]{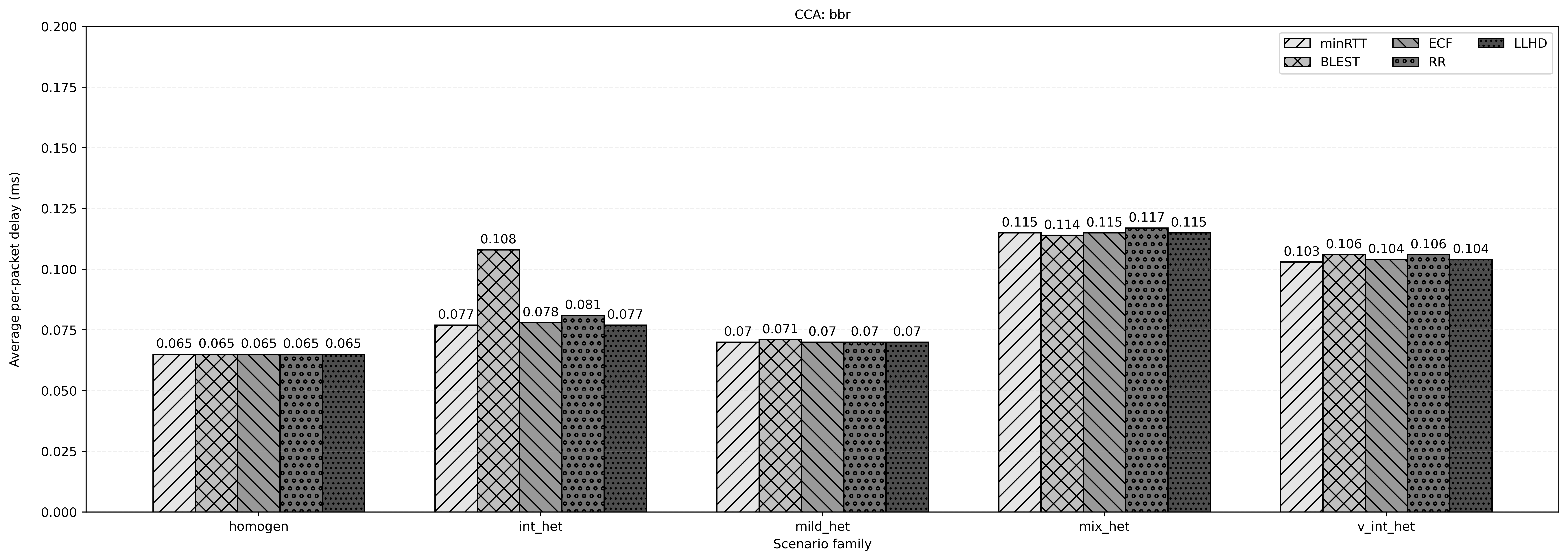}
  \caption{PPD of each PSA per scenario family for BBR}
  \label{fig:PPD_Numer_BBR}
  \Description[PPD\_Numer\_BBR]{This figure provides a numerical view of each packet scheduling algorithm's (PSA) per-packet delay (PPD) per scenario family when combined with BBR.}
\end{figure}

\begin{figure}[htb!]
  \centering
  \includegraphics[width=\linewidth]{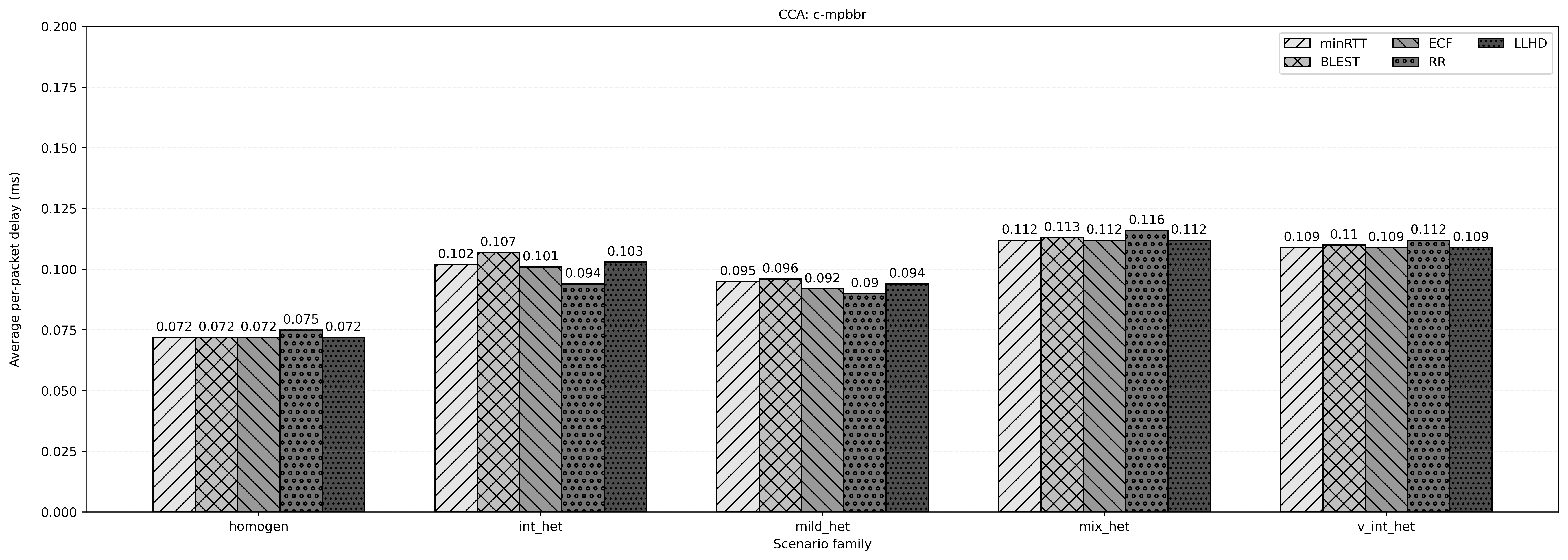}
  \caption{PPD of each PSA per scenario family for C-MPBBR}
  \label{fig:PPD_Numer_C-MPBBR}
  \Description[PPD\_Numer\_C-MPBBR]{This figure provides a numerical view of each packet scheduling algorithm's (PSA) per-packet delay (PPD) per scenario family when combined with C-MPBBR.}
\end{figure}

\begin{figure}[htb!]
  \centering
  \includegraphics[width=\linewidth]{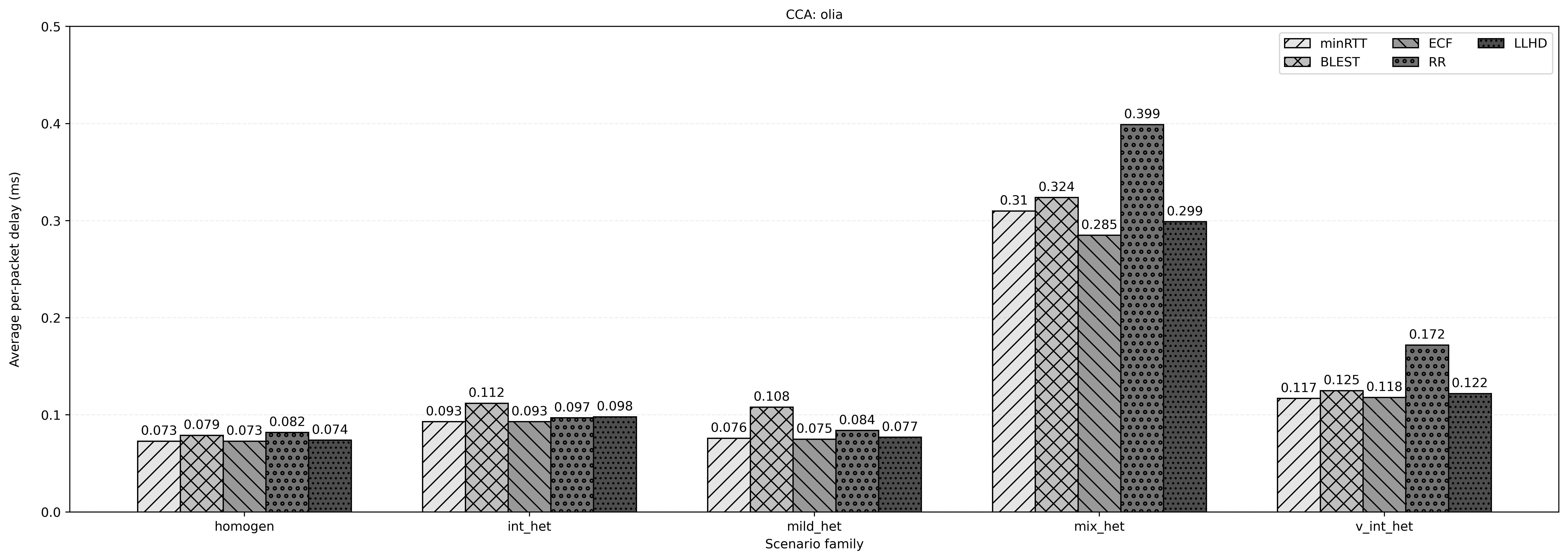}
  \caption{PPD of each PSA per scenario family for OLIA}
  \label{fig:PPD_Numer_OLIA}
  \Description[PPD\_Numer\_OLIA]{This figure provides a numerical view of each packet scheduling algorithm's (PSA) per-packet delay (PPD) per scenario family when combined with OLIA.}
\end{figure}

Besides the evaluation in terms of goodput, we found it also useful to provide insight on how each packet scheduling algorithm would perform in case of interactive applications. A key metric for this type of applications is the experienced \textit{per-packet delay (PPD)}. For this purpose, we provide a high-level view of each packet scheduling algorithm's\footnote{"Packet Scheduling Algorithm" is abbreviated as \textit{PSA} within figure captions.} performance in terms of per-packet delay in Figure \ref{fig:PPD_Qualitative}; a more fine-grained numerical evaluation of each packet scheduling algorithms per-packet delay for each of the best-three congestion control algorithms (i.e., BBR, C-MPBBR, and OLIA) can be found in Figures \ref{fig:PPD_Numer_BBR}, \ref{fig:PPD_Numer_C-MPBBR}, and \ref{fig:PPD_Numer_OLIA}, respectively.

\begin{figure}[htb!]
  \centering
  \includegraphics[width=\linewidth]{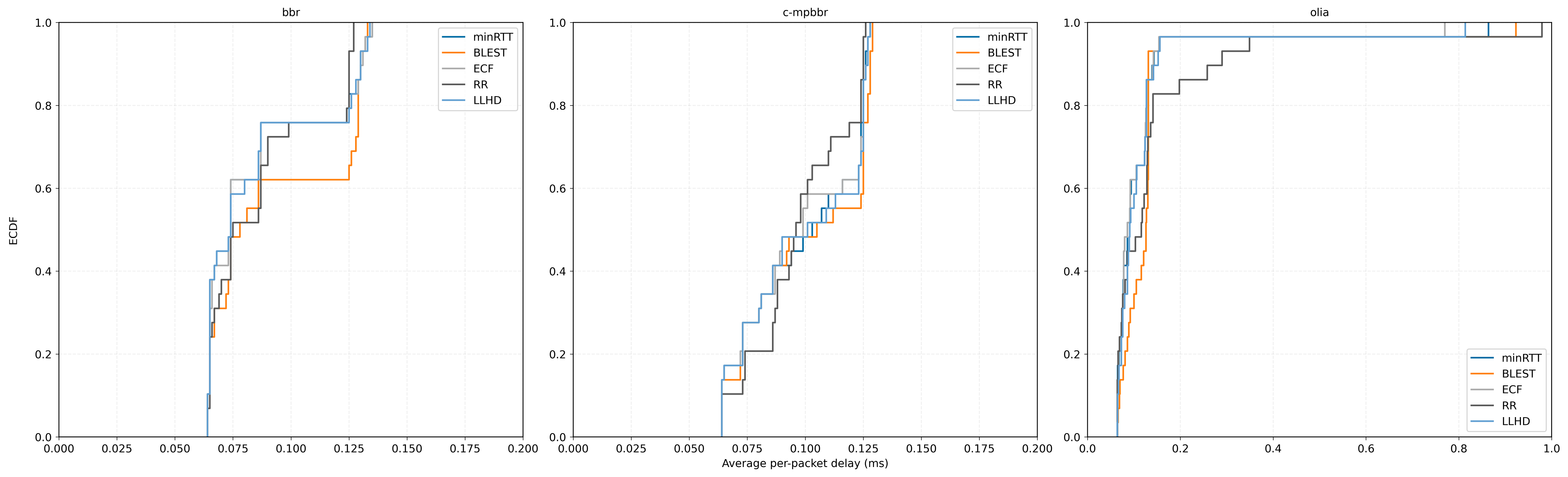}
  \caption{ECDF depicting packet schedulers' PPD probability across scenario families, for BBR, C-MPBBR, and OLIA.}
  \label{fig:ECDF_PPD_overall}
  \Description[ECDF\_PPD\_overall]{This figure illustrates ECDF which depicts packet scheduler per-packet delay probability across scenario families, for BBR, C-MPBBR, and OLIA.}
\end{figure}

Finally, Figure \ref{fig:ECDF_PPD_overall} depicts the \textit{Empirical Cumulative Distribution function (ECDF)}, which provides a metric of scheduler's probability to achieve below a specific level of average per-packet delay, for each one of the top-three CCAs (i.e., BBR, C-MPBBR, and OLIA); ECDF considers the \textit{average per-packet delay} across all scenario families and sub-scenarios.

\section{Discussion}
\label{sec:discussion}
The performance results derived by this extended experimentation campaign indicate that the there is no single congestion control and packet scheduling algorithm combination that fits well all underlying network conditions. As a result, it would be recommended to select an algorithmic scheme combination according to the level of heterogeneity experienced in the network, as illustrated through the aforementioned results. On the other hand, adjusting the protocol configuration for fine-tuning can be a cumbersome procedure since network changes can be of dynamic, and in most cases stochastic, nature. It is then reasonable to seek for a stable configuration, able to cope with all potential network characteristics. Toward this objective, the initial factor which determines the overall performance outcome, is the appropriate selection of the congestion control algorithm. In this direction, our analysis indicated that \textit{BBR}, \textit{C-MPBBR}, and \textit{OLIA} could be potentially applied as generic-use congestion control schemes. With any one of these three CCA's, our performance results indicate that \textit{ECF} and the \textit{MPTCP default (minRTT)} packet scheduling algorithms exhibit the highest goodput and lowest per-packet delay performance, across almost all scenarios. \textit{RR} turns out to be useful only for experimentation purposes, and should rather be avoided in production environments, since it experiences the worst results among the examined schedulers; RR's performance outcome across the diverse scenarios confirm the recommendation included within \cite{mptcporg}. \textit{LLHD} is also considered a stable, in terms of goodput and per-packet delay, packet scheduling algorithm whose results are very close to these of the first two. \textit{BLEST} is also a highly-sophisticated algorithm which performs quite well in case of homogeneous and mild heterogeneity; however, when the underlying network's path heterogeneity becomes severe, BLEST decides to skip the non-beneficial paths, underutilizing the available network resources. Doing so, diminishes its performance results compared to the other best-performing schedulers (except RR which performs the worst). Nevertheless, even with this kind of resource underuse, BLEST does not experience any significant performance inefficiency, and can thus be included within the group of top-four. An important observation here is the fact that if the optimization goal is application-level smoothness instead of raw throughput maximization, BLEST could be considered superior since it actively avoids the impaired path to prevent HoL blocking. This attribute also stresses the fact that the "best" scheduler is ultimately application-dependent.

One limitation of the current experimentation test-set is the fact that our traffic resembled bulk-traffic conditions; thus, schedulers that might lag behind in this specific use case, might reach better results under other types of traffic, such as short web-page loads, or traffic generated by video streaming applications.


\section{Related Work}
\label{sec:related_work}
There are different directions in literature around the performance assessment of MPTCP and its underlying packet scheduling and congestion control algorithms. Some works maintain the default MPTCP algorithm configuration (e.g., minRTT packet scheduler and (O)LIA congestion control) and examine the impact of the various (MP)TCP protocol parameters (e.g., send/receive buffer size, SACK, TCP auto-tuning, etc.) on the overall MPTCP performance. Another direction is focusing on a specific research area (e.g., on packet scheduling) and assessing how the various algorithms of this particular domain perform, while keeping the default configuration in other protocol areas (e.g., default congestion control and/or default (MP)TCP parameters). Finally, there is a limited number of works assessing how the various algorithm combinations affect MPTCP protocol operation. Our work is positioned to this latter case, coupling the theoretical aspects with an extensive performance assessment of the state-of-art packet scheduling and congestion control algorithms. Our manuscript's motivation and distinctive aspects are described hereinafter.

The performance evaluation conducted within \cite{barre2011implementation} has been one of the first attempts to quantify the impact that various factors may impose on MPTCP performance (i.e., the buffer size, packet loss, the MSS-size). Besides evaluating the impact of a single parameter (i.e., BW, RTT delay, or packet loss) on MPTCP performance, our work considers also a mixture of these parameters and examines how these affect factors other than merely goodput, such as the SRTT, the OFO-queue size, and the number of retransmissions. Moreover, our experiments go beyond the default MPTCP packet scheduling (minRTT), and Reno / Coupled (LIA) congestion control schemes, assessing a variety of schedulers and congestion control algorithm combinations. \cite{paasch2014experimental} implements the first framework which enables loading packet scheduling and congestion control algorithms as modules, allowing algorithm alternations at runtime. Experimental evaluation there, is conducted using minRTT scheduler, both for bulk transfer and application-limited traffic. On the contrary, our work examines a bunch of modern packet scheduling and congestion control schemes. \cite{HammarMScKarlstad} is a comprehensive work trying to measure the extent to which MPTCP is beneficial compared to single-path TCP under different scenarios. The authors consider minRTT scheduler combined with Coupled (LIA) CCA for MPTCP, and Cubic CCA for SP-TCP, while our work examines multiple packet scheduling and congestion control algorithm combinations. Another difference is the fact that the experiments conducted throughout \cite{HammarMScKarlstad} consider variable traffic size and higher 5G bandwidths close to 1Gbps; our work seeks to provide foundational details behind the operation of schedulers and CCA combinations, thus selected to keep a static length-based traffic approach (iperf traffic tries to fill in resources for 30 sec. test duration) and cap bandwidth to 100Mbps per subflow in order to be able to identify basic discrepancies and inefficiencies within the algorithms, which might be invisible in high data rates. \cite{hurtig2018low} focuses on the development of a new scheduler (STTF) while assessing it against other packet scheduling algorithms (i.e., minRTT, BLEST, ECF, DAPS \cite{kuhn2014daps}, OTIAS \cite{yang2014out}) under various traffic conditions. However, packet schedulers are compared to one another using only the default MPTCP congestion control algorithm, which was Coupled (LIA). \cite{polese2019survey} provides a high level overview on the advancements of transport layer protocols, a classification of the various congestion control schemes and a short description of them, as well as details around the various multipath protocols. In a different direction, our work is focused on a limited number of packet scheduling and congestion control schemes, providing an in-depth analysis of their internal operation as well as the detailed algorithmic steps. Furthermore \cite{polese2019survey} is a theoretical survey, while our work is an experimental manuscript, combining theory with experimental evaluation of the algorithms under review. As mentioned earlier, part of our work has been inspired by the work conducted in \cite{cech2020analyzing}, especially regarding the visibility of the results in a combined figure including goodput, SRTT, OFO queue size, and the number of retransmissions. \cite{cech2020analyzing} also provides a short description as well as the algorithmic steps of some packet scheduling algorithms (i.e., minRTT, RR, MuSher \cite{saha2019musher}, BLEST, ECF, STTF \cite{hurtig2018low}), and examines their performance in conjunction with some congestion control schemes (i.e., LIA, OLIA, BALIA, wVegas). However, it modifies the send and receive buffers to identify their impact on performance, while we selected to maintain the default values in order to focus mainly on the underlying algorithms' performance, and to avoid blending many factors which could further affect performance results. In addition, our work includes the latest BBR and C-MPBBR congestion control schemes and tries to further analyze the results by introducing novel metrics, in an attempt to directly assess all possible combinations and extract the best-performing ones. Another extensive experimental survey has been provided through the work conducted in \cite{kimura2020packet}. The authors of this work provide a comprehensive analysis of packet scheduling schemes along with a classification based on their use cases and the number of scheduling criteria considered by each algorithm. Their work has also influenced the design of our own experimentation strategy, and considers a nearly-exhaustive list of packet schedulers and CCAs. However, the authors of that work examine different packet scheduling and a subset of the congestion control algorithms we use in our own experiments (i.e., $PSAs=\{minRTT, RR, LWS, LTS, HSR\}$ and $CCAs=\{LIA, BALIA, OLIA, wVegas\}$, respectively).


\section{Conclusion}
\label{sec:conclusion}

This work has been conceived during our own research on multipath protocols and more specifically on MPTCP. While anyone interested to conduct research in this field may need to skim through multiple sources, we thought it would be useful for the reader to have a single-point of reference where all the basic information around MPTCP protocol is collected. To this end, Section \ref{sec:background} has been dedicated to providing the necessary details for a preliminary introduction to the MPTCP protocol, its structure and architectural overview, as well as the main modules determining its operation. This section provides also detailed information on the various packet scheduling and congestion control schemes, including both their theoretical and algorithmic aspects. To this end, five packet scheduling (i.e., minRTT, BLEST, ECF, RR, and LLHD) and seven congestion control algorithms (i.e., Cubic, Coupled/LIA, OLIA, BALIA, wVegas, BBR, and C-MPBBR) have been examined, in an attempt to cover the theoretical aspects before moving on with their performance evaluation. Section \ref{sec:exper_method} provided the overall experimentation methodology, our Mininet environment topology, as well as the underlying protocol and kernel configuration details, which we considered when preparing our experimentation testbed. The same section included also details of the traffic methodology, the design of the experimentation scenarios, and the performance metrics used to assess the various algorithm combinations. Section \ref{sec:evaluation} presented the experimentation results of the various packet scheduling and congestion control algorithm combinations. The score metrics conceived to assist our evaluation as well as the main evaluation criteria (e.g., goodput and per-packet delay), have been also presented within this section. Section \ref{sec:discussion} concluded the core part of this work by providing a short summary on the best-performing packet scheduling and congestion control schemes, while Section \ref{sec:related_work} provided details on the related work present in literature. 

While this work serves as a starting point for anyone interested in studying the preliminaries of MPTCP, such as the basic details around the protocol operation, as well as the available packet scheduling and congestion control schemes, it is not and cannot be a single point of study; however, appropriate references are provided to the original manuscripts and IETF RFCs, which provide the entire picture.

Finally, the list of packet scheduling and congestion control algorithms that have been selected for our theoretical analysis and the experiments, contains the ones considered as foundational and best-performing for MPTCP, laying essentially the groundwork for the development of many other algorithms, either unique or variations of existing ones.

\begin{acks}
    This research has been partially supported by the EU HORIZON SNS-JU projects: 6G-SANDBOX (GA: 101096328), 6G-BRICKS (GA: 101096954), and SUNRISE-6G (GA: 101139257).
\end{acks}

\bibliographystyle{ACM-Reference-Format.bst}
\bibliography{cite.bib}

\clearpage

\appendix

\section{Packet Scheduling Algorithms}



\SetKwInput{KwInput}{Input}
\SetKwInput{KwOutput}{Output}
\SetKwComment{Comment}{/* }{ */}
\RestyleAlgo{algoruled}
\SetAlgoLongEnd
\begin{algorithm}[H]    
\caption{minRTT}\label{alg:minRTT}
\KwInput{Set of available subflows: $\mathcal{S}$,\, set of paths: $\mathcal{R}$,\, $sf_{i} \in \mathcal{S},\, i \in \mathcal{R} $}
\KwOutput{Best subflow, $sf_{best} \in \mathcal{S} $}
$ srtt_{min} \gets 0xffffffff $ \;
$ sf_{best} \gets NULL $ \;
\For {each\,  $ sf_{i} \in \mathcal{S} $}
{
    \If{$ srtt_{sf_{i}} < srtt_{min} $}
    {
        $ srtt_{min} \gets srtt_{sf_{i}} $ \;
        $ sf_{best} \gets sf_{i} $ \;
    }
}
\textbf{return} $ sf_{best} $
\end{algorithm}
\BlankLine
The theoretical aspects of \textit{minRTT} packet scheduling algorithm are provided in \ref{subsub:minrtt}.
\BlankLine
\BlankLine



\SetKwInput{KwInput}{Input}
\SetKwInput{KwOutput}{Output}
\SetKwComment{Comment}{/* }{ */}
\RestyleAlgo{algoruled}
\SetAlgoLongEnd
\begin{algorithm}[H]
\caption{BLEST}\label{alg:BLEST}
\KwInput{Set of available subflows: $\mathcal{S}$,\, set of paths: $\mathcal{R}$,\, $sf_{i} \in \mathcal{S},\, i \in \mathcal{R} $,\, $srtt_{f} < srtt_{s}$}
\KwOutput{Best subflow, $sf_{best} \in \mathcal{S} $}
$ x_{f} \gets minRTT() $ \; 
\uIf{fastest subflow $x_{f}$ is available}
{
    $ sf_{best} \gets x_{f} $ \;
}
\uElseIf{slower subflow $x_{s}$ is available}
{
    $rtt_{s} = srtt_{s} / srtt_{f} $ \;
    $X = MSS_{f} * (CWND_{f} + (rtt_{s}-1)/2) * rtt_{s} $ \;
    \uIf{$X * \lambda \le MPTCP_{SW} - MSS_{s} * (inflight_{s}+1) $}
    {
        $ sf_{best} \gets x_{s} $ \;
    }
    \Else
    {
        $ sf_{best} \gets NULL $ \;
    }
}
\Else 
{
    $ sf_{best} \gets NULL $  \Comment*[r]{no available subflow}
}
\textbf{return} $ sf_{best} $
\end{algorithm}
\BlankLine
The theoretical aspects of \textit{BLEST} packet scheduling algorithm are provided in \ref{subsub:blest}.
\BlankLine
\BlankLine



\SetKwInput{KwInput}{Input}
\SetKwInput{KwOutput}{Output}
\SetKwComment{Comment}{/* }{ */}
\RestyleAlgo{algoruled}
\SetAlgoLongEnd
\begin{algorithm}[H]
\caption{ECF}\label{alg:ECF}
\KwInput{Set of available subflows: $\mathcal{S}$,\, set of paths: $\mathcal{R}$,\, $sf_{i} \in \mathcal{S},\, i \in \mathcal{R} $,\, $srtt_{f} < srtt_{s}$}
\KwOutput{Best subflow, $sf_{best} \in \mathcal{S} $}
$ x_{f} \gets minRTT() $ \Comment*[r]{find the fastest subflow $x_{f}$ using minRTT algorithm}
\uIf {fastest subflow $x_{f}$ is available}
{
    $ sf_{best} \gets x_{f} $ \;
}
\Else 
{
    $ x_{s} \gets minRTT() $  \Comment*[r]{find the second fastest subflow \(x_{s}\) using minRTT algorithm}
    $ n=1+\frac{k}{CWND_{f}} $ \;
    $ \delta = max(\sigma_{f}, \sigma_{s}) $ \;
    \uIf{$ n * RTT_{f} < (1+waiting*\beta)(RTT_{s} + \delta) $}
    {
        \uIf{$ \frac{k}{CWND_{s}} * RTT_{s} \ge 2*RTT_{f} + \delta $}
        {
            $ waiting = 1 $  \Comment*[r]{Wait for $x_{f}$}
            $ sf_{best} \gets NULL $ \Comment*[r]{no available subflow}
        }
        \Else
        {
            $ sf_{best} \gets x_{s} $ \;
        }
    }
    \Else
    {
        $ waiting = 0 $ \;
        $ sf_{best} \gets x_{s} $ \;
    }
}
\textbf{return} $ sf_{best} $
\end{algorithm}
\BlankLine
The theoretical aspects of \textit{ECF} packet scheduling algorithm are provided in \ref{subsub:ecf}.
\BlankLine
\BlankLine



\SetKwInput{KwInput}{Input}
\SetKwInput{KwOutput}{Output}
\SetKwComment{Comment}{/* }{ */}
\RestyleAlgo{algoruled}
\SetAlgoLongEnd
\begin{algorithm}[H]
\caption{RR (num\_segments $\ge$ 1, cwnd\_limited=true)}\label{alg:RR}
\KwInput{Set of available subflows: $\mathcal{S}$,\, set of paths: $\mathcal{R}$,\, $sf_{i} \in \mathcal{S},\, i \in \mathcal{R} $}
Get \textit{num\_segments} from MPTCP send buffer \;
\For {each\,  $ sf_{i} \in \mathcal{S} $}
{
    \uIf{$ {sf_{i}}\_cwnd - {sf_{i}}\_inflight \ge num\_segments * MSS_{i} $}
    {
        send ($num\_segments$ * $MSS_{i}$) bytes on ${sf_{i}}$ \Comment*[r]{$sf_{i}$ has sufficient cwnd space}
    }
    \Else
    {
        continue \Comment*[r]{skip $sf_{i}$ in case of insufficient cwnd space}
    }
}
\end{algorithm}
\BlankLine
The theoretical aspects of \textit{RR} packet scheduling algorithm are provided in \ref{subsub:rr}.
\BlankLine
\BlankLine



\makeatletter
\newcommand{\nosemic}{\renewcommand{\@endalgocfline}{\relax}}
\newcommand{\dosemic}{\renewcommand{\@endalgocfline}{\algocf@endline}}
\newcommand{\pushline}{\Indp}
\newcommand{\popline}{\Indm\dosemic}
\let\oldnl\nl
\newcommand{\nonl}{\renewcommand{\nl}{\let\nl\oldnl}}
\makeatother

\SetKwInput{KwInput}{Input}
\SetKwInput{KwOutput}{Output}
\SetKwInput{KwInit}{Initialization}
\SetKwComment{Comment}{/* }{ */}
\RestyleAlgo{algoruled}
\SetAlgoLongEnd
\begin{algorithm}[H]
\caption{LLHD}\label{alg:LLHD}
\KwInput{Set of subflows: $\mathcal{S}$,\, set of paths: $\mathcal{R}$,\, $sf_{i} \in \mathcal{S},\, i \in \mathcal{R} $}
\KwOutput{Best subflow, $sf_{best} \in \mathcal{S} $}
\KwInit{\\
    $sf_{best} = NULL$ \;
    $\gamma_{max} = 0$ \;
    $\beta = 0.001$  \Comment*[r]{$\beta = 0.001$ in paper \cite{lubna2022low}; however, actual value in code is $\beta = 10$ \cite{githubllhd}}
    \nonl $RTT_{max}=9999999$ \;
    \nonl $GP_{max}=9999$ \;
    \BlankLine
}
\nonl Upon reception of ACK:\\
\For {each\,  $ sf_{i} \in \mathcal{S} $}
{
    \If{${sf_{i}}$ is backup}
    {
        $sf_{i} = NULL$ \;
        $sf_{best} \gets sf_{i}$ \;
    }
    \If{${sf_{i}}$ is unavailable}
    {
        $sf_{i} = NULL$ \;
        $sf_{best} \gets sf_{i}$ \;
    }
    \If{${sf_{i}}$ is temp\_unavailable}
    {
        $sf_{i} = NULL$ \;
        $sf_{best} \gets sf_{i}$ \;
    }
    $\gamma\_curr = ({GP}\_sf_{i} / GP_{max}) + \beta \times (RTT_{max} / RTT\_sf_{i}) $ \;
    \If{$sf_{i}\_CWND\_available$ \textbf{\upshape{and}} $\gamma\_curr > \gamma\_max $}
    {
        $\gamma_{max} \gets \gamma\_curr$ \;
        $sf_{best} \gets sf_{i}$ \;
    }
}
\textbf{return} $ sf_{best} $
\end{algorithm}
\BlankLine
The theoretical aspects of \textit{LLHD} packet scheduling algorithm are provided in \ref{subsub:llhd}.
\BlankLine
\BlankLine



\SetKwInput{KwInput}{Input}
\SetKwInput{KwOutput}{Output}
\SetKwInput{KwInit}{Initialization}
\SetKwComment{Comment}{/* }{ */}
\RestyleAlgo{algoruled}
\SetAlgoLongEnd
\begin{algorithm}[H]
\caption{ReMP TCP (redundant)}\label{alg:ReMPTCP}
\KwInput{Set of subflows: $\mathcal{S}$,\, set of paths: $\mathcal{R}$,\, $sf_{i} \in \mathcal{S},\, i \in \mathcal{R} $}
\KwOutput{$sf_{i},\, sf_{j} \in \mathcal{S} $}

\For {each segment to be transmitted}
{
    \For {each\,  $ sf_{k} \in \mathcal{S} $}
    {
        \If{${sf_{k}}$ is available}
        {
            $sf_{i} \gets sf_{k}$ \;
            \textbf{return} $sf_{i}$ \;
        }
    }
    \For {each\,  $ sf_{k} \in \mathcal{S} $}
    {
        \If{${sf_{k}} \ne sf_{i}$ \textbf{\upshape{and}} ${sf_{k}}$ is available}
        {
            $sf_{j} \gets sf_{k}$ \;
            \textbf{return} $sf_{j}$ \;
        }
    }
}
redundant\textbf{return} NULL \Comment*[r]{either one or both subflows unavailable}
\end{algorithm}
\BlankLine
The theoretical aspects of \textit{ReMP TCP} packet scheduling algorithm are provided in \ref{subsub:redundant}.
\BlankLine
\BlankLine


\section{Congestion Control Algorithms}



\makeatletter
\newcommand{\AlgoResetCount}{\renewcommand{\@ResetCounterIfNeeded}{\setcounter{AlgoLine}{0}}}
\newcommand{\AlgoNoResetCount}{\renewcommand{\@ResetCounterIfNeeded}{}}
\newcounter{AlgoSavedLineCount}
\newcommand{\AlgoSaveLineCount}{\setcounter{AlgoSavedLineCount}{\value{AlgoLine}}}
\newcommand{\AlgoRestoreLineCount}{\setcounter{AlgoLine}{\value{AlgoSavedLineCount}}}
\makeatother

\SetKwComment{Comment}{/* }{ */}
\SetAlgoVlined

\begin{algorithm}[H]
\caption{CUBIC (v2.2)}\label{alg:CUBIC}

\nonl Initialization: \\
\Indp   \nonl $tcp\_friendliness \longleftarrow 1, \,\, \beta \longleftarrow 0.2 $ \\
        \nonl $fast\_convergence \longleftarrow 1, \,\, C \longleftarrow 0.4 $ \\
        \nonl $cubic\_reset()$ \\
\Indm
\nonl On each ACK: \\
\nonl \Begin{
     \nosemic \lIf{$dMin$} {$dMin \longleftarrow min(dMin, RTT) $}
     \nosemic \lElse{$dMin\longleftarrow RTT$}
     \nosemic \lIf{$cwnd \le ssthresh $} {$cwnd \longleftarrow cwnd+1$}
     \Else
     {
        $cnt \longleftarrow cubic\_update()$ \\
        \If{$cwnd\_cnt > cnt$} 
        {
            $cwnd \longleftarrow cwnd+1$, \,\, $cwnd\_cnt \longleftarrow 0$ \\
        }
        \lElse
        {
            $cwnd\_cnt \longleftarrow cwnd\_cnt+1$
        }
     }
}

\end{algorithm}
\newpage
\AlgoNoResetCount
\begin{algorithm}

\nonl Packet loss: \\
\nonl \Begin{
    $epoch\_start \longleftarrow 0$ \\
    \nosemic \If{$cwnd < W_{last\_max}$ \textbf{\upshape{and}} $fast\_convergence$} 
    {
        $W_{last\_max} \longleftarrow cwnd*\frac{(2-\beta)}{2}$ 
    }
    \nosemic \lElse
    {
        $W_{last\_max} \longleftarrow cwnd$
    }
    $ssthresh \longleftarrow cwnd \longleftarrow cwnd*(1-\beta)$
}

\nonl Timeout: \\
\nonl \Begin{
    $cubic\_reset()$
}
$cubic\_update():$ \\
\nonl \Begin{
    $ack\_cnt \longleftarrow ack\_cnt+1$ \\
    \If{$epoch\_start \le 0$}
    {
        $epoch\_start \longleftarrow tcp\_time\_stamp$ \\
        \nosemic \If{$cwnd < W_{last\_max}$}
        {
            $K \longleftarrow \sqrt[3]{\frac{W_{last\_max}-cwnd}{C}}$ \\
            $origin\_point \longleftarrow W_{last\_max}$ 
        }
        \Else
        {
            $K \longleftarrow 0$ \\
            $origin\_point \longleftarrow cwnd$
        }
        $ack\_cnt \longleftarrow 1$ \\
        $W_{tcp} \longleftarrow cwnd$
    }
    $t \longleftarrow tcp\_time\_stamp + dMin - epoch\_start$ \\
    $target \longleftarrow origin\_point+C(t-K)^3$ \\
    \nosemic \lIf{$target > cwnd$}
    {
        $cnt \longleftarrow \frac{cwnd}{target-cwnd}$
    }
    \lElse
    {
        $cnt \longleftarrow 100*cwnd$
    }
    \lIf{$tcp\_friendliness$}
    {
        $cubic\_tcp\_friendliness()$
    }
}

$cubic\_tcp\_friendliness():$ \\
\nonl \Begin{
    $W_{tcp} \longleftarrow W_{tcp}+\frac{3\beta}{2-\beta}*\frac{ack\_cnt}{cwnd}$ \\
    $ack\_cnt \longleftarrow 0$ \\
    \If{$W_{tcp} > cwnd$}
    {
        $max\_cnt \longleftarrow \frac{cwnd}{W_{tcp}-cwnd}$ \\
        \nosemic \lIf{$cnt > max\_cnt$}
        {
            $cnt \longleftarrow max\_cnt$
        }
    }
}
$cubic\_reset():$ \\
\nonl \Begin{
    $W_{last\_max} \longleftarrow 0, \,\, epoch\_start \longleftarrow 0, \,\, origin\_point \longleftarrow 0$ \\
    $dMin \longleftarrow 0, \,\, W_{tcp} \longleftarrow 0, \,\, K \longleftarrow 0, \,\, ack\_cnt \longleftarrow 0$
}
\end{algorithm}
\AlgoResetCount
\BlankLine
The theoretical aspects of \textit{Cubic} congestion control algorithm are provided in \ref{subsub:cubic}.
\BlankLine
\BlankLine



\SetKwInput{KwInit}{Initialization}
\SetKwComment{Comment}{/* }
\SetAlgoLongEnd
\begin{algorithm}[H]
\caption{Coupled (LIA) - Congestion Avoidance phase (AIMD)}\label{alg:Coupled}

\KwInput{Set of subflows: $\mathcal{S}$,\, set of paths: $\mathcal{R}$,\, $sf_{i} \in \mathcal{S},\, i \in \mathcal{R},\, cwnd_i,\, cwnd_{total},\, RTT_i,\, MSS_i$}
\KwOutput{$cwnd_{i}$}

\For {$sf_{i} \in \mathcal{S}$}
{
    \nonl \textbf{On each ACK received on} $sf_i$: \\
        \Indp
        $\alpha \longleftarrow cwnd_{total}\, *\, \frac{max(cwnd_i / RTT_i^2)}{(\sum_{i}{(cwnd_i/RTT_i)})^2}$ \\
        \Indm
        \BlankLine
        \Indp
        $cwnd_i\, \longleftarrow cwnd_i + \min(\frac{\alpha\, *\, bytes\_{acked}\, *\, MSS_i }{cwnd_{total}},\, \frac{bytes\_{acked}\, *\, MSS_i}{cwnd_i}) $ \\
        \Indm
        \BlankLine
        \nosemic \Comment*[r]{\parbox[t]{2.6in}{\raggedright in the above min(A, B) computation, A: denotes the computed increase of the multipath subflow, B: denotes the increase TCP would get on the same path i{ */}}  }

    \BlankLine
    \nonl \textbf{On packet loss event on} $sf_i$: \\
    \Indp
    $cwnd_i \longleftarrow \frac{cwnd_i}{2}$ \\
    $fast\_retransmit();$ \nosemic \Comment*[r]{\parbox[t]{2.6in}{\raggedright legacy TCP NewReno { */}}}
    $fast\_recovery();$ \\
    \Indm
}
\textbf{return} $cwnd_i$
\end{algorithm}
\BlankLine
The theoretical aspects of \textit{Coupled (LIA)} congestion control algorithm are provided in \ref{subsub:lia}.
\BlankLine
\BlankLine



\SetKwInput{KwInit}{Initialization}
\SetKwComment{Comment}{/* }
\SetAlgoLongEnd
\begin{algorithm}[H]
\caption{Opportunistic Linked-Increases Algorithm (OLIA) - Congestion Avoidance phase (AIMD)}\label{alg:OLIA}

\KwInput{\\ 
    \Indp
        set of paths available to user u:\, ${R}_u$, \, $r \in R_u$ \nosemic \Comment*[r]{\parbox[t]{2.6in}{\raggedright $R_u$:  the set of \textbf{all\_paths}, $|R_u|$: the number of paths available to user $u$ at time $t$ { */}}}
        
        \BlankLine
        \nonl set of paths with the largest window sizes:\, $\mathcal{M}$, \, \nosemic \Comment*[r]{\parbox[t]{2.6in}{\raggedright $\mathcal M$:  the set of \textbf{max\_w\_paths} of user $u$ at time $t$, $|\mathcal M|$: the number of paths in $\mathcal M$ { */}}}

        \BlankLine
        \nonl set of presumably best paths:\, $\mathcal{B}$, \, \nosemic \Comment*[r]{\parbox[t]{2.6in}{\raggedright $\mathcal B$:  the set of presumably \textbf{best\_paths} for user $u$ at time $t$, $|\mathcal B|$: the number of paths in $\mathcal B$ { */}}}

        \BlankLine
        \nonl set of collected paths:\, $\mathcal{B} \textbackslash \mathcal{M}$, \, \nosemic \Comment*[r]{\parbox[t]{2.6in}{\raggedright $\mathcal B \textbackslash \mathcal{M}$:  the set of \textbf{collected\_paths} for user $u$ at time $t$ -- all paths belonging to best\_paths, but not among those with largest $cwnd$ { */}}}

        \BlankLine
        \nonl congestion window of path $r$:\, $w_r$

        \BlankLine
        \nonl round-trip time of path $r$:\, $rtt_r$

        \BlankLine
        \nonl a subflow established over path $r$:\, $sf_r$ 

        \BlankLine
        \nonl set of subflows established:\, $S$,\, where\, $sf_r \in S$

        \BlankLine
    \Indm
}

\end{algorithm}
\newpage
\AlgoNoResetCount
\begin{algorithm}[h]

\KwOutput{$w_r$}

\For {$sf_{r} \in S$}
{
    \nonl \textbf{On each ACK received on} $sf_r$: \\
        \Indp
            $a_r = 
            \begin{cases}
            \frac{1/|R_u|}{|\mathcal B \textbackslash \mathcal M|}, & \text{if} \,\,\, r \in \mathcal B \textbackslash \mathcal M \ne \emptyset \\
            - \frac{1/|R_u|}{|\mathcal M|}, & \text{if} \,\,\, r \in \mathcal M \,\, \text{and} \,\, \mathcal B \textbackslash \mathcal M \ne \emptyset \\
            0, & \text{otherwise.}
            \end{cases}
            $  
            \nosemic \Comment*[r]{\parbox[t]{2.6in}{\raggedright calculate the increase parameter $a_r$ 
            // also note that $\displaystyle \sum_{r \in R_u} a_r=0 $ { */}}}

            \BlankLine
            $w_r \longleftarrow w_r + \left ( \frac{w_r/rtt_r^2}{\left (\displaystyle \sum_{p \in R_u} w_p/rtt_p \right)^2} + \frac{a_r}{w_r} \right ) 
            $  
            \nosemic \Comment*[r]{\parbox[t]{2.6in}{\raggedright window increase measured in packets // multiply by $(MSS_r \,*\, bytes\_acked) $ to measure in bytes { */}}}

            \BlankLine
        \Indm

    \BlankLine
    \nonl \textbf{On packet loss event on} $sf_r$: \\
    \Indp
        $w_r \longleftarrow \frac{w_r}{2}$ \\
        $fast\_retransmit();$ \nosemic \Comment*[r]{\parbox[t]{2.6in}{\raggedright legacy TCP NewReno { */}}}
        $fast\_recovery();$ \\
    \Indm
}
\textbf{return} $w_r$
\end{algorithm}
\AlgoResetCount
\BlankLine
The theoretical aspects of \textit{OLIA} congestion control algorithm are provided in \ref{subsub:olia}.
\BlankLine
\BlankLine



\SetKwInput{KwInit}{Initialization}
\SetKwComment{Comment}{/* }
\SetAlgoLongEnd
\begin{algorithm}[h]
\caption{Balanced Linked Adaptation (BALIA) - Congestion Avoidance phase (AIMD)}\label{alg:BALIA}

\KwInput{\\ 
    \Indp
        set of available paths:\, ${R}$, \, $r \in R$

        \BlankLine
        \nonl congestion window of path $r$:\, $w_r$

        \BlankLine
        \nonl round-trip time of path $r$:\, $rtt_r$

        \BlankLine
        \nonl a subflow established over path $r$:\, $sf_r$ 

        \BlankLine
        \nonl set of subflows established:\, $S$,\, where\, $sf_r \in S$

        \BlankLine
    \Indm
}
\KwOutput{$w_r$}
\For {$sf_{r} \in S$}
{
    \nonl \textbf{On each ACK received on} $sf_r$: \\
        \Indp
            $a_r \coloneq \frac{\max\{x_k\}}{x_r} $
            \nosemic \Comment*[r]{\parbox[t]{2.6in}{\raggedright calculate the increase parameter $a_r$ // $x_r \coloneq \frac{w_r}{rtt_r}$
            // also note that in case of single path $a_r=1 $ { */}}}

            \BlankLine
            $w_r \longleftarrow w_r + \frac{x_r}{rtt_r \left (\sum x_k \right)^2} \left (\frac{1\, + \, a_r}{2} \right) \left (\frac{4\, + \, a_r}{5} \right)
            $ 
            \BlankLine
        \Indm

    \nonl \textbf{On packet loss event on} $sf_r$: \\
    \Indp
        $w_r \longleftarrow w_r \, - \, \frac{w_r}{2} \min\{a_r, 1.5\} $ \\
        $fast\_retransmit();$ \nosemic \Comment*[r]{\parbox[t]{2.6in}{\raggedright legacy TCP NewReno { */}}}
        $fast\_recovery();$ \\
    \Indm
}
\textbf{return} $w_r$
\end{algorithm}
\BlankLine
The theoretical aspects of \textit{BALIA} congestion control algorithm are provided in \ref{subsub:balia}.
\BlankLine



\makeatletter
\long\def\KwInitBlock#1{%
  \noindent\KwSty{Initialization:}~#1%
}
\makeatother

\SetKwInput{KwInit}{Initialization}
\SetKwComment{Comment}{/* }{ */}
\RestyleAlgo{algoruled}
\SetAlgoVlined
\SetAlgoLongEnd
\SetKwBlock{Indent}{}{}
\begin{algorithm}[h]
\caption{Weighted Vegas (wVegas)}\label{alg:wVegas}

\KwInitBlock{
    \Indent{
        \nl $total\_alpha \gets 10$ \tcc*[r]{namely $\alpha_s$: preconfigured parameter coupling subflows of flow $s$}
        \For {$r \in R_s$}
        {
            $alpha[r] \gets 2$ \;
            $equilibrium\_rates[r] \gets 0$ \;
            $queue\_delays[r] \gets 0$ \;
        }
    }
}
\BlankLine
\nonl \text{\textbf{On the end of round for subflow $r$:}}
    \Indent{
        \nosemic \Comment*[l]{average RTT estimated in the last round - used instead of smoothed RTT for faster reaction to congestion}
        \dosemic
        $rtt \gets sampled\_rtts[r] / sampled\_num[r] $ \;
        $diff \gets cwnd[r] \times (rtt - baseRTT[r])/rtt$ \;
        \nosemic \Comment*[l]{tweak weights and alphas}
        \dosemic
        \If{$diff \ge alpha[r]$}
        {
            $equilibrium\_rates[r] \gets cwnd[r]/rtt$ \;
            \text{Adjust\_Weights()} \;
            $alpha[r] \gets weights[r] \times total\_alpha$ \;
            $alpha[r] \gets \max\{2, alpha[r]\}$ \tcp*[l]{lower bound}
        }
        \nosemic \Comment*[l]{window adjustment}
        \dosemic
        \If{$diff < alpha[r]$}
        {
            $cwnd[r] \gets cwnd[r] + 1$ \;
        }
        \ElseIf{$diff > alpha[r]$}
        {
            $cwnd[r] \gets cwnd[r] - 1$ \;
        }

        \nosemic \Comment*[l]{try to drain link queues if needed}
        \dosemic
        $q \gets rtt-baseRTT[r]$ \tcp*[l]{current queuing delay}
        \If{$queue\_delays[r]=0$ \textbf{\upshape{or}} $queue\_delays[r] > q$}
        { 
            $queue\_delays[r] \gets q$ \;
        }
        \If(\nosemic \tcc*[f]{cwnd backoff once queuing delay exceeds threshold}){$q \ge 2 \times queue\_delays[r]$}
        {
            $backoff\_factor \gets 0.5 \times baseRTT[r]/rtt$ \;
            $cwnd[r] \gets cwnd[r] \times backoff\_factor$ \; 
            $queue\_delays[r] \gets 0$ \;
        }
        $cwnd[r] \gets \max\{2,cwnd[r]\}$ \tcp*[l]{lower bound}
    }
\BlankLine
\nonl \text{\textbf{Adjust\_Weights():}}
    \Indent{
        $total\_rate \gets \sum equilibrium\_rates$ \;
        \For {$r \in R_s$}
        {
            \If{$equilibrium\_rates[r] \ne 0$}
            {
                $weights[r] \gets equilibrium\_rates[r] / total\_rate$ \;
            }
        }
    }
\BlankLine
\nonl \text{\textbf{On packet loss for subflow $r$:}}
    \Indent{
        $equilibrium\_rates[r] \gets 0$ \;
        $queue\_delays[r] \gets 0$ \;
    }
\end{algorithm}
\BlankLine
The theoretical aspects of \textit{wVegas} congestion control algorithm are provided in \ref{subsub:wvegas}.



\makeatletter
\long\def\KwInitBlock#1{%
  \noindent\KwSty{Initialization:}~#1%
}
\makeatother

\SetKwInput{KwInit}{Initialization}
\SetKwComment{Comment}{/* }{ */}
\RestyleAlgo{algoruled}
\SetAlgoVlined
\SetAlgoLongEnd
\SetKwBlock{Indent}{}{}
\begin{algorithm}[htb!]
\caption{Bottleneck Bandwidth and Round-trip propagation time (BBR) - [theoretical part in \ref{subsub:bbr}]}\label{alg:BBRv1}

\KwInitBlock{
    \Indent{
        BBR\_On\_Connection\_Init():
        \Indent{BBR\_Init()\;}
    }
}

\nonl \text{\textbf{On ACK arrival:}}
\Indent{
    BBR\_Update\_On\_ACK():
    \Indent{
        $BBR\_Update\_Model\_And\_State()$ \;
        $BBR\_Update\_Control\_Parameters()$ \;
    }
    \BlankLine
    BBR\_Update\_Model\_And\_State():
    \Indent{
        $BBR\_Update\_BtlBw()$ \;
        $BBR\_Check\_Cycle\_Phase()$ \;
        $BBR\_Check\_Full\_Pipe()$ \;
        $BBR\_Check\_Drain()$ \;
        $BBR\_Update\_RTprop()$ \;
        $BBR\_Check\_Probe\_RTT()$ \;
    }
    \BlankLine
    BBR\_Update\_Control\_Parameters():
    \Indent{
        $BBR\_Set\_Pacing\_Rate()$ \;
        $BBR\_Set\_Send\_Quantum()$ \;
        $BBR\_Set\_Cwnd()$ \;
    }
}

\nonl \text{\textbf{On packet transmission:}}
\Indent{
    BBR\_On\_Transmit():
    \Indent{
        $BBR\_Handle\_Restart\_From\_Idle()$ \;
    }
}

\nonl \text{\textbf{Initialization function definitions:}}
\Indent{
    BBR\_Init():  
    \Indent{
           $init\_windowed\_max\_filter(filter=BBR.BtlBwFilter,\, value=0,\, time=0)$ \;
           $BBR.rtprop = SRTT\,\, ?\,\, SRTT\, :\, Inf$ \;
           $BBR.rtprop\_stamp = Now()$ \;
           $BBR.probe\_rtt\_done\_stamp = 0$ \;
           $BBR.probe\_rtt\_round\_done = false$ \;
           $BBR.packet\_conservation = false$ \;
           $BBR.prior\_cwnd = 0$ \;
           $BBR.idle\_restart = false$ \;
           $BBR\_Init\_Round\_Counting()$ \;
           $BBR\_Init\_Full\_Pipe()$ \;
           $BBR\_Init\_Pacing\_Rate()$ \;
           $BBR\_Enter\_Startup()$ \;
    }
    BBR\_Init\_Round\_Counting():
    \Indent{
           $BBR.next\_round\_delivered = 0$ \;
           $BBR.round\_start = false$ \;
           $BBR.round\_count = 0$ \;
    }
}
\end{algorithm}

\newpage
\AlgoNoResetCount
\begin{algorithm}[htb!]

\nonl \text{\textbf{Initialization function definitions (cont.):}}
\Indent{
    BBR\_Init\_Full\_Pipe():
    \Indent{
           $BBR.filled\_pipe = false$ \;
           $BBR.full\_bw = 0$ \;
           $BBR.full\_bw\_count = 0$ \;
    }
    BBR\_Init\_Pacing\_Rate():
    \Indent{
           $nominal\_bandwidth = InitialCwnd / (SRTT\,\, ?\,\, SRTT\, :\, 1ms)$ \;
           $BBR.pacing\_rate =  BBR.pacing\_gain\, *\, nominal\_bandwidth$ \;
    }    
    BBR\_Enter\_Startup():
    \Indent{
           $BBR.state = Startup$ \;
           $BBR.pacing\_gain = BBRHighGain$ \;
           $BBR.cwnd\_gain = BBRHighGain$ \;
    }
}

\nonl \text{\textbf{ACK-arrival function definitions (1/4):}}
\Indent{
    BBR\_Update\_BtlBw():
    \Indent{
       $BBR\_Update\_Round()$ \;
       \If{$rs.delivery\_rate \ge BBR.BtlBw$\, \textbf{\upshape{or}}\, \textbf{\upshape{not}} $rs.is\_app\_limited$}
        {
           $BBR.BtlBw = update\_windowed\_max\_filter(
                         filter=BBR.BtlBwFilter,
                         value=rs.delivery\_rate,$ \\ 
                         \nonl $time=BBR.round\_count,
                         window\_length=BtlBwFilterLen) $ \;
        }
    }
    
    BBR\_Update\_Round():
    \Indent{
        $BBR.delivered += packet.size$ \;
        \If{$packet.delivered \ge BBR.next\_round\_delivered$}
        {
            $BBR.next\_round\_delivered = BBR.delivered$ \;
            $BBR.round\_count++$ \;
            $BBR.round\_start = true$ \;
        }
        \Else  
        {
            $BBR.round\_start = false$ \;
        }
    }

    BBR\_Check\_Cycle\_Phase():
    \Indent{
        \If{$BBR.sate == ProbeBW$ \,\textbf{\upshape{and}}\, $BBR\_Is\_Next\_Cycle\_Phase()$}
        {
            $BBR\_Advance\_Cycle\_Phase()$ \;
        }
    }

    BBR\_Is\_Next\_Cycle\_Phase():
    \Indent{
        $is\_full\_length = (Now() - BBR.cycle\_stamp) > BBR.RTprop$\;
        \If{$BBR.pacing\_gain == 1$}
        {
            \textbf{return} $is\_full\_length$ \;
        }
        \If{$BBR.pacing\_gain > 1$}
        {
            \textbf{return} $is\_full\_length$\, \textbf{and}\, $(packets\_lost > 0$\, \textbf{or}\, $prior\_inflight \ge BBR\_Inflight(BBR.pacing\_gain))$ \;
        }
        \Else (\nosemic \tcc*[f]{(BBR.pacing\_gain < 1)})
        {
            \dosemic \textbf{return} $is\_full\_length$\, \textbf{or}\, $prior\_inflight \le BBR\_Inflight(1)$ \;
        }
    }
}
\end{algorithm}

\newpage
\AlgoNoResetCount
\begin{algorithm}[h]

\nonl \text{\textbf{ACK-arrival function definitions (cont. 2/4):}}
\Indent{
    BBR\_Advance\_Cycle\_Phase():
    \Indent{
        $BBR.cycle\_stamp = Now()$ \;
        $BBR.cycle\_index = (BBR.cycle\_index + 1)\,\, \% \,\, BBRGainCycleLen$\;
        $pacing\_gain\_cycle = [5/4, 3/4, 1, 1, 1, 1, 1, 1]$ \;
        $BBR.pacing\_gain = pacing\_gain\_cycle[BBR.cycle\_index]$ \;
    }

    BBR\_Check\_Full\_Pipe():
    \Indent{
        \If{$BBR.filled\_pipe$\, \textbf{\upshape{or}}\, \textbf{\upshape{not}}\,$BBR.round\_start$\, \textbf{\upshape{or}}\, $rs.is\_app\_limited$}
        {
            \textbf{return}  \tcc*[r]{no need to check for a full pipe now}
        }
        \If(\tcc*[f]{BBR.BtlBw still growing?}){$BBR.BtlBw \ge BBR.full\_bw \,*\, 1.25$}
        {
            $BBR.full\_bw = BBR.BtlBw$ \tcc*[r]{record new baseline level}
            $BBR.full\_bw\_count = 0$ \;
            \textbf{return} \;
        }
        $BBR.full\_bw\_count++$ \tcc*[r]{another round w/o much growth}
        \If{$BBR.full\_bw\_count \ge 3$}
        {
            $BBR.filled\_pipe = true$ \;
        }
    }

    BBR\_Check\_Drain():
    \Indent{
        \If{$BBR.state == Startup$  \,\textbf{\upshape{and}}\, $BBR.filled\_pipe$}
        {
            $BBR\_Enter\_Drain()$ \;
        }
        \If{$BBR.state == Drain$ \,\textbf{\upshape{and}}\, $packets\_in\_flight \le BBR\_Inflight(1.0)$}
        {
            $BBR\_Enter\_ProbeBW()$ \tcc*[r]{we estimate queue is drained}
        }
    }

    BBR\_Update\_RTprop():
    \Indent{
        $BBR.rtprop\_expired = Now() > BBR.rtprop\_stamp + RTpropFilterLen$ \;
        \If{$packet.rtt \ge 0$ \,\textbf{\upshape{and}}\, $(packet.rtt \le BBR.RTprop$ \,\textbf{\upshape{or}}\, $BBR.rtprop\_expired)$}
        {
            $BBR.RTprop = packet.rtt$ \;
            $BBR.rtprop\_stamp = Now()$ \;
        }
    }

    BBR\_Check\_Probe\_RTT():
    \Indent{
        \If{$BBR.state \,!=\, ProbeRTT$ \,\textbf{\upshape{and}}\, $BBR.rtprop\_expired$ \,\textbf{\upshape{and}}\, \textbf{\upshape{not}} $BBR.idle\_restart$}
        {
            $BBR\_Enter\_Probe\_RTT()$ \; 
            $BBR\_Save\_Cwnd()$ \;
            $BBR.probe\_rtt\_done\_stamp = 0$ \;
            \If{$BBR.state == ProbeRTT$}
            {
                $BBR\_Handle\_Probe\_RTT()$ \;
            }
            $BBR.idle\_restart = false$ \;
        }
    }

    BBR\_Enter\_Probe\_RTT():
    \Indent{
        $BBR.state = ProbeRTT$ \;
        $BBR.pacing\_gain = 1$ \;
        $BBR.cwnd\_gain = 1$ \;
    }

}

\end{algorithm}
\newpage
\AlgoNoResetCount
\begin{algorithm}[h]

\nonl \text{\textbf{ACK-arrival function definitions (cont. 3/4):}}
\Indent{
    BBR\_Handle\_Probe\_RTT():
    \Indent{
        \nosemic \tcc*[l]{Ignore low rate samples during ProbeRTT:}
        \dosemic
        $C.app\_limited = (BW.delivered + packets\_in\_flight) \,?\, :\, 1$ \;
        \If{$BBR.probe\_rtt\_done\_stamp == 0$ \,\textbf{\upshape{and}}\, $packets\_in\_flight \le BBRMinPipeCwnd$}
        {
            $BBR.probe\_rtt\_done\_stamp = Now() + ProbeRTTDuration$ \;
            $BBR.probe\_rtt\_round\_done = false$ \;
            $BBR.next\_round\_delivered = BBR.delivered$ \;
        }
        \ElseIf{$BBR.probe\_rtt\_done\_stamp \, != \, 0$}
        {
            \If{$BBR.round\_start$}
            {
                $BBR.probe\_rtt\_round\_done = true$ \;
            }
            \If{$BBR.probe\_rtt\_round\_done$  \,\textbf{\upshape{and}}\, $Now() > BBR.probe\_rtt\_done\_stamp$}
            {
                $BBR.rtprop\_stamp = Now()$ \;
                $BBR\_Restore\_Cwnd()$ \;
                $BBR\_Exit\_Probe\_RTT()$ \;
            }
        }
    }

    BBR\_Exit\_Probe\_RTT():
    \Indent{
        \If{$BBR.filled\_pipe$}
        {
            $BBR\_Enter\_Probe\_BW()$ \;
        }
        \Else
        {
            $BBR\_Enter\_Startup()$ \;
        }
    }

    BBR\_Enter\_Probe\_BW():
    \Indent{
        $BBR.state = ProbeBW$ \;
        $BBR.pacing\_gain = 1$ \;
        $BBR.cwnd\_gain = 2$ \;
        $BBR.cycle\_index = BBRGainCycleLen - 1 - random\_int\_in\_range(0..6)$ \;
        $BBR\_Advance\_Cycle\_Phase()$ \;
    }

    BBR\_Set\_Pacing\_Rate():
    \Indent{
        $BBR\_Set\_Pacing\_Rate\_With\_Gain(BBR.pacing\_gain)$ \;
    }

    BBR\_Set\_Pacing\_Rate\_With\_Gain(pacing\_gain):
    \Indent{
        $rate = pacing\_gain \,*\, BBR.BtlBw$ \;
        \If{$BBR.filled\_pipe$ \,\textbf{\upshape{or}}\, $rate > BBR.pacing\_rate$}
        {
            $BBR.pacing\_rate = rate$ \;
        }
    }

    BBR\_Set\_Send\_Quantum():
    \Indent{
        \If{$BBR.pacing\_rate < 1.2 \,Mbps$}
        {
            $BBR.send\_quantum = 1 \,* MSS$ \;
        }
        \ElseIf{$BBR.pacing\_rate < 24 \,Mbps$}
        {
            $BBR.send\_quantum  = 2 \,*\, MSS$ \;
        }
        \Else
        {
            $BBR.send\_quantum  = min(BBR.pacing\_rate \,*\, 1ms,\, 64KBytes)$ \;
        }
    }
}
\end{algorithm}

\newpage
\AlgoNoResetCount
\begin{algorithm}[h]
\nonl \text{\textbf{ACK-arrival function definitions (cont. 4/4):}}
\Indent{

    BBR\_Set\_Cwnd():
    \Indent{
        $BBR\_Update\_Target\_Cwnd()$ \;
        $BBR\_Modulate\_Cwnd\_For\_Recovery()$ \;
        \If{\textbf{\upshape{not}}\, $BBR.packet\_conservation$}
        {
            \If{$BBR.filled\_pipe$}
            {
                $cwnd = min(cwnd + packets\_delivered, BBR.target\_cwnd)$ \;
            }
            \ElseIf{$cwnd < BBR.target\_cwnd$ \,\textbf{\upshape{or}}\, $BBR.delivered < InitialCwnd$}
            {
                $cwnd = cwnd + packets\_delivered$ \;
            }
            $cwnd = max(cwnd, BBRMinPipeCwnd)$ \;
        }
        $BBR\_Modulate\_Cwnd\_For\_Probe\_RTT()$ \;
    }

    BBR\_Update\_Target\_Cwnd():
    \Indent{
        $BBR.target\_cwnd = BBR\_Inflight(BBR.cwnd\_gain)$ \;
    }

    BBR\_Inflight(gain):
    \Indent{
        \If{$BBR.RTprop == Inf$}
        {
            \textbf{return} $InitialCwnd$  \tcc*[r]{no valid RTT samples yet}
        }
        $quanta = 3 \,*\, BBR.send\_quantum$ \;
        $estimated\_bdp = BBR.BtlBw \,*\, BBR.RTprop$ \;
        \textbf{return} $gain \,*\, estimated\_bdp + quanta$ \;
    }

    BBR\_Modulate\_Cwnd\_For\_Recovery():
    \Indent{
        \If{$packets\_lost > 0$}
        {
            $cwnd = max(cwnd - packets\_lost, 1)$ \;
        }
        \If{$BBR.packet\_conservation$}
        {
            $cwnd = max(cwnd, packets\_in\_flight + packets\_delivered)$ \;
        }
    }

    BBR\_Modulate\_Cwnd\_For\_ProbeRTT():
    \Indent{
        \If{$BBR.state == ProbeRTT$}
        {
            $cwnd = min(cwnd, BBRMinPipeCwnd)$ \;
        }
    }
}

\nonl \text{\textbf{Packet-transmission function definitions:}}
\Indent{
    BBR\_Handle\_Restart\_From\_Idle():
    \Indent{
        \If{$packets\_in\_flight == 0$  \,\textbf{\upshape{and}}\, $C.app\_limited$}
        {
            $BBR.idle\_start = true$ \;
            \If{$BBR.state == ProbeBW$}
            {
                $BBR\_Set\_Pacing\_Rate\_With\_Gain(1)$ \;
            }
        }
    }
}
\nonl \text{\textbf{Save and Restore cwnd function definitions:}}
\Indent{
    BBR\_Save\_Cwnd():
    \Indent{
        \If{\textbf{\upshape{not}}\, $In\_Loss\_Recovery()$ \,\textbf{\upshape{and}}\, $BBR.state != ProbeRTT$}
        {
            \textbf{return} $cwnd$ \;
        }
        \Else
        {
            \textbf{return} $max(BBR.prior\_cwnd, cwnd)$ \;
        }
    }

    BBR\_Restore\_Cwnd():
    \Indent{
        $cwnd = max(cwnd, BBR.prior\_cwnd)$ \;
    }
}
\end{algorithm}
\AlgoResetCount




\makeatletter
\long\def\KwInitBlock#1{%
  \noindent\KwSty{Initialization:}~#1%
}
\makeatother

\SetKwInput{KwInit}{Initialization}
\SetKwComment{Comment}{/* }
\SetAlgoVlined
\SetAlgoLongEnd
\SetKwBlock{Indent}{}{}
\begin{algorithm}[htb]
\caption{Coupled Multipath BBR (C-MPBBR) - [theoretical part in \ref{subsub:cmpbbr}]}\label{alg:C-MPBBR}

\nonl \text{\textbf{Fulfilling Goal 1 \, (MPTCP incentive):}}
    \Indent{
        $\beta = 40$, \, $Total\_Del\_Rt = 0$ \;
        $lowest\_bw\_among\_all\_SFs = 999999999$ \;
        $highest\_bw\_among\_all\_SFs = 0$ \;
        $total\_number\_of\_SFs = 0$ \;
        \If{$cmpbbr \to mode = CMPBBR\_PROBE\_BW$ \,\textbf{\upshape{and}}\, $cmpbbr \to cycle\_index = 3$}
        {
            \For {\textbf{all} $SF_i$}
            {
                \textbf{Calculate: $Del\_Rt\_of\_SF_i$} \;
                $total\_number\_of\_SF_s++$ \;
                $Total\_Del\_Rt \,+=\, Del\_Rt\_of\_SF_i $ \;
                \If{$lowest\_bw\_among\_all\_SFs > bw\_of\_SF_i$}
                {
                    $lowest\_bw\_among\_all\_SFs = bw\_of\_SF_i$ \;
                }
                \If{$highest\_bw\_among\_all\_SFs < bw\_of\_SF_i$}
                {
                    $highest\_bw\_among\_all\_SFs = bw\_of\_SF_i$ \;
                }
            }
        }
        $threshold\_bw\_for\_stopping\_lowest\_bw\_SF = highest\_bw\_among\_all\_SFs \,*\, (1 - \frac{\beta}{100})$ \;
        \If{$threshold\_bw\_for\_stopping\_lowest\_bw\_SF > Total\_Del\_Rt$ \,\textbf{\upshape{and}}\, $cmpbbr \to last\_number\_of\_SFs\_in\_btlneck < 2$ \,\textbf{\upshape{and}}\, $total\_number\_of\_SFs > 1$ \,\textbf{\upshape{and}}\, $lowest\_bw\_among\_all\_SFs \ne highest\_bw\_among\_all\_SFs$}
        {
            $cmpbbr \to stop\_lowest\_bw\_SF\_count++$ \;
        }
        \Else
        {
            $cmpbbr \to stop\_lowest\_bw\_SF\_count = 0$ \;
        }
        \If{$cmpbbr \to stop\_lowest\_bw\_SF\_count \ge 5$ \,\textbf{\upshape{and}}\, $total\_number\_of\_SFs > 1$ \,\textbf{\upshape{and}}\, $bw\_of\_SF_{this} = lowest\_bw\_among\_all\_SFs$ }
        {
            $cmpbbr \to stop\_lowest\_bw\_SF\_count = 5$ \;
            \textbf{Close:} $SF_{this}$  \nosemic \Comment*[r]{\parbox[t]{3.6in}{\raggedright Close $SF_{this}$ whose $Total\_Del\_Rt < \beta * highest\_bw\_among\_all\_SFs$ for five successive ProbeBw states{ */}}} \dosemic
        }
    }

\nonl \text{\textbf{Fulfilling Goal 2 \, (MPTCP fairness):}}
    \Indent{
        $\alpha = 20$, \, $number\_of\_SFs\_in\_btlneck = 0$ \;
        \If{$cmpbbr \to mode = CMPBBR\_PROBE\_BW$ \,\textbf{\upshape{and}}\, $cmpbbr \to cycle\_index = 3$}
        {
            $bw\_lower\_limit = bw\_of\_SF_{this} \,*\, (1 - \frac{\alpha}{100})$ \;
            $bw\_upper\_limit = bw\_of\_SF_{this} \,*\, (1 + \frac{\alpha}{100})$ \;
            \For {\textbf{all} $SF_i$}
            {
                \If{$bw\_of\_SF_i \ge bw\_lower\_limit$ \,\textbf{\upshape{and}}\, $bw\_of\_SF_i \le bw\_upper\_limit$}
                {
                    $number\_of\_SFs\_in\_btlneck++$ \;
                }
            }
            \If{$number\_of\_SFs\_in\_btlneck > 1$ \,\textbf{\upshape{and}}\, $cmpbbr \to last\_number\_of\_SFs\_in\_btlneck > 1$}
            {
                $final\_number\_of\_SFs\_in\_btlneck = number\_of\_SFs\_in\_btlneck$ \;
            }
            \ElseIf{$number\_of\_SFs\_in\_btlneck = 1$ \,\textbf{\upshape{and}}\, $cmpbbr \to last\_number\_of\_SFs\_in\_btlneck > 1$}
            {
                $final\_number\_of\_SFs\_in\_btlneck = cmpbbr \to last\_number\_of\_SFs\_in\_btlneck$ \;
            }
            \Else
            {
                $final\_number\_of\_SFs\_in\_btlneck = 1$ \;
            }
            $cmpbbr \to last\_number\_of\_SFs\_in\_btlneck = number\_of\_SFs\_in\_btlneck$ \;
        }
        $bw\_of\_SF_{this} = bw\_of\_SF_{this} \,/\, final\_number\_of\_SFs\_in\_btlneck$ \;
    }
\end{algorithm}

\end{document}